\documentclass[a4paper,11pt]{article}
\pdfoutput=1
\usepackage{jheppub}
\usepackage{subfigure}
\usepackage{relsize}
\usepackage{graphicx}

\def  \bcen   {\begin{center}}
\def  \ecen   {\end{center}}
\def  \beq    {\begin{equation}}
\def  \eeq    {\end{equation}}
\def  \beqa   {\begin{eqnarray}}
\def  \eeqa   {\end{eqnarray}}
\def  \nn     {\nonumber }
\def\bea{\begin{eqnarray}}
\def\eea{\end{eqnarray}}

\title{Corrections for tribimaximal, bimaximal and democratic neutrino mixing matrices
}

\author[a]{Sumit K. Garg,}
\author[a]{Shivani Gupta}

\emailAdd{sumit@yonsei.ac.kr}
\emailAdd{shivani@yonsei.ac.kr}

\abstract{In this work we analyze the corrections to tribimaximal (TBM), bimaximal (BM) and democratic (DC) mixing matrices for explaining large reactor mixing angle $\theta_{13}$ 
and checking the consistency with other neutrino mixing angles. The corrections are parameterized in terms of small orthogonal rotations (R) with corresponding modified PMNS matrix of the form 
$R_{ij}\cdot U \cdot R_{kl}$ where $R_{ij}$ is rotation in ij sector and  U is 
any one of these special matrices. We showed the rotations $R_{13}\cdot U \cdot R_{23}$, 
$R_{12}\cdot U \cdot R_{13}$  for BM and $R_{13}\cdot U \cdot R_{13}$ for TBM  perturbative case successfully fit all neutrino mixing angles within $1\sigma$ range.
The perturbed PMNS matrix $R_{12}\cdot U \cdot R_{13}$ for DC, TBM and $R_{23}\cdot U \cdot R_{23}$ for TBM case is successful in producing mixing
angles at 2$\sigma$ level.  The other rotation schemes are either excluded or successful in producing mixing angles at $3\sigma$ level.} 


\begin{document}
\maketitle
\section{Introduction}
It is now well established fact from the  solar, atmospheric and reactor neutrino experiments \cite{T2K, Dayabay, Doublechooz, Minos, RENO} 
that neutrino switches flavor while traveling because of
their tiny mass and flavor mixing. The understanding of this mixing is one of much explored question in Particle physics. 
There are three type of well explored neutrino mixing patterns: tribimaximal \cite{scott}, bimaximal \cite{BM} and democratic mixing \cite{DM}. 
These special structures can originate from various discrete symmetries like $A_4$ \cite{A4}, $S_4$ \cite{S4} etc which are extensively discussed in literature.  
All these mixing scenarios have same predictions for the reactor mixing angle viz
$\theta_{13}=0$. The atmospheric mixing angle, $\theta_{23}=45^{\circ}$ for BM and TBM while for DC it takes the value 54.7$^{\circ}$.
The solar mixing angle is maximal (i.e. $45^{\circ}$) for BM and DC while takes the value of 35.3$^{\circ}$ for TBM mixing.

However, long baseline neutrino oscillation T2K experiment \cite{T2K} 
observed the events corresponding to $\nu_{\mu}\rightarrow \nu_e$ transition 
which is consistent with non zero $\theta_{13}$ in a three flavor scenario. The  value of 1-3 mixing angle consistent with data 
at 90\% CL is reported to be  in the range $ 5^\circ(5.8^\circ) < \theta_{13} < 16^\circ(17.8^\circ)$ for Normal (Inverted) neutrino mass hierarchy. These results goes
in well agreement with other oscillation experiments like Daya Bay \cite{Dayabay}, Double Chooz \cite{Doublechooz}, Minos \cite{Minos} and RENO \cite{RENO}. Moreover it is
evident from recent global fit \cite{GonzalezGarcia:2012sz} for 
neutrino masses and mixing angles (given in Table \ref{parameters}) that these mixing scenarios can't be taken at their face value and thus should be
investigated for possible perturbations. This issue has been taken up many times \cite{largeth13, models} in literature. In particular, ref.~\cite{dcpertbs} and 
ref.~\cite{tbmpertbs} discussed the perturbations that are parametrized in terms of mixing angles to DC and TBM mixing respectively for 
obtaining large $\theta_{13}$. The ref.~\cite{Chaoetal} in similar tune looked into detail the analytical correlations between neutrino mixing angles
that originated from different corrections to TBM, BM and DC mixing schemes. These modifications were parametrized in terms of orthogonal rotation matrices and
various cases were studied by invoking one or two rotations at a time in some particular sector of unperturbed PMNS matrix. 
Here we continue with similar approach and study the perturbations with corresponding modified PMNS matrix of the form $R_{ij}\cdot U \cdot R_{kl}$ where $R_{ij}$ is rotation in 
ij sector and  U is any one of these special matrices. Since the form of PMNS matrix is given by $U_{PMNS} = U_l^{\dagger} U_\nu$ so these corrections may originate from charged lepton and neutrino
sector. Here we would like to mention that CP violation can have a deep impact on these studies. However including CP violation for
the perturbations characterized by two mixing angles is a elaborative task which needs separate investigation. So we leave the discussion
of corrections coming from non zero CP phase for future investigation. Unlike previous studies \cite{dcpertbs, tbmpertbs, Chaoetal}, we used $\chi^2$  approach to investigate 
the situation of mixing angle data fitting in parameter space and thus 
capture essential information about the magnitude of corresponding correction parameters. Here we confined ourselves to small rotation case which 
in turn justify them to pronounce as perturbative corrections. 
The various numerical results are presented in terms of $\chi^2$ vs perturbation parameters and mixing angle correlations. We also looked into
few previously considered rotations \cite{Chaoetal} which were reported to be either excluded or highly unfavorable with one of the mixing angle fixed to 
its central value. We showed when all three angles vary in their permissible 3$\sigma$ ranges then these cases are still viable at 2-3$\sigma$ level. These results may help in
restricting vast number of possible models and thus provide a guideline for the neutrino model building physics. It would also be interesting to inspect the 
origin of these perturbations. However we are leaving the investigation of all such issues for future consideration.

The main outline of the paper is as follows. In section 2 we will give 
general discussion about our work. In section 3 we will present our numerical results. Finally in 
section 4 we will give the summary and conclusions of our study.

\begin{table} 

\begin{tabular}{lccc}

\hline

\hline

Parameter & Best fit & $1\sigma$ & $3\sigma$ \\

\hline

$\Delta m^2_{\odot}/10^{-5}~\mathrm{eV}^2 $ (NH or IH) & 7.50 & 7.31 -- 7.68 &7.0 -- 8.09 \\

\hline

$\Delta m^2_{A}/10^{-3}~\mathrm{eV}^2 $ (NH) & 2.47 & 2.406 -- 2.543 & 2.276 -- 2.695\\

\hline

$\theta_{12}^o$ & 33.33  &32.58 -- 34.14 & 31.09 -- 35.89\\

\hline

$\theta_{13}^o$ & 8.72  & 8.25 -- 9& 7.19 -- 9.96\\

\hline

$\theta_{23}^o$ & 40 (50)  & 38.52 -- 42.13 (49.1 -- 51.7)& 35.79 -- 54.8   \\

\hline

\label{parameters} 

\end{tabular}
\vspace{-1cm}
\begin{center}

\caption{The experimental constraints on neutrino oscillation parameters \cite{GonzalezGarcia:2012sz}.}

 \end{center}

\end{table}

\section{General Setup}

The form of mixing matrix for the three mixing scenarios under consideration is given as follows
\begin{eqnarray} \nn
U_{\rm TBM} = \left ( \begin{array}{rrr}
\sqrt{2\over 3}&\sqrt{1\over 3}&0\\
-\sqrt{{1\over 6}}&\sqrt{{1\over 3}}&\sqrt{1\over 2}\\
-\sqrt{{1\over 6}}&\sqrt{{1\over 3}}&-\sqrt{{1\over 2}}
\end{array}
\right )\; , \hspace{0.2cm}
U_{\rm BM}=\left(
\begin{array}{rrr}
\sqrt{1\over 2 } & \sqrt{1\over 2 } & 0 \\
-{1 \over 2 }& {1 \over 2 } & \sqrt{1\over 2 } \\
{1 \over 2 } &  -{ 1 \over 2 } &\sqrt{1\over 2 }
\end{array}\right) \; , \hspace{0.2cm}
U_{\rm DC} = \left ( \begin{array}{rrr}
\sqrt{\frac{1}{2}}&\sqrt{\frac{1}{2}}&0\\
\sqrt{\frac{1}{6}}&-\sqrt{\frac{1}{6}}&-\sqrt{\frac{2}{3}}\\
-\sqrt{\frac{1}{3}}&\sqrt{\frac{1}{3}}&-\sqrt{\frac{1}{3}}
\end{array}
\right )\;. \label{vtri}
\end{eqnarray}

All the scenarios under consideration predict vanishing value of $\theta_{13}=0^{\circ}$. While $\theta_{23}$ is maximal in TBM and BM scenarios,
it takes the larger value of 54.7 $^{\circ}$ for DC mixing. The value of solar mixing angle ($\theta_{12}$) is maximal in BM and DC scenarios while its
 value is $35.3^{\circ}$ for TBM case. These mixing angles are in conflict with recent experimental observations which provide best fit values at 
 $\theta_{13}=9^{\circ}$, $\theta_{12}=33.3^{\circ}$ and $\theta_{23}=40^{\circ}(50^{\circ})$.
Thus these well studied structures need corrections \cite{chrgdleptcrrs, neutrinocrrs, bthsctrcrrs} in order to be consistent with current neutrino mixing angles data.\\
In this study we consider possible modifications of the 
form $ R_Y \cdot U \cdot R_X$ where $R_X$ and $R_Y$
denote generic perturbation matrices and U is any of these special matrix. Since PMNS matrix is of the form $U_{l}^\dagger U_{\nu}$
so these corrections may originate from charged lepton and neutrino sector.
The perturbation matrices $R_X$ and $R_Y$ can be expressed in terms of mixing matrices
as $R_X(R_Y)= \{ R^{}_{23}, R^{}_{13}, R^{}_{12} \}$ in general, where
$R^{}_{23}$, $R^{}_{13}$ and $R^{}_{12}$ represent the rotations in 23, 13 and 12 sector respectively and are given by
\begin{eqnarray} \nn
&&R^{}_{12} = \left (\begin{array}{ccc}
\cos \alpha & \sin \alpha &0\\
-\sin \alpha &\cos \alpha &0\\
0&0&1
\end{array}
\right )\;,  R^{}_{23} = \left (\begin{array}{ccc}
1&0&0\\
0&\cos \beta  &\sin \beta \\
0&-\sin \beta & \cos \beta
\end{array}\right )\;, R^{}_{13} = \left ( \begin{array}{ccc}
\cos \gamma &0&\sin \gamma  \\
0&1&0\\
-\sin \gamma  &0& \cos \gamma
\end{array}
\right )\; \label{vb}
\end{eqnarray}
where $\alpha$, $\beta$, $\gamma$ denote rotation angles. The corresponding PMNS
matrix thus becomes:
\begin{eqnarray}
&& U^{TBM}_{\rm ijkl}= R_{ij}^{} \cdot U_{TBM}^{} \cdot R_{kl}^{} \; , \label{p1}\\
&& U^{BM}_{\rm ijkl}= R_{ij}^{} \cdot U_{BM}^{} \cdot R_{kl}^{} \; ,\label{p2} \\
&& U^{DC}_{\rm ijkl}= R_{ij}^{} \cdot U_{DC}^{} \cdot R_{kl}^{} \; ,\label{p3} 
\end{eqnarray}
where  $(ij), (kl) =(12), (13),
(23)$ respectively. From theoretical point of view, 
neutrino mixing matrix U comes from the mismatch between diagonalization of charged lepton and neutrino mass matrix and  is given as \\
\beq \nn
U = U_l^{\dagger} U_\nu
\eeq
where U$_l$ and U$_\nu$ are the unitary matrices that diagonalizes the charged lepton (M$_l$) and neutrino mass matrix (M$_\nu$).
Now to get neutrino mixing angles the above matrices
are compared with the standard PMNS matrix.

The three light neutrino mixing is described by the Pontecorvo, Maki, Nakagawa, Sakata (PMNS) 3 $\times$3 unitary mixing matrix \cite{upmns}
which in the standard form is given as
\begin{eqnarray}
U = \left( \begin{array}{ccc} c^{}_{12} c^{}_{13} & s^{}_{12}
c^{}_{13} & s^{}_{13} e^{-i\delta} \\ -s^{}_{12} c^{}_{23} -
c^{}_{12} s^{}_{13} s^{}_{23} e^{i\delta} & c^{}_{12} c^{}_{23} -
s^{}_{12} s^{}_{13} s^{}_{23} e^{i\delta} & c^{}_{13} s^{}_{23} \\
s^{}_{12} s^{}_{23} - c^{}_{12} s^{}_{13} c^{}_{23} e^{i\delta} &
-c^{}_{12} s^{}_{23} - s^{}_{12} s^{}_{13} c^{}_{23} e^{i\delta} &
c^{}_{13} c^{}_{23} \end{array} \right) P
,\label{standpara}
\end{eqnarray}
where $c_{ij}\equiv \cos\theta_{ij}$, $s_{ij}\equiv \sin\theta_{ij}$ and $\delta$ is the Dirac CP violating phase.
P is the diagonal phase matrix (P $\equiv diag \{1,e^{i\rho},e^{i\sigma}\}$) which comprises of two additional phases
$\rho$ and $\sigma$ and is relevant if neutrinos are the Majorana particles. The Majorana phases however do not affect the neutrino oscillations
and are not accessible to experimental scrutiny at present. Here we will consider the case where all the CP violating phases
i.e. $\delta, \rho, \sigma$ are zero. The effects  of CP violation will be discussed somewhere else.

To investigate numerically the effect of these perturbations we define
a function $\chi^2$ which is a measure of deviation from the central value of mixing angles:
\begin{equation}
   \chi^2 = \mathlarger{\mathlarger{‎‎\sum}}_{i=1}^{3‎} \{ \frac{\theta_i(P)-\theta_i}{\delta \theta_i} \}^2
\end{equation}
with $\theta_i(P)$ are the theoretical value of mixing angles which are functions  of perturbation parameters ($\alpha,\beta, \gamma$).
$\theta_i$ are the experimental value of neutrino mixing angles with corresponding $1\sigma$ uncertainty $\delta \theta_i$. For TBM, BM
and DC mixing matrix the corresponding $\chi^2$ value is 109.9, 354.5 and 413.6 respectively.

\section{Numerical Section}
In this section we will discuss the various numerical results of our study. We
investigate the role of these perturbations in producing large $\theta_{13}$ \cite{largeth13} and fitting other mixing angles.
In Figs.~\ref{fig.1}-\ref{fig.15} we present the $\chi^2$ over perturbation parameters (left figs.) and
$\theta_{13}$ over $\theta_{23}-\theta_{12}$ neutrino mixing angles (right figs.) for different studied cases.
The numerical value of correction parameters $\alpha$, $\beta$ and $\gamma$ are confined in the
range [-0.5, 0.5] so as to keep them in perturbative limits. We enforced the condition $\chi^2 < \chi^2_{i}$ (i = TBM, DC and BM) for selecting the 
data points. In plots of $\chi^2$ vs perturbation parameters ($\theta_1, \theta_2$) red, blue and light green color 
regions corresponds to $\chi^2$ value in the interval $[0, 3]$,  $[3, 10]$  and  $ > 10$ respectively. 
In figures of  mixing angles, light green band corresponds to $1\sigma$ and full color band to $3\sigma$ values
of $\theta_{13}$. A good mixing angle fit prefers much lower value of $\chi^2$. In order to see the correlations between left and right figures we marked 
the $\chi^2 < 3, [3, 10]$ regions 
in mixing angle plots. The white region corresponds to $ 3 < \chi^2 < 10 $ while yellow region belongs to $\chi^2 < 3$. 
Horizontal and vertical dashed and black lines corresponds to $1\sigma$ and $3\sigma$ ranges of the other two mixing angles. 
Here we also give the approximate form of mixing angles for various possible cases in order to understand the numerical results
in small rotation limit. However we used full analytical results in our numerical investigation. Now we will take up the case of various possible 
forms of rotations one by one:

\section{Rotations-$R_{ij}.U.R_{kl}(ij \neq kl)$}

Here we first consider the perturbations for which $ij \neq kl$ and investigate their role in fitting the neutrino
mixing data. 

\subsection{12-13 Rotation}

This case corresponds to rotations in 12 and 13 sector of  these special matrices. 
Since for small rotation $\sin\theta \approx \theta$ and $\cos\theta \approx 1-\theta^2$, so the neutrino mixing angles
truncated at order O ($\theta^2$) for these rotations are given by

\beqa
 \sin\theta_{13} &\approx&  |\alpha U_{23} + \gamma U_{11} + \alpha \gamma U_{21} |,\\
 \sin\theta_{23} &\approx& |\frac{ U_{23} + \gamma U_{21}-(\alpha^2 + \gamma^2) U_{23} -\alpha\gamma U_{11} }{\cos\theta_{13}}|,\\
 \sin\theta_{12} &\approx& |\frac{U_{12} +\alpha U_{22} -\alpha^2 U_{12}}{\cos\theta_{13}}|.
\eeqa

\begin{figure}[!t]\centering
\begin{tabular}{c c} 
\includegraphics[angle=0,width=80mm]{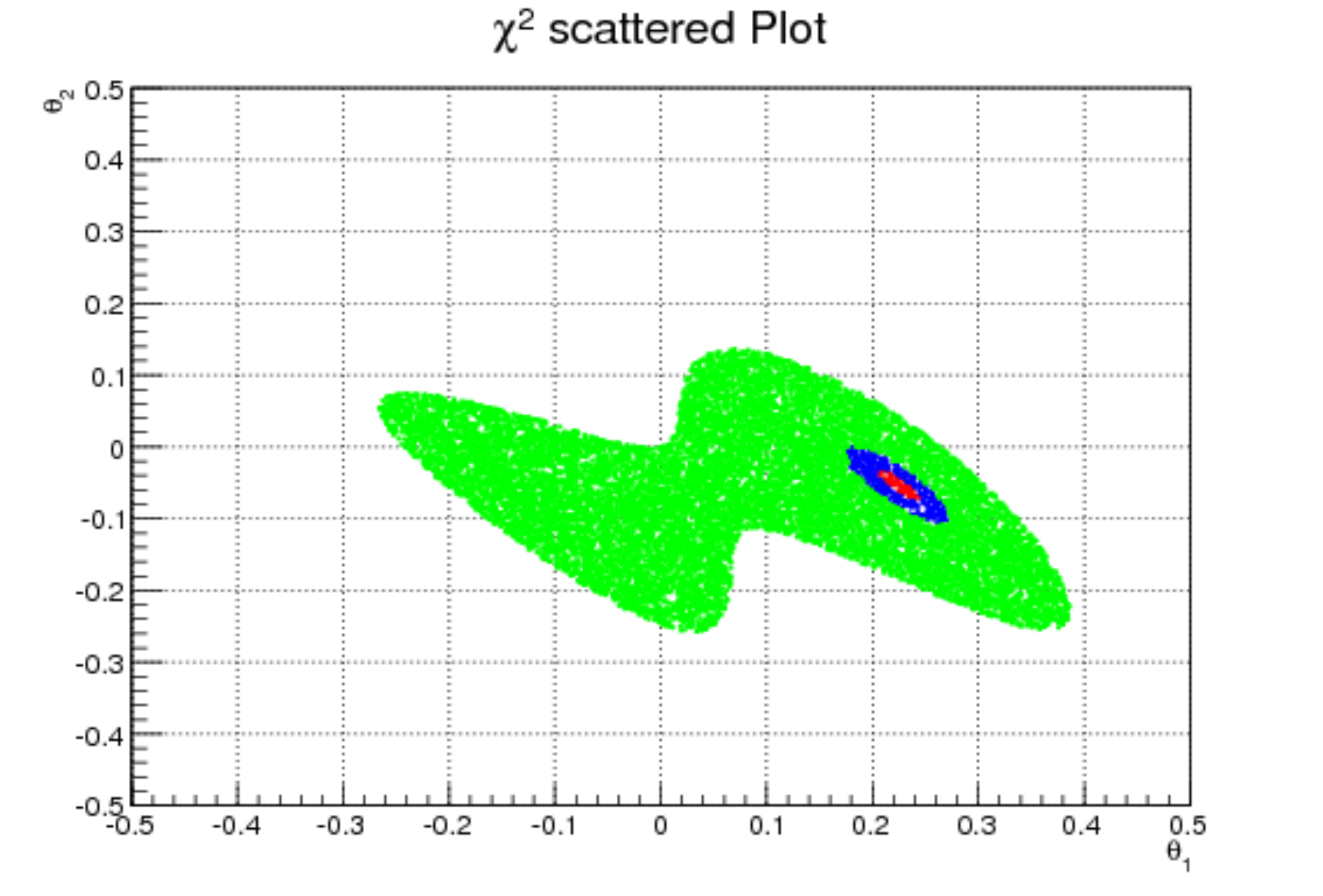} &
\includegraphics[angle=0,width=80mm]{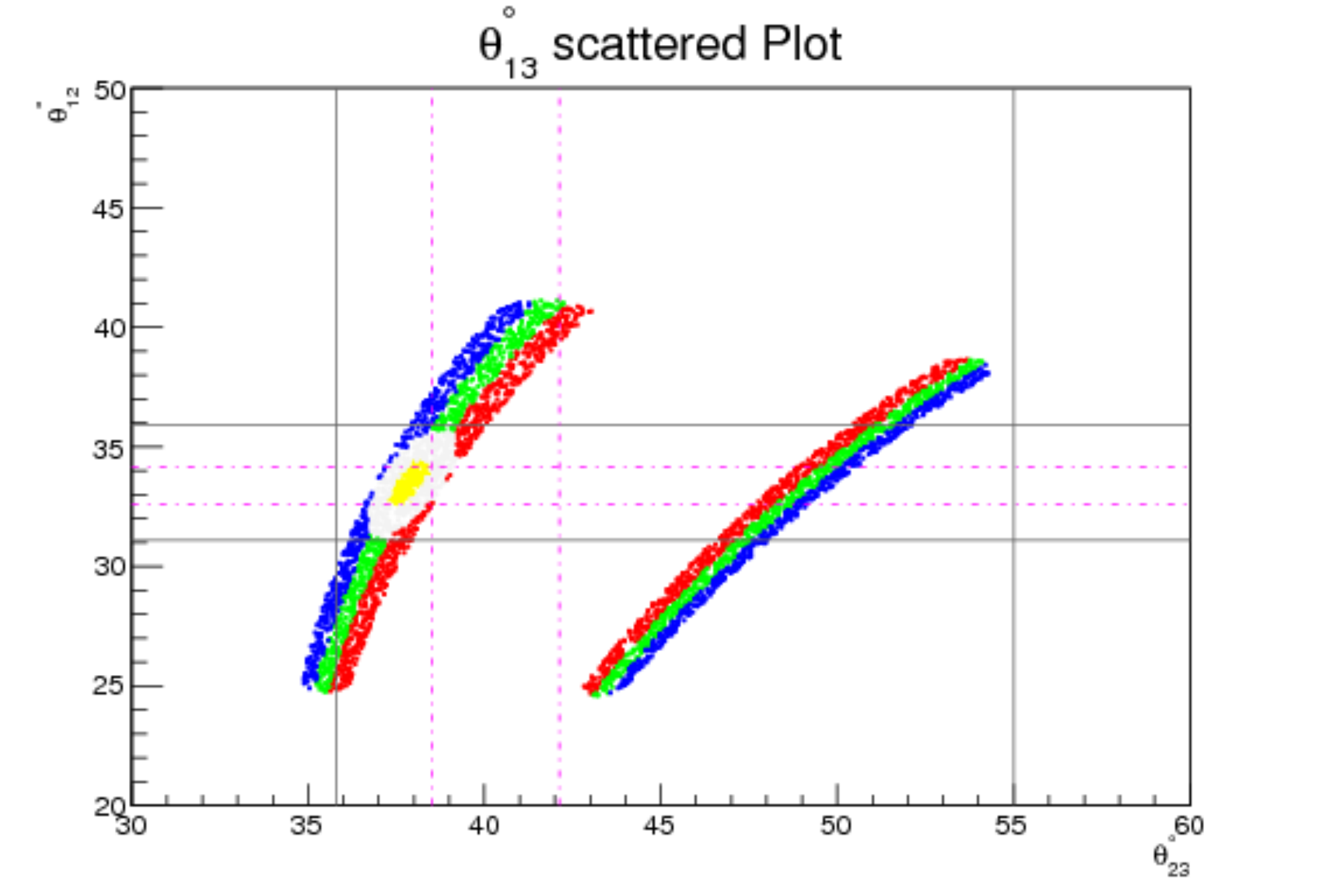}\\
\end{tabular}
\caption{$U_{TBM}^{1213}$ scatter plot of $\chi^2$ (left fig.) over $\alpha-\gamma$ (in radians) plane and $\theta_{13}$ (right fig.) 
over $\theta_{23}-\theta_{12}$ (in degrees) plane. The information about other color coding and various horizontal, vertical lines in right fig. is given in text. }
\label{fig.1}
\end{figure}

\begin{figure}[!t]\centering
\begin{tabular}{c c} 
\includegraphics[angle=0,width=80mm]{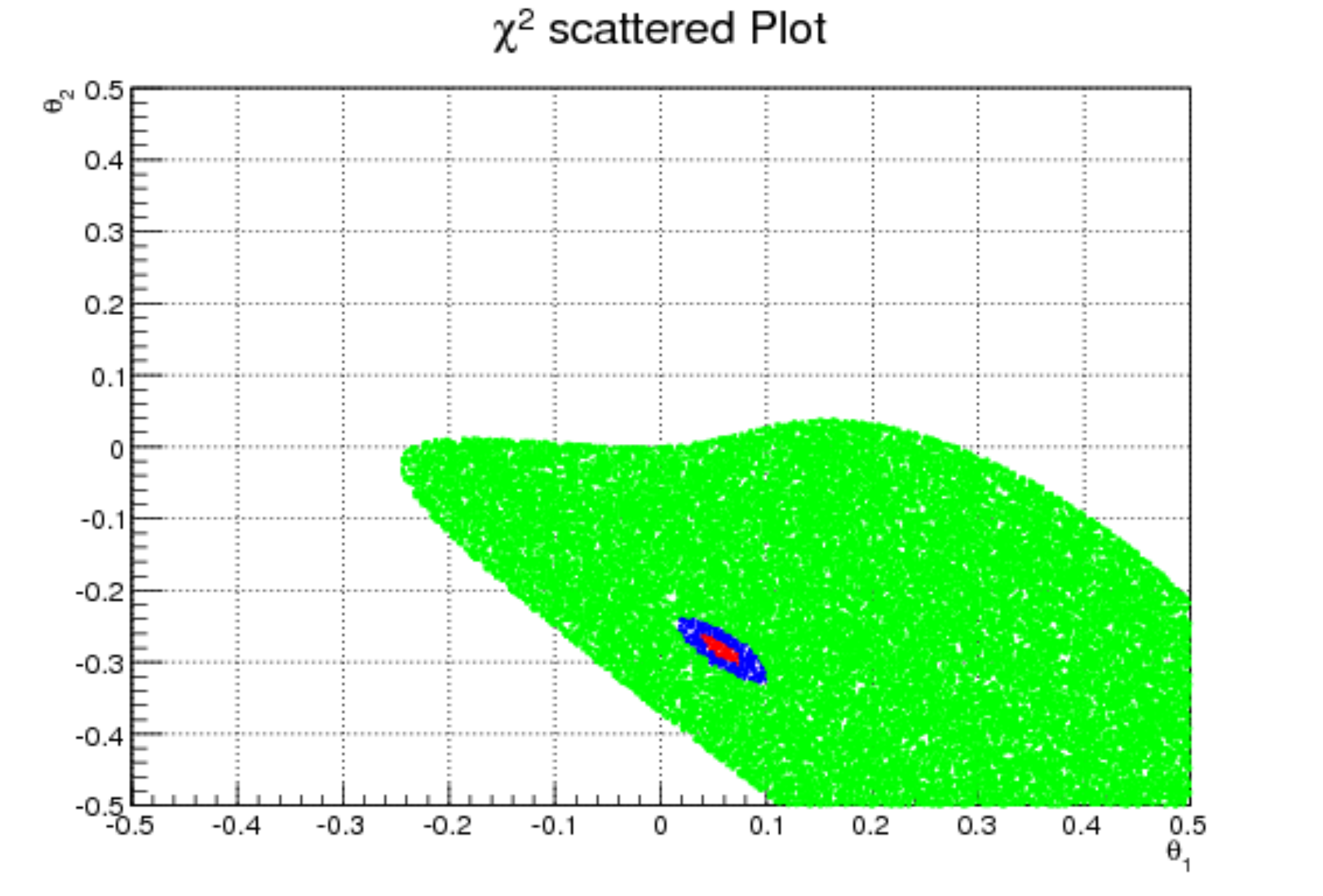} &
\includegraphics[angle=0,width=80mm]{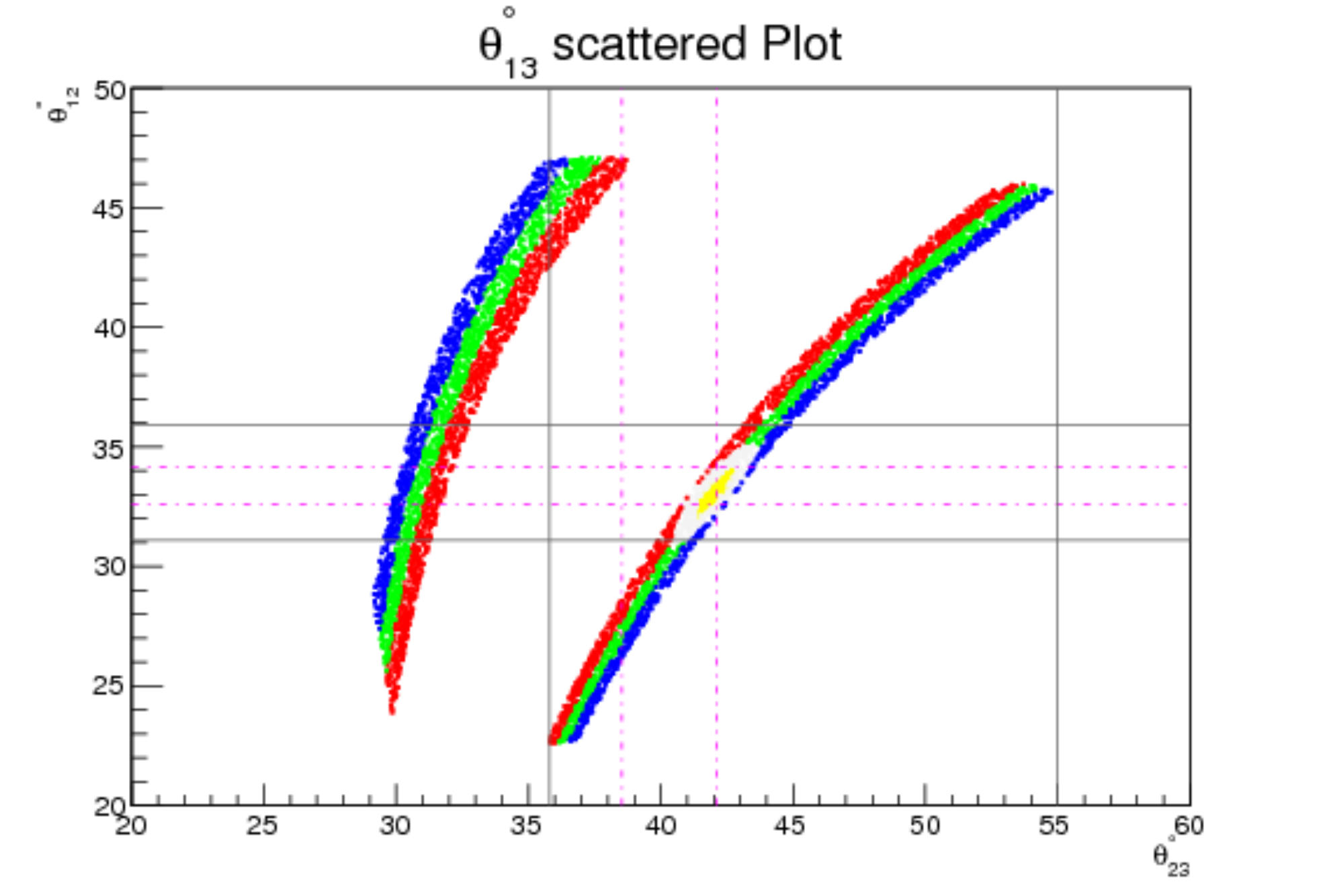}\\
\end{tabular}
\caption{$U_{BM}^{1213}$ scatter plot of $\chi^2$ (left fig.) over $\alpha-\gamma$ (in radians) plane and $\theta_{13}$ (right fig.) 
over  $\theta_{23}-\theta_{12}$ (in degrees) plane. The information about color coding and various horizontal, vertical lines in right fig. is given in text.}
\label{fig.2}
\end{figure}

\begin{figure}[!t]\centering
\begin{tabular}{c c} 
\includegraphics[angle=0,width=80mm]{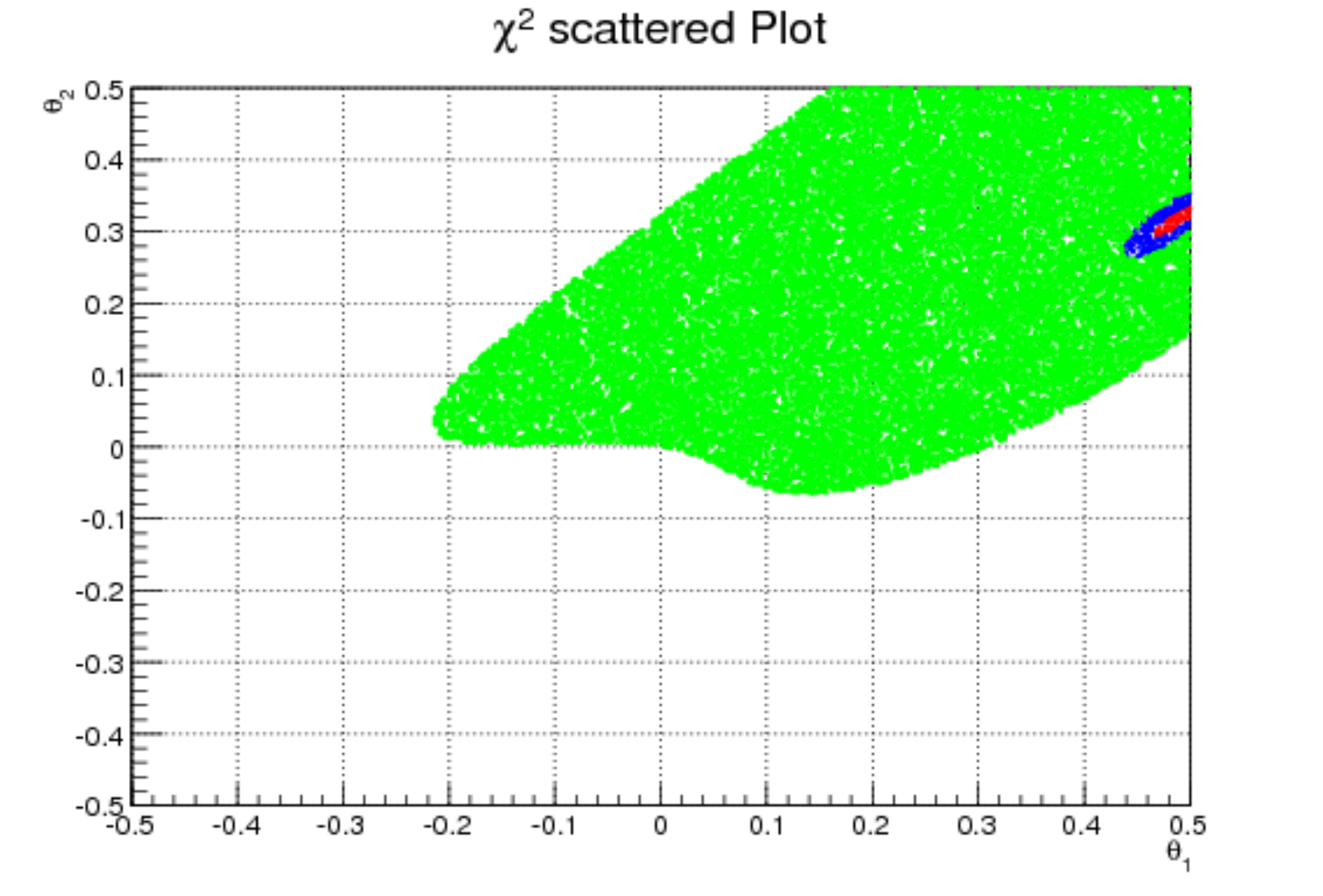} &
\includegraphics[angle=0,width=80mm]{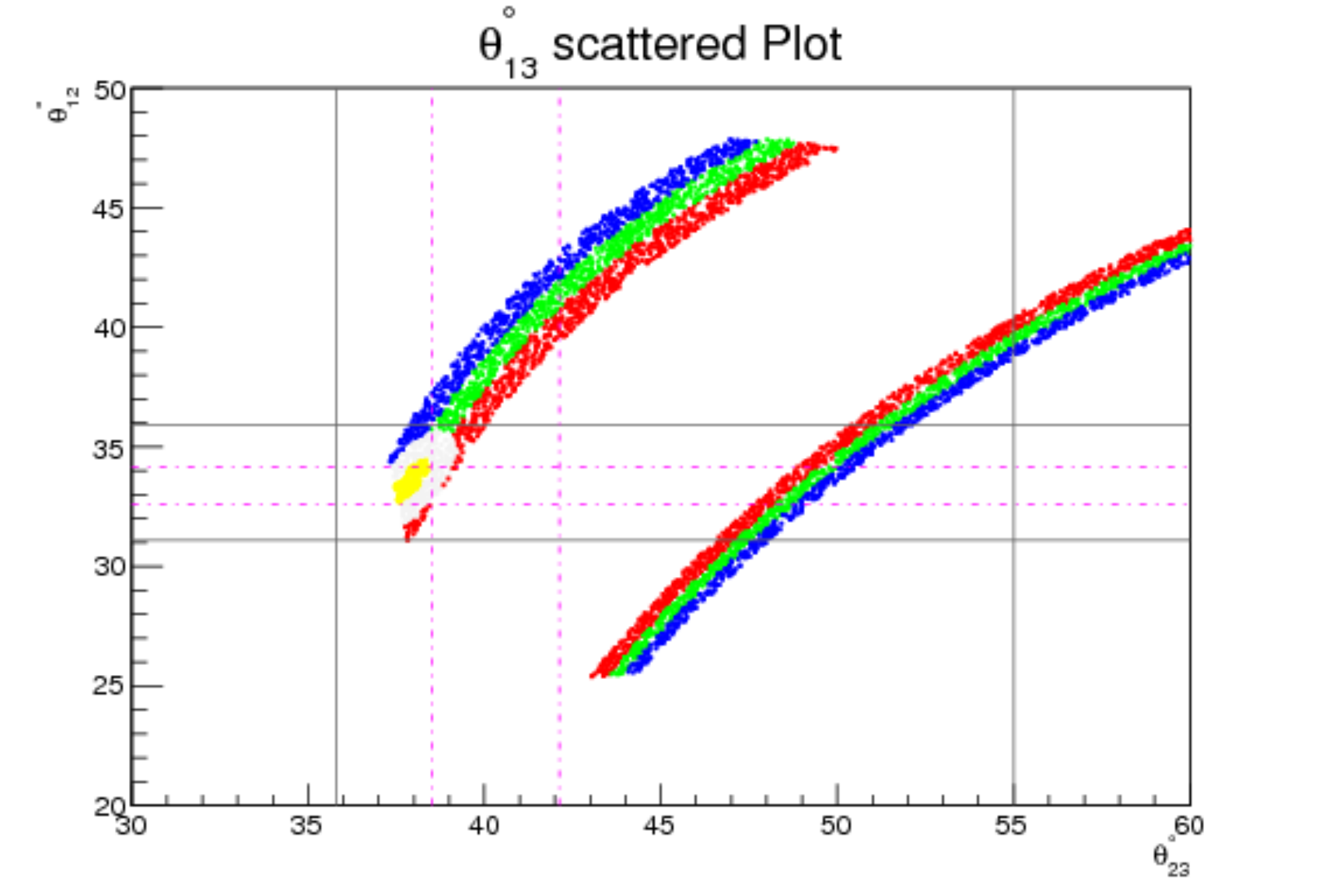}\\
\end{tabular}
\caption{$U_{DC}^{1213}$ scatter plot of $\chi^2$ (left fig.) over $\alpha-\gamma$ (in radians) plane and $\theta_{13}$ (right fig.) 
over  $\theta_{23}-\theta_{12}$ (in degrees) plane. The information about color coding and various horizontal, vertical lines in right fig. is given in text.}
\label{fig.3}
\end{figure}

From the above equation it is clear that the perturbation parameters ($\alpha, \gamma$) enters into these mixing angles at leading order
and thus exhibit good correlations among themselves. In Figs.~\ref{fig.1}-\ref{fig.3} 
we present the numerical results corresponding to TBM, BM and DC case  with $\theta_1 = \gamma$ and $\theta_2 = \alpha$.
As evident from figures it is possible to get $\chi^2 < 3$ in a very tiny region of parameter space for all cases. 
Here parameter space prefers two regions for mixing angles. In TBM and DC case for first viable region $\theta_{23} \in [36^\circ, 40^\circ]$ 
while  for 2nd it lies in the interval $[46^\circ, 52^\circ]$. In BM case lower $\theta_{23}$ region remains outside
$3\sigma$ band while for 2nd region $\theta_{23}$ varies in the interval $[40^\circ, 45^\circ]$. 
However only in BM case its possible to fit all mixing angles at $1\sigma$ level while other two mixing schemes can only be best consistent at
2$\sigma$ level.

\subsection{12-23 Rotation}

This case corresponds to rotations in 12 and 23 sector of  these special matrices.  
The neutrino mixing angles for small perturbation 
parameters $\alpha$ and $\beta$ are given by

\beqa
 \sin\theta_{13} &\approx&  |\alpha U_{23} + \beta U_{12} + \alpha\beta U_{22} |,\\
 \sin\theta_{23} &\approx& |\frac{ U_{23} + \beta U_{22} -(\alpha^2 + \beta^2)U_{23} -\alpha\beta U_{12}}{\cos\theta_{13}}|,\\
 \sin\theta_{12} &\approx& |\frac{U_{12} +\alpha U_{22} -(\alpha^2 +\beta^2) U_{12}-\alpha\beta U_{23}}{\cos\theta_{13}}|.
\eeqa

\begin{figure}[!t]\centering
\begin{tabular}{c c} 
\includegraphics[angle=0,width=80mm]{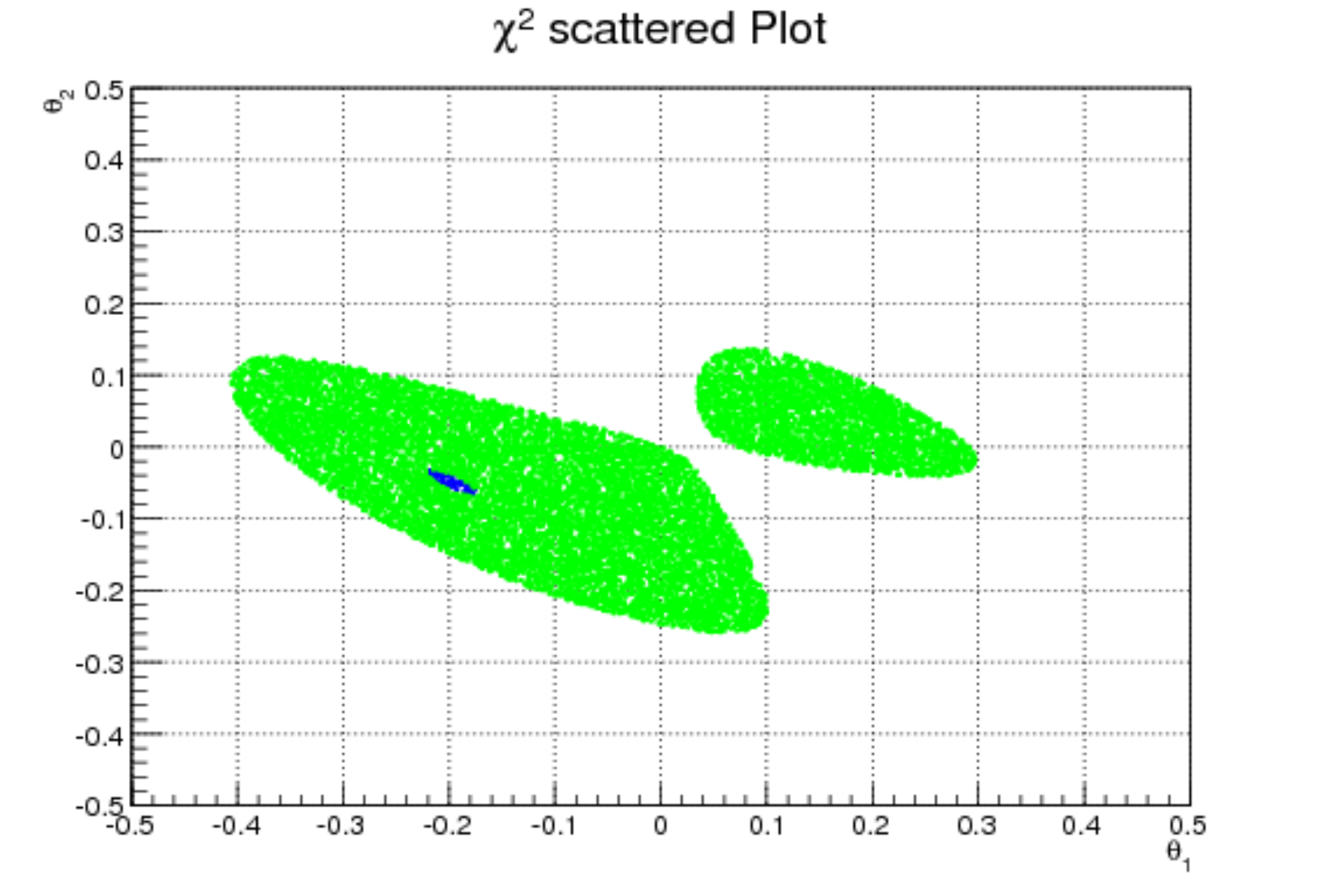} &
\includegraphics[angle=0,width=80mm]{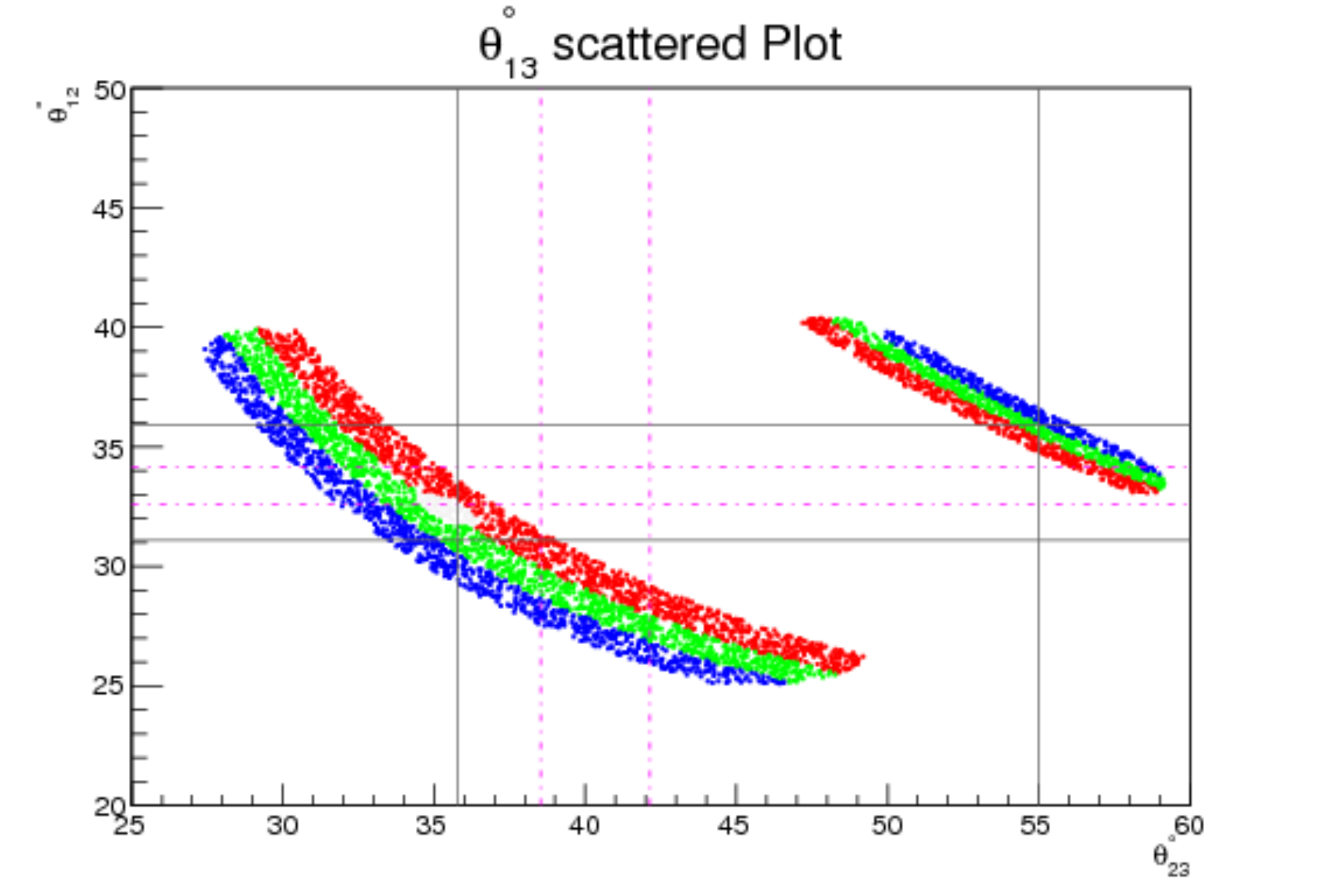}\\
\end{tabular}
\caption{$U_{TBM}^{1223}$ scatter plot of $\chi^2$ (left fig.) over $\alpha-\beta$ (in radians) plane and $\theta_{13}$ (right fig.) 
over  $\theta_{23}-\theta_{12}$ (in degrees) plane. The information about color coding and various horizontal, vertical lines in right fig. is given in text.}
\label{fig.4}
\end{figure}

\begin{figure}[!t]\centering
\begin{tabular}{c c} 
\includegraphics[angle=0,width=80mm]{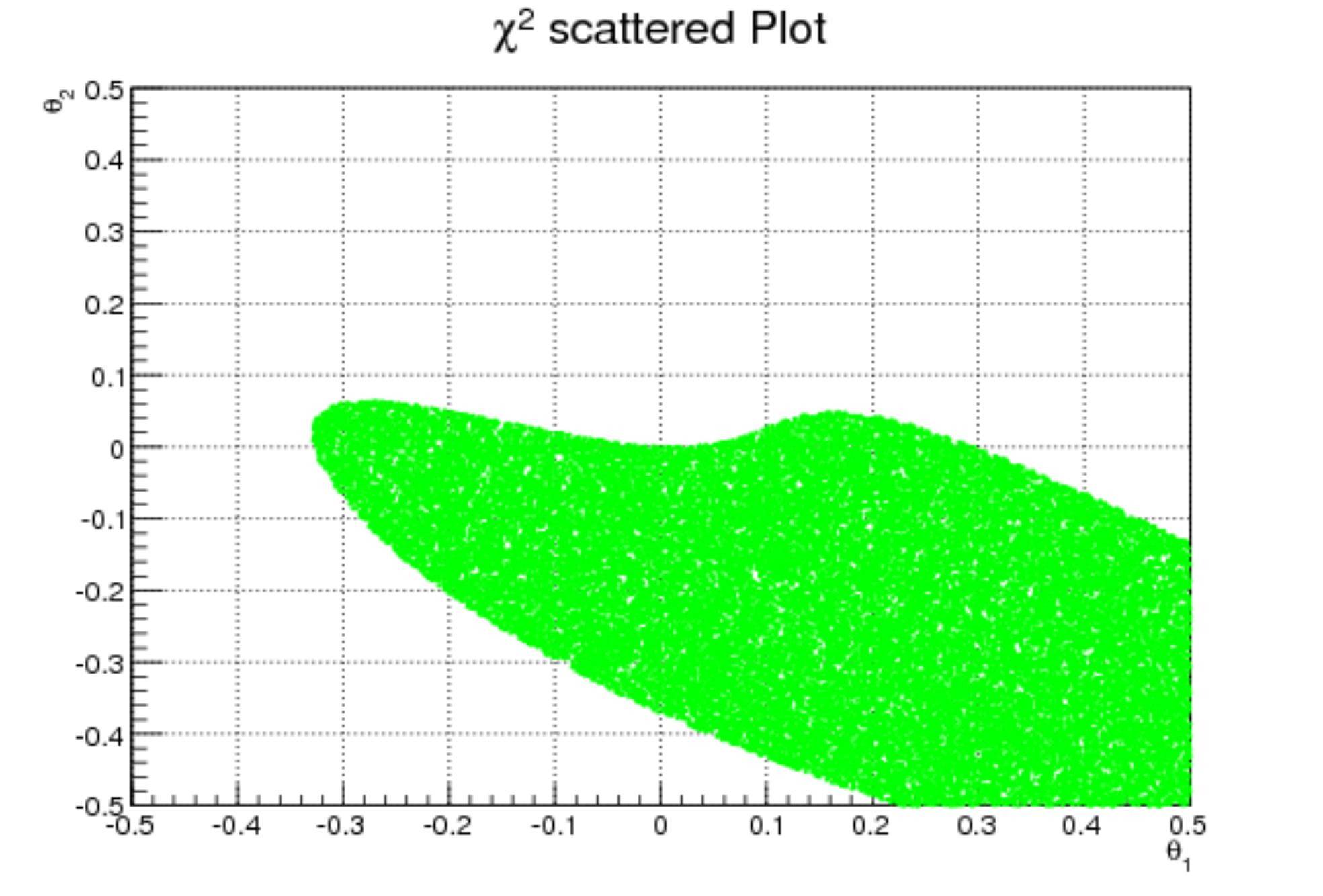} &
\includegraphics[angle=0,width=80mm]{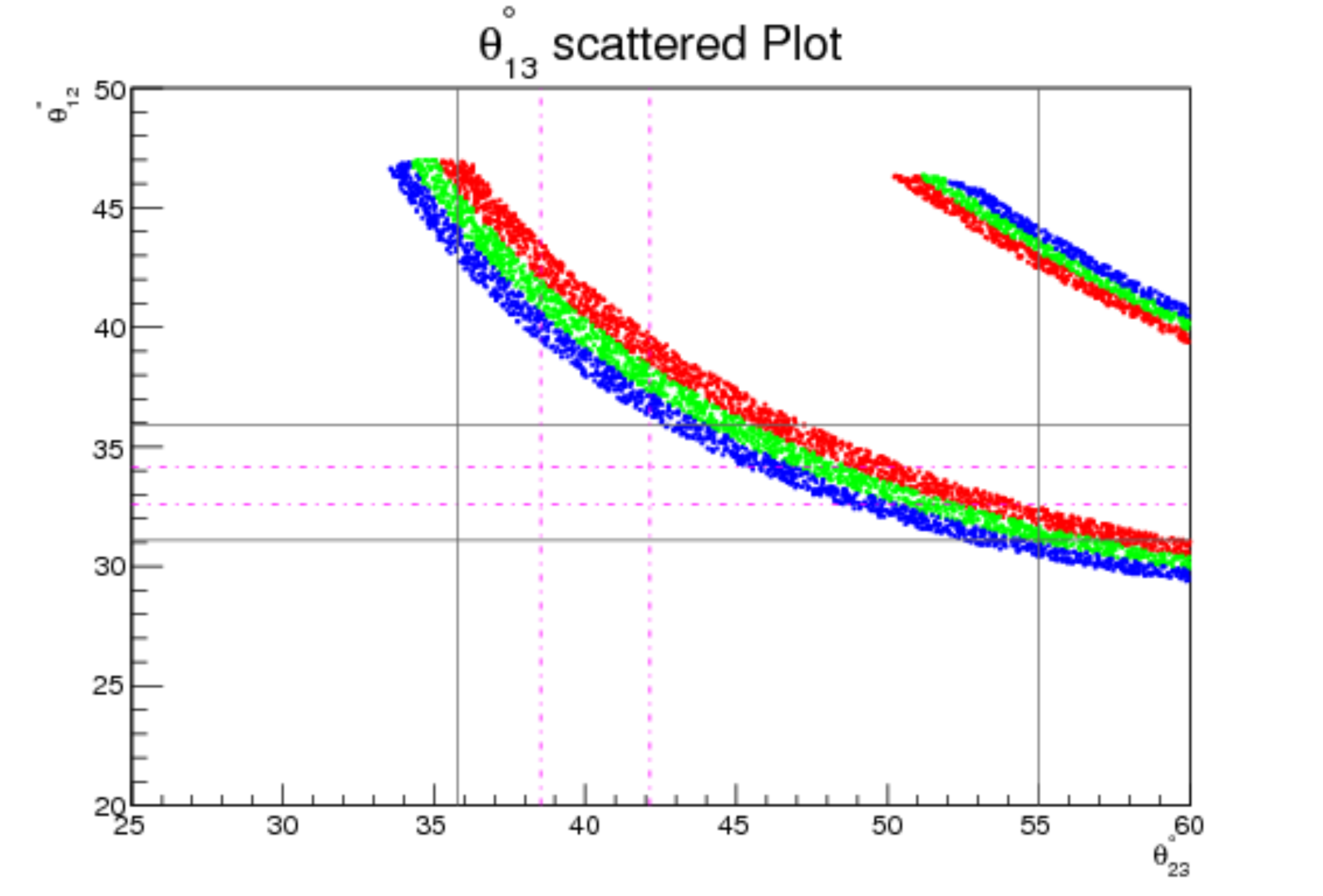}\\
\end{tabular}
\caption{$U_{BM}^{1223}$ scatter plot of $\chi^2$ (left fig.) over $\alpha-\beta$ (in radians) plane and $\theta_{13}$ (right fig.) 
over  $\theta_{23}-\theta_{12}$ (in degrees) plane. The information about color coding and various horizontal, vertical lines in right fig. is given in text.}
\label{fig.5}
\end{figure}

\begin{figure}[!t]\centering
\begin{tabular}{c c} 
\includegraphics[angle=0,width=80mm]{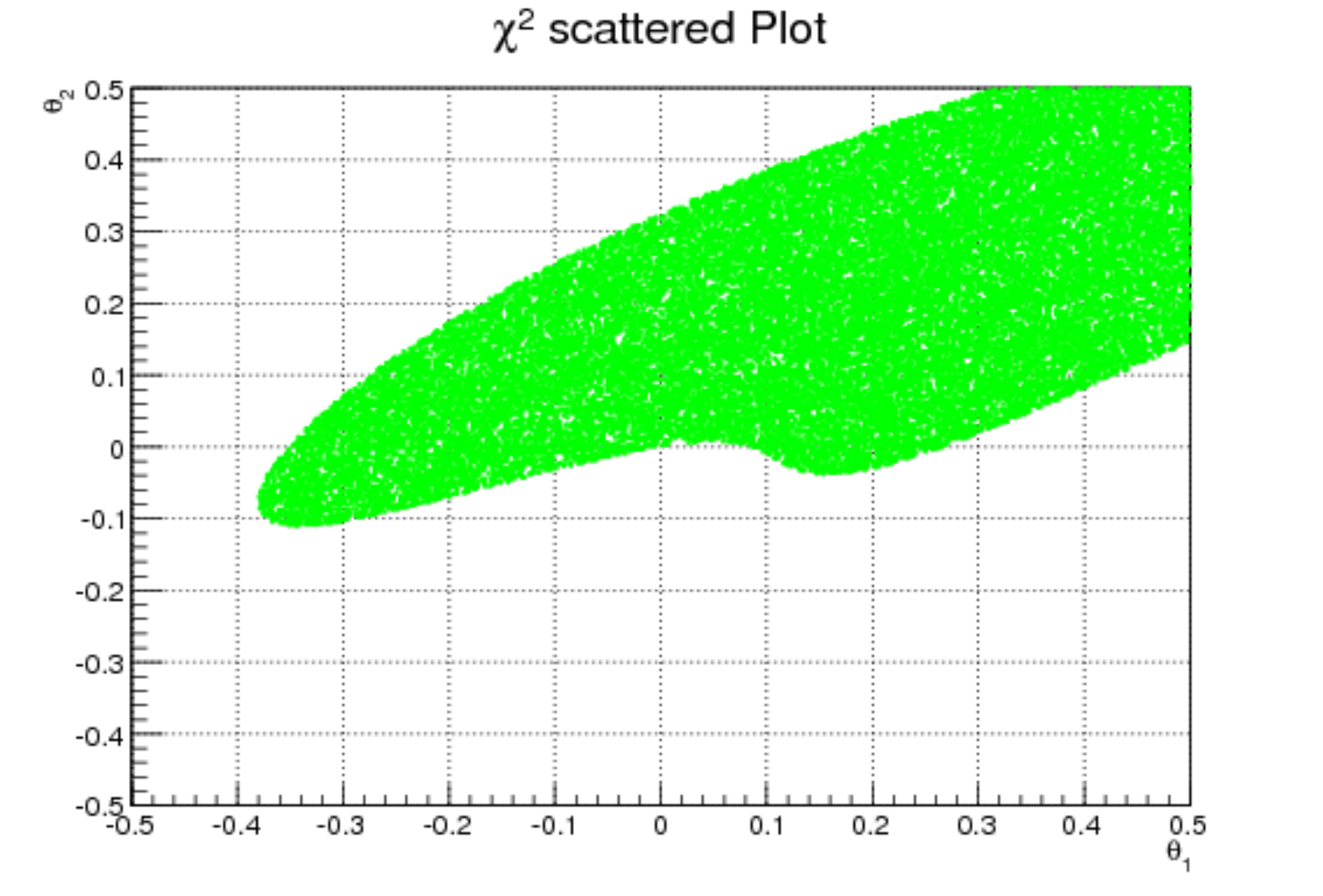} &
\includegraphics[angle=0,width=80mm]{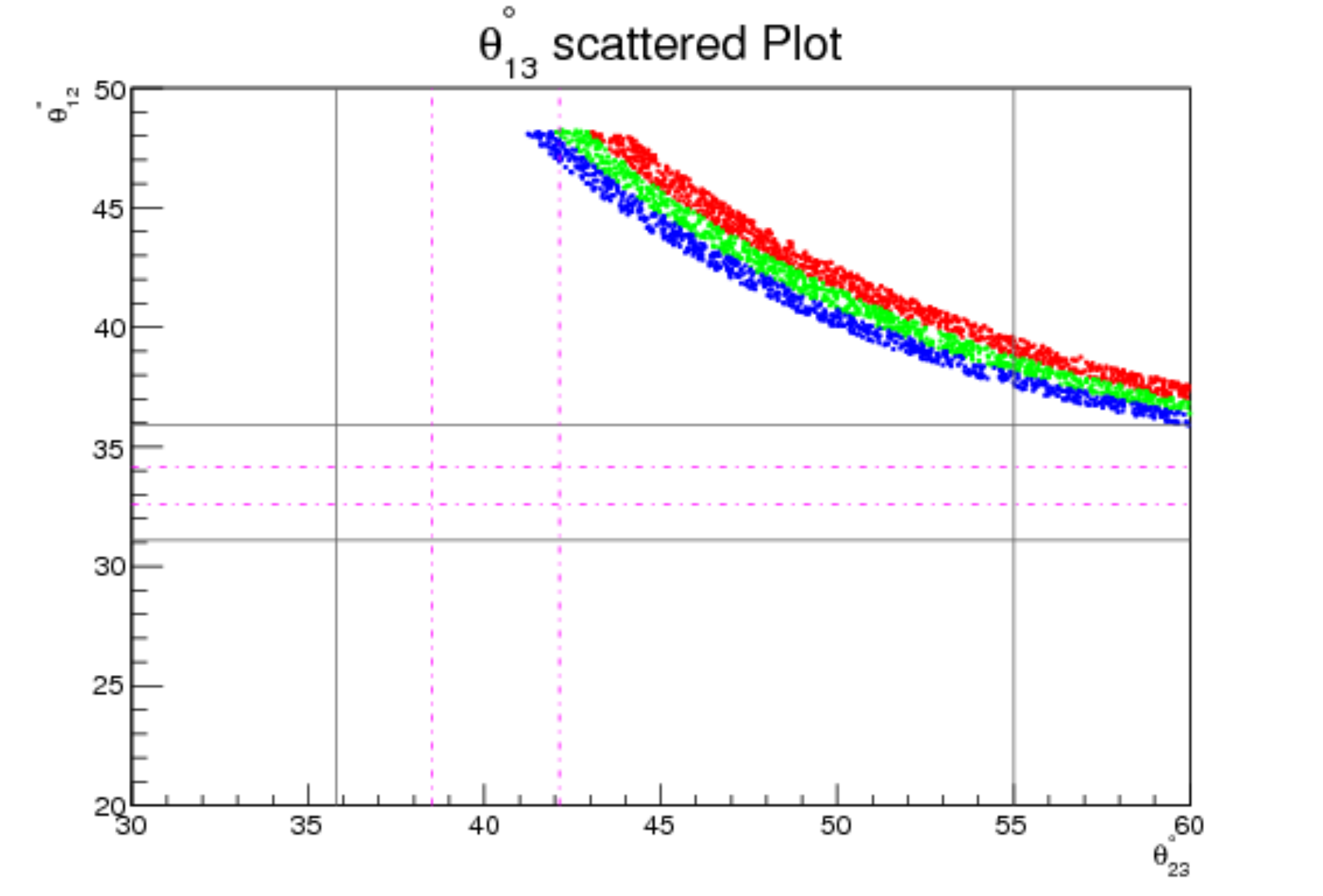}\\
\end{tabular}
\caption{$U_{DC}^{1223}$ scatter plot of $\chi^2$ (left fig.) over $\alpha-\beta$ (in radians) plane and $\theta_{13}$ (right fig.) 
over  $\theta_{23}-\theta_{12}$ (in degrees) plane. The information about color coding and various horizontal, vertical lines in right fig. is given in text.}
\label{fig.6}
\end{figure}

Like previous case, perturbation parameters enters at leading order into these mixing angles and hence show interesting correlations among themselves.
Figs.~\ref{fig.4}-\ref{fig.6} corresponds to TBM, BM and DC case respectively with $\theta_1 = \beta$ and $\theta_2 = \alpha$. 
As visible from figures only for BM case its possible to get $\chi^2$ in the interval
[3, 10] while for other two cases $\chi^2 > 10$ always. In TBM case there exist two tiny viable 3$\sigma$ regions: first one prefers $\theta_{23}$ to be
in the interval $[36^\circ, 38^\circ]$ while for 2nd it lies in the range $[53^\circ, 55^\circ]$. In BM there is only one region which comes under $3\sigma$ band with
$\theta_{23} \in [42^\circ, 55^\circ]$. In DC case all values lie outside $3\sigma$ band and thus this case is excluded.

\subsection{13-12 Rotation}

This case corresponds to rotations in 13 and 12 sector of  these special matrices. 
The neutrino mixing angles for small perturbation parameters $\alpha$ and $\gamma$ are given by

\beqa
 \sin\theta_{13} &\approx&  |\gamma U_{33} |,\\
  \sin\theta_{23} &\approx& |\frac{U_{23} }{\cos\theta_{13}}|,\\
  \sin\theta_{12} &\approx& |\frac{ U_{12} + \alpha U_{11} + \gamma U_{32} -(\alpha^2 + \gamma^2) U_{12} + \alpha\gamma U_{31} }{\cos\theta_{13}}|.
\eeqa

\begin{figure}[!t]\centering
\begin{tabular}{c c} 
\includegraphics[angle=0,width=80mm]{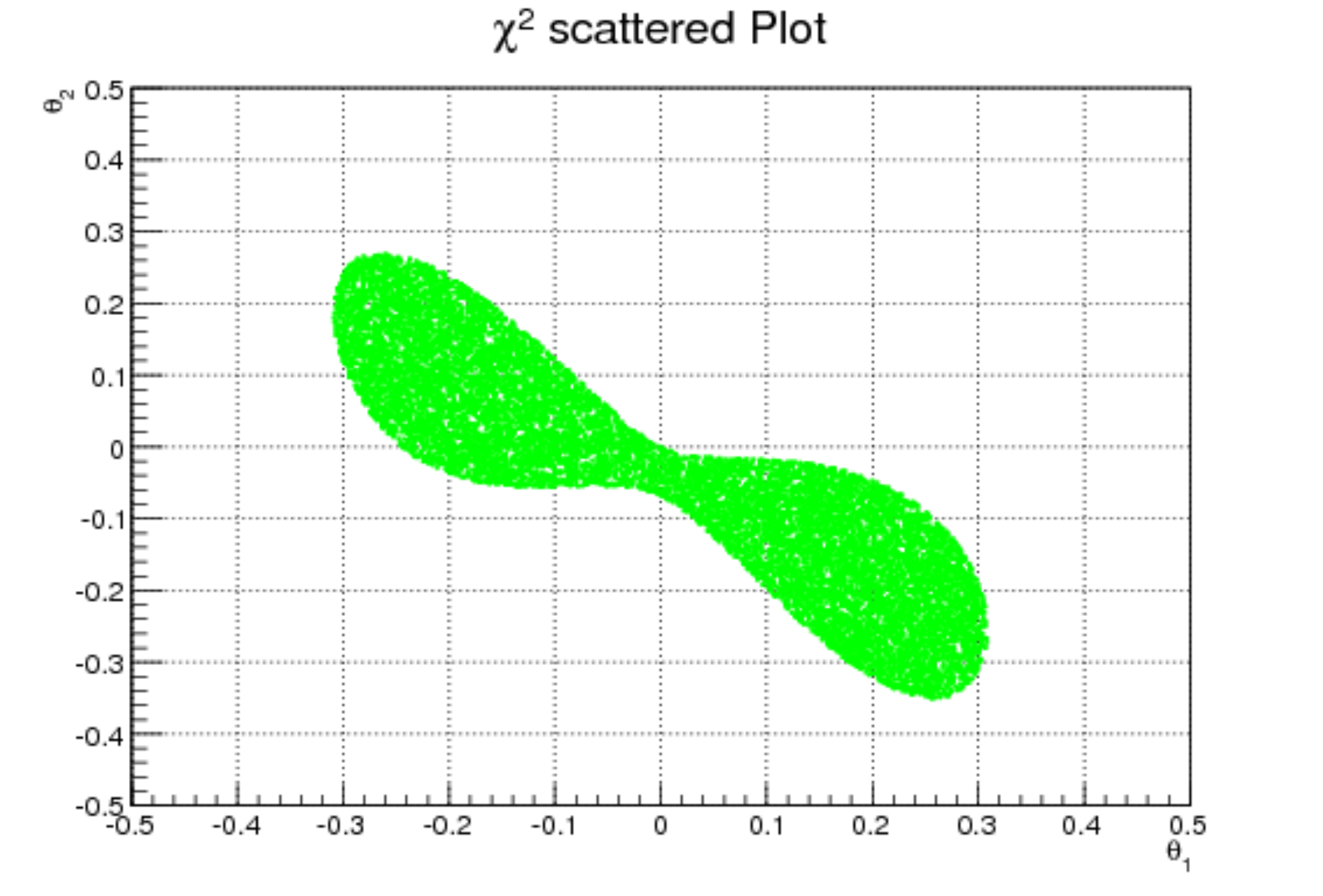} &
\includegraphics[angle=0,width=80mm]{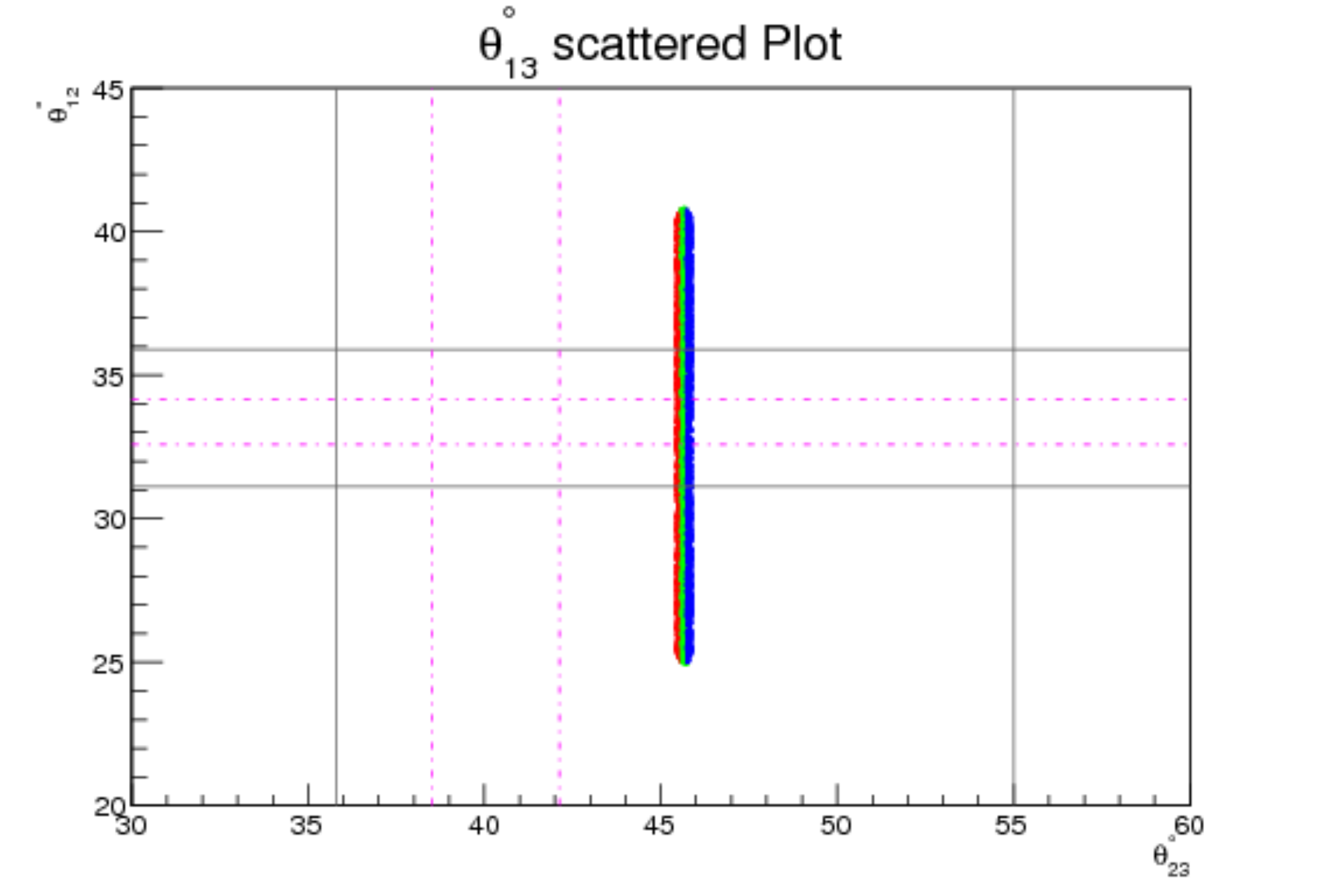}\\
\end{tabular} 
\caption{$U_{TBM}^{1312}$ scatter plot of $\chi^2$ (left fig.) over $\gamma-\alpha$(in radians) plane and $\theta_{13}$ (right fig.) 
over  $\theta_{23}-\theta_{12}$ (in degrees) plane. The information about color coding and various horizontal, vertical lines in right fig. is given in text.}
\label{fig.7} 
\end{figure}

\begin{figure}[!t]\centering
\begin{tabular}{c c} 
\includegraphics[angle=0,width=80mm]{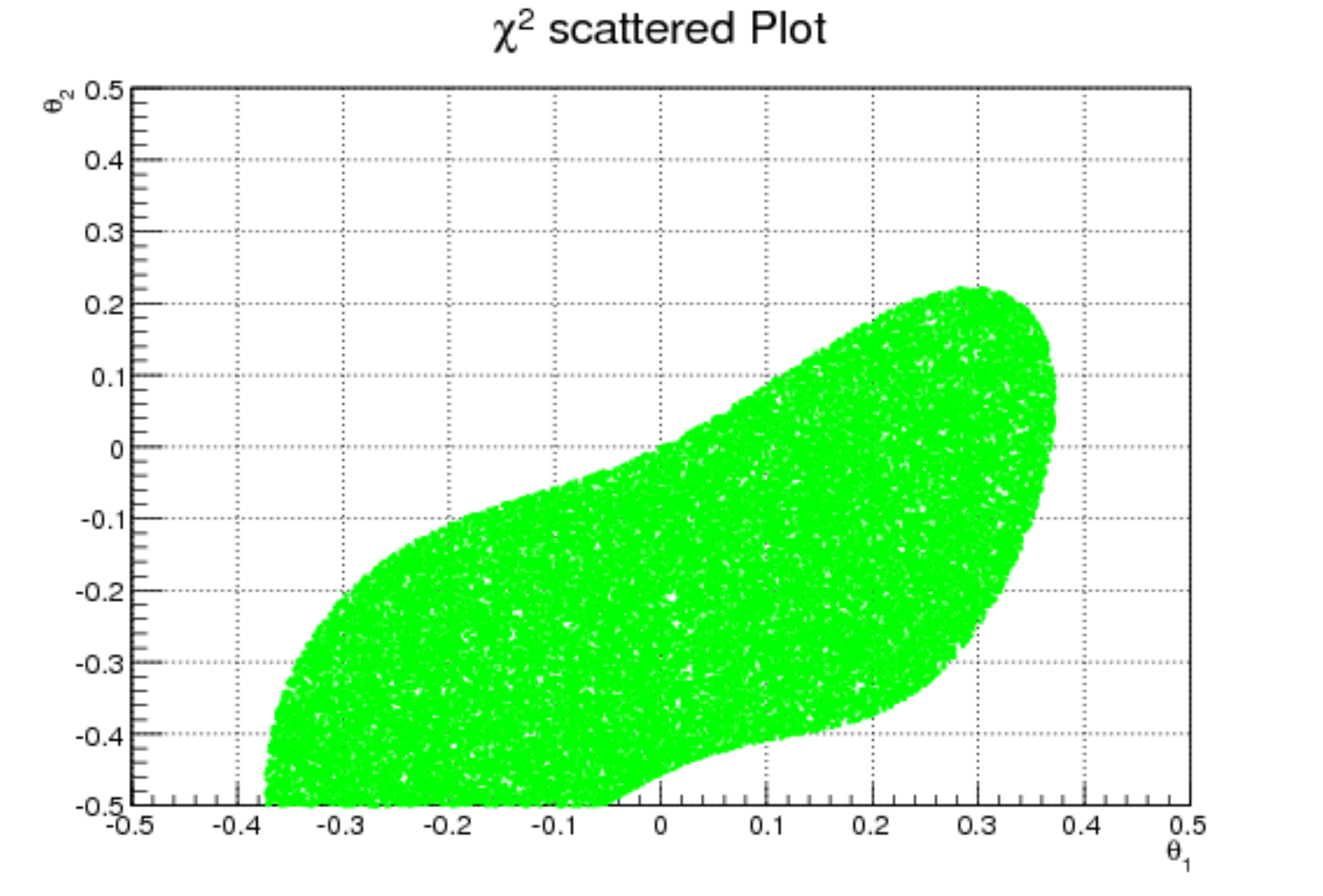} &
\includegraphics[angle=0,width=80mm]{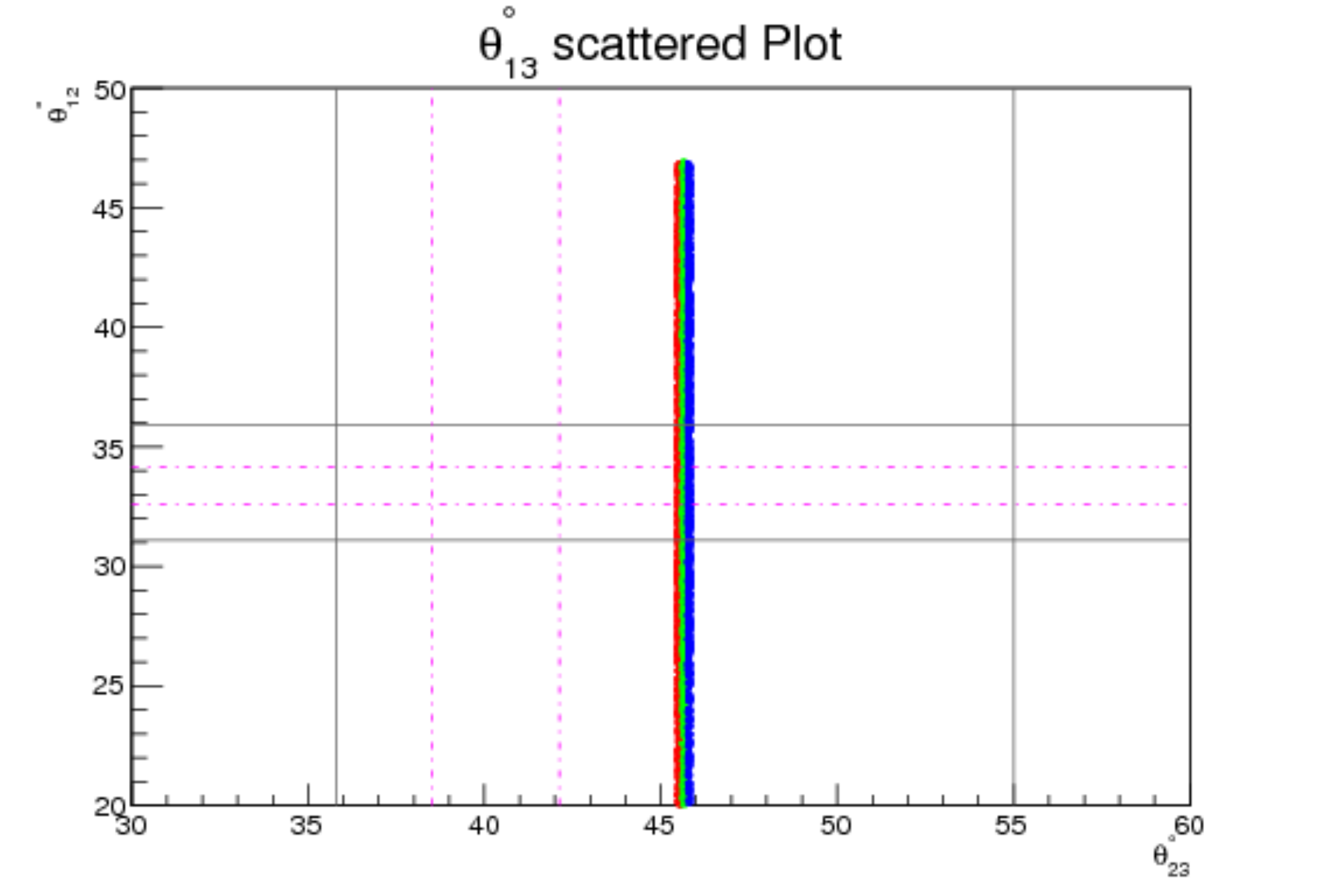}\\
\end{tabular}
\caption{$U_{BM}^{1312}$ scatter plot of $\chi^2$ (left fig.) over $\gamma-\alpha$ (in radians) plane and $\theta_{13}$ (right fig.) 
over  $\theta_{23}-\theta_{12}$ (in degrees) plane. The information about color coding and various horizontal, vertical lines in right fig. is given in text.}
\label{fig.8}
\end{figure}

\begin{figure}[!t]\centering
\begin{tabular}{c c} 
\includegraphics[angle=0,width=80mm]{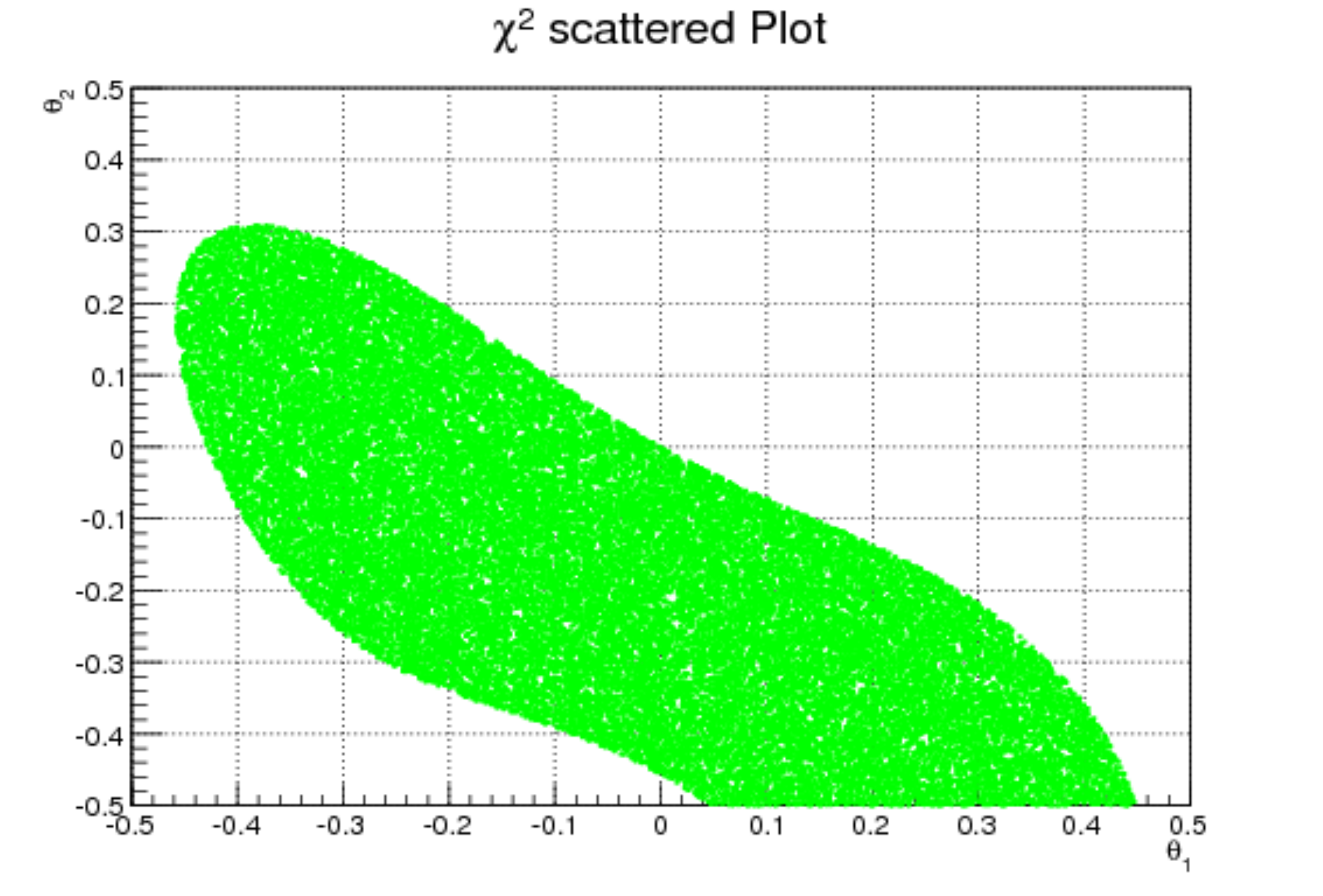} &
\includegraphics[angle=0,width=80mm]{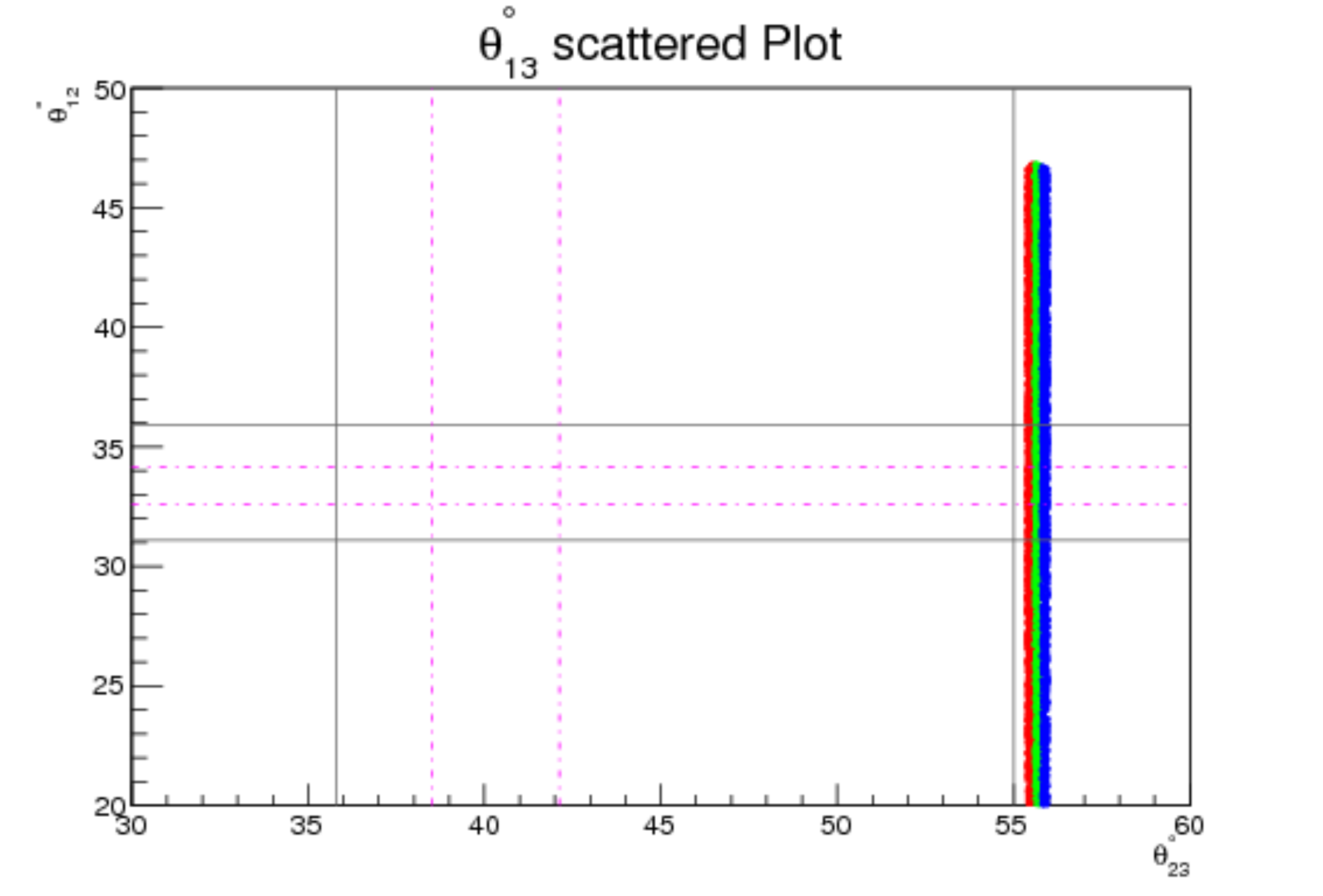}\\
\end{tabular}
\caption{$U_{DC}^{1312}$ scatter plot of $\chi^2$ (left fig.) over $\gamma-\alpha$ (in radians) plane and $\theta_{13}$ (right fig.) 
over  $\theta_{23}-\theta_{12}$ (in degrees) plane. The information about color coding and various horizontal, vertical lines in right fig. is given in text.}
\label{fig.9}
\end{figure}

Figs.~\ref{fig.7}-\ref{fig.9} corresponds to TBM, BM and DC case respectively with $\theta_1 = \gamma$ and $\theta_2 =\alpha$. 
For this rotation its not possible to get $\chi^2 < 10$ in viable parameter space for all the three mixing schemes. Since mixing angle 
$\theta_{23}$ doesn't receive any leading order corrections in this perturbation scheme so its value remain close to its unperturbed value.  The value of $\gamma$ in BM and TBM case is 
$|\gamma|\approx 0.175-0.245$ for $\theta_{13}$ to be in its 3$\sigma$ range. Thus for these cases $\theta_{23}$ lies in a vary narrow range
$[45.4^\circ, 45.8^\circ]$ which is consistent with its 3$\sigma$ CL. However corresponding value for DC case is $\theta_{23}\approx 56^\circ$ which
lies outside its $3\sigma$ allowed band and thus this case is excluded.

\subsection{13-23 Rotation}

This case corresponds to rotations in 13 and 23 sector of  these special matrices. 
The neutrino mixing angles for small perturbation parameters $\gamma$ and $\beta$ are given by

\beqa
 \sin\theta_{13} &\approx&  |\beta U_{12} + \gamma U_{33} + \beta\gamma U_{32}|,\\
 \sin\theta_{23} &\approx& |\frac{ U_{23} + \beta U_{22}-\beta^2 U_{23} }{\cos\theta_{13}}|,\\
 \sin\theta_{12} &\approx& |\frac{U_{12} +\gamma U_{32} -(\gamma^2 +\beta^2) U_{12}-\beta\gamma U_{33}}{\cos\theta_{13}}|.
\eeqa

\begin{figure}[!t]\centering
\begin{tabular}{c c} 
\includegraphics[angle=0,width=80mm]{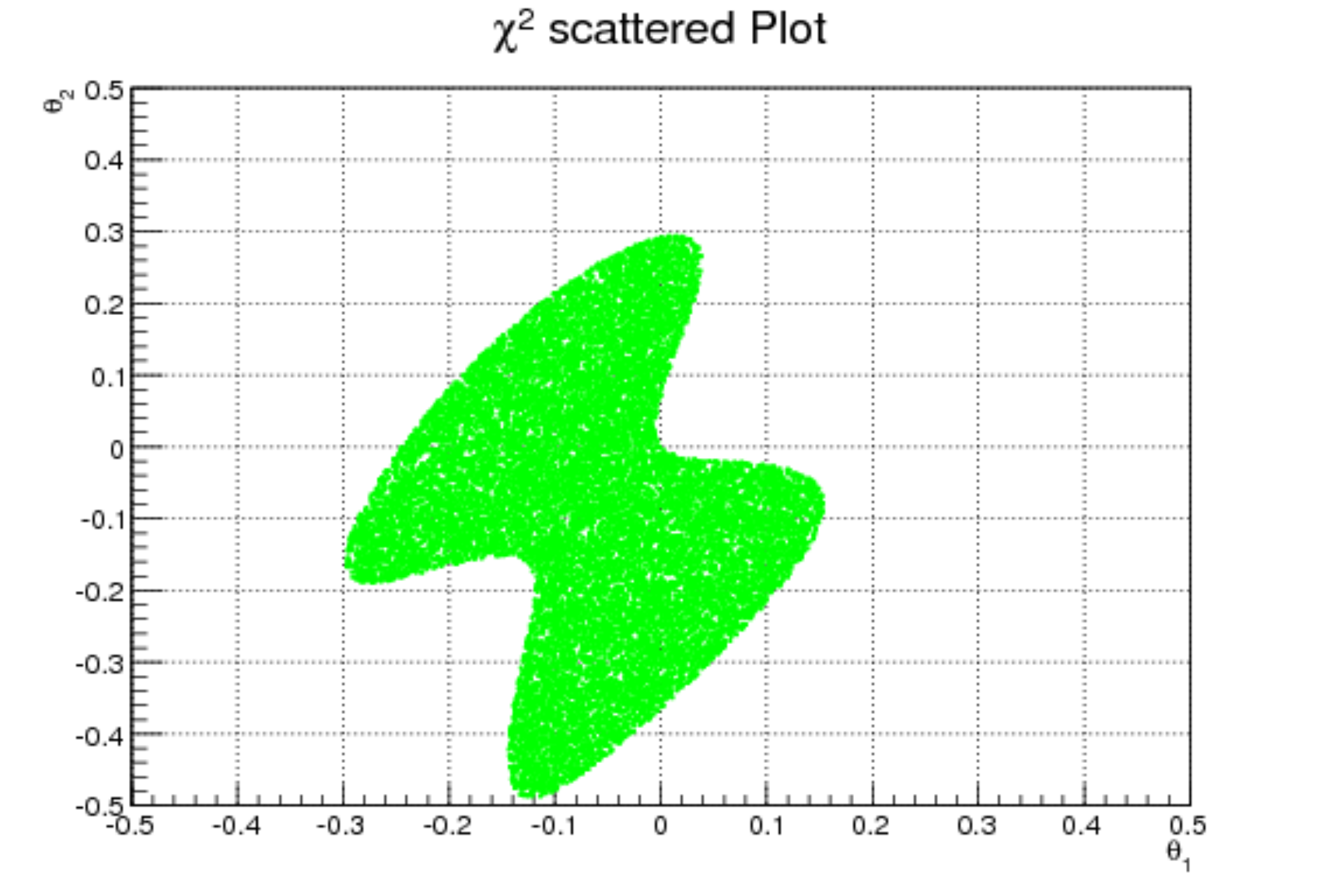} &
\includegraphics[angle=0,width=80mm]{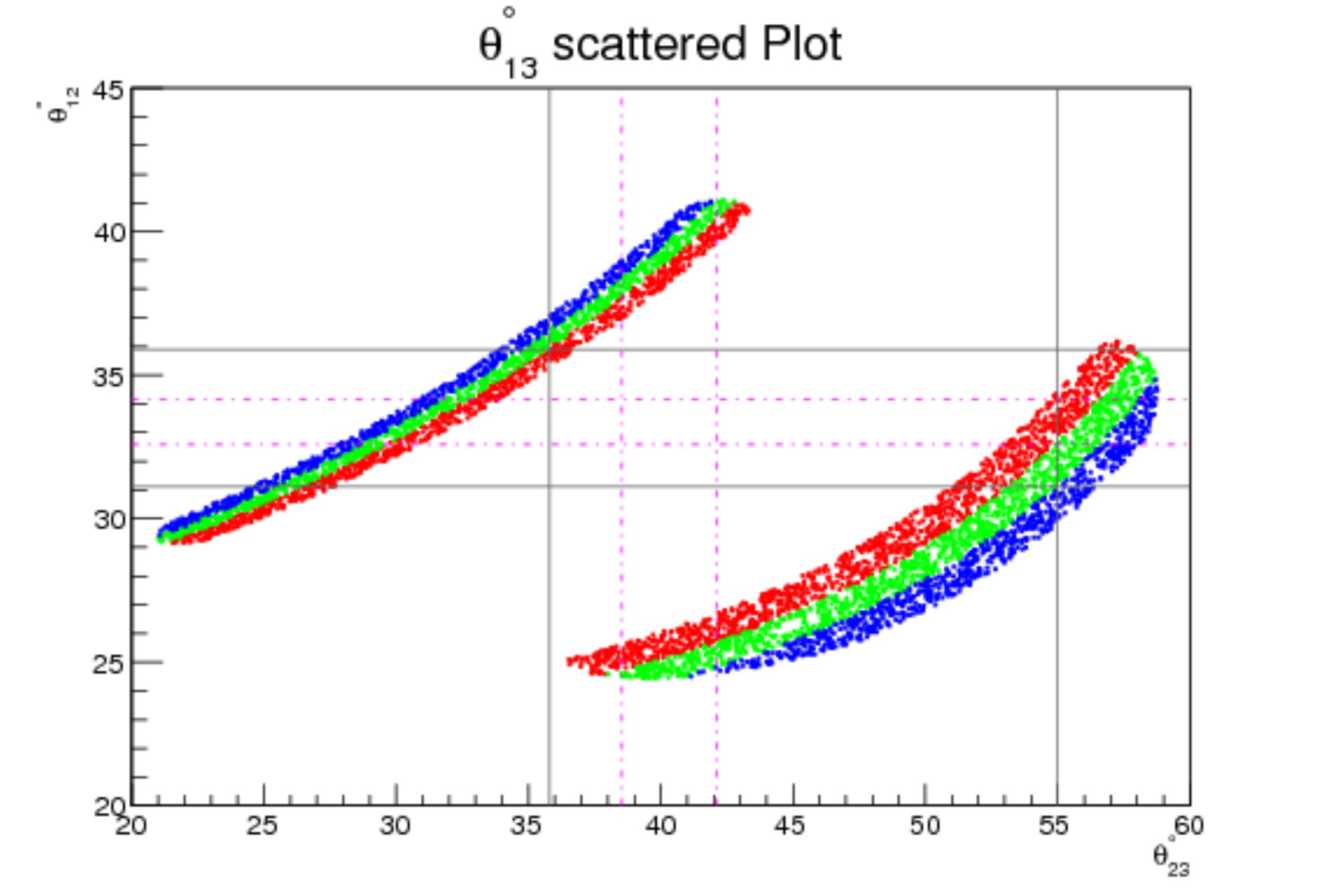}\\
\end{tabular}
\caption{$U_{TBM}^{1323}$ scatter plot of $\chi^2$ (left fig.) over $\gamma-\beta$ (in radians) plane and $\theta_{13}$ (right fig.)
over  $\theta_{23}-\theta_{12}$ (in degrees) plane. The information about color coding and various horizontal, vertical lines in right fig. is given in text.}
\label{fig.10}
\end{figure}

\begin{figure}[!t]\centering
\begin{tabular}{c c} 
\includegraphics[angle=0,width=80mm]{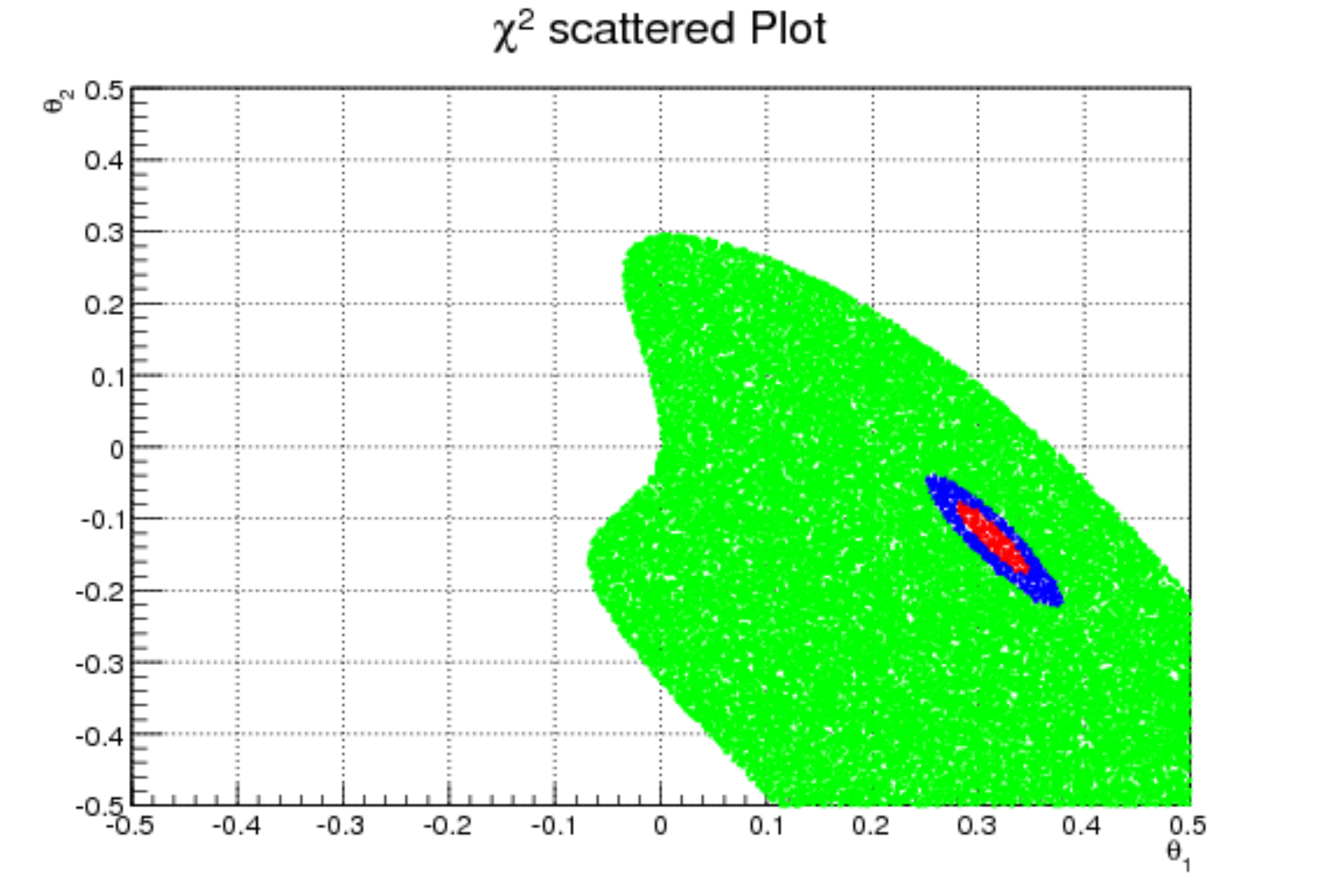} &
\includegraphics[angle=0,width=80mm]{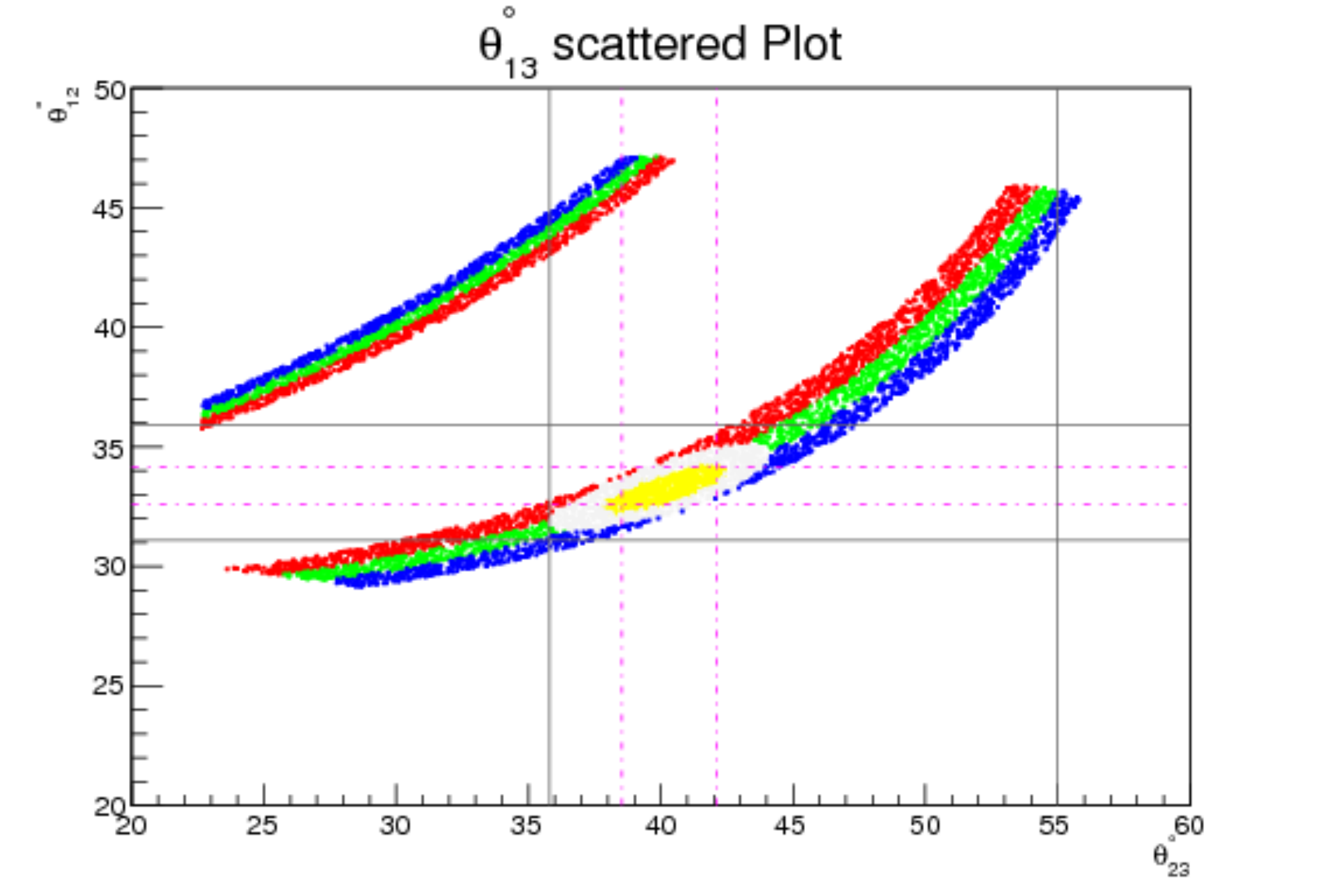}\\
\end{tabular}
\caption{$U_{BM}^{1323}$ scatter plot of $\chi^2$ (left fig.) over $\gamma-\beta$ (in radians) plane and $\theta_{13}$ (right fig.) 
over  $\theta_{23}-\theta_{12}$ (in degrees) plane. The information about color coding and various lines is given in text.}
\label{fig.11}
\end{figure}

\begin{figure}[!t]\centering
\begin{tabular}{c c} 
\includegraphics[angle=0,width=80mm]{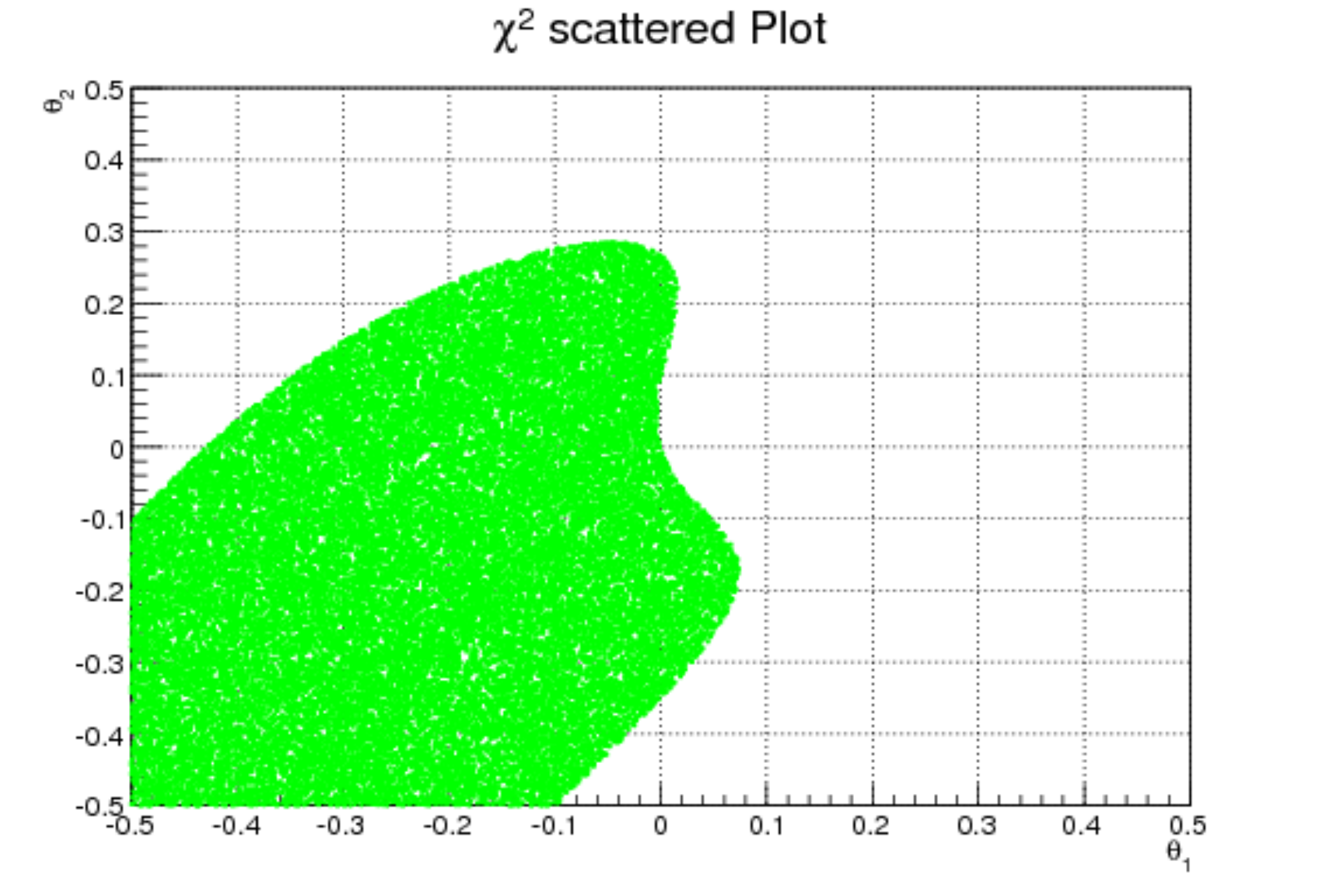} &
\includegraphics[angle=0,width=80mm]{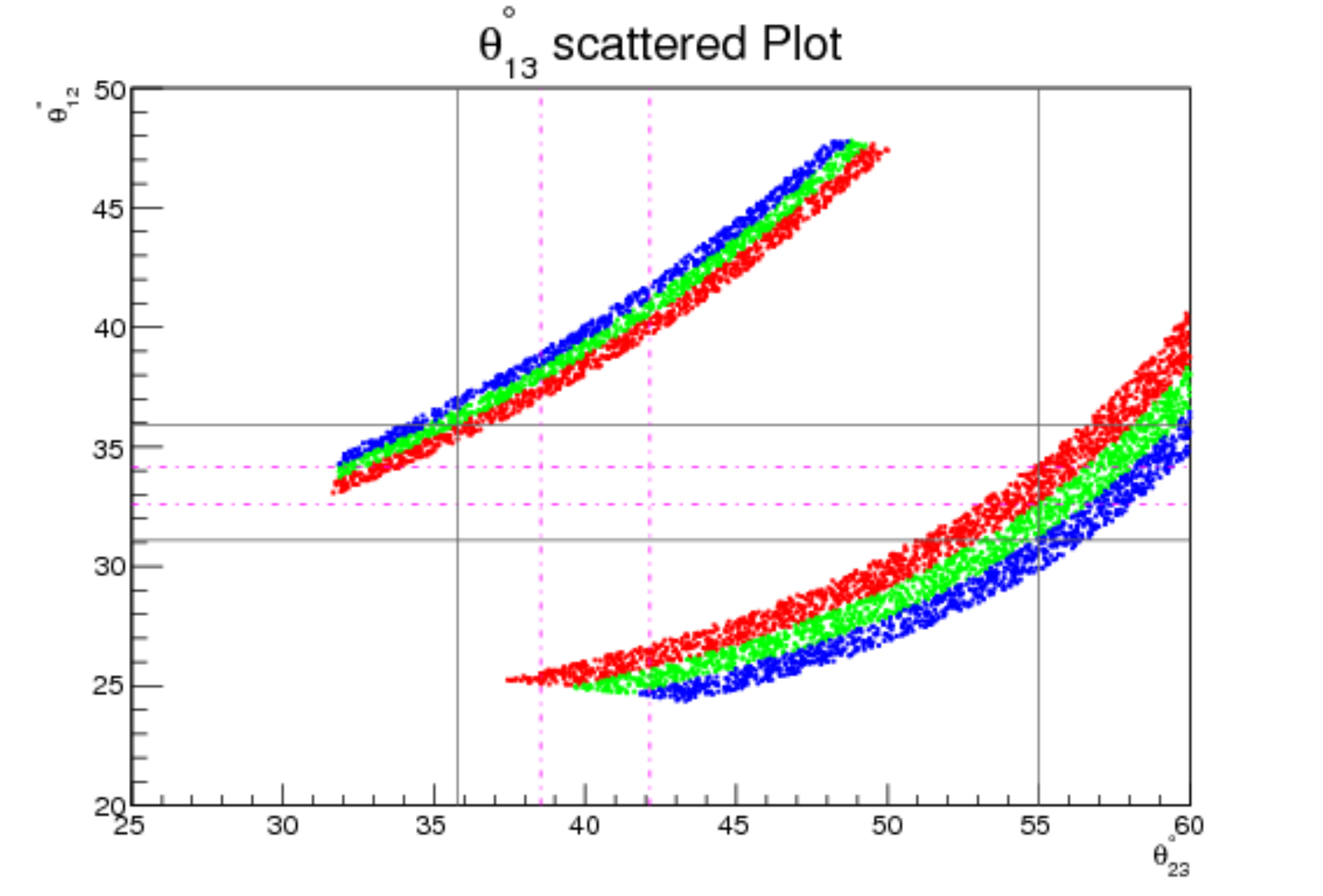}\\
\end{tabular}
\caption{$U_{DC}^{1323}$ scatter plot of $\chi^2$ (left fig.) over $\gamma-\beta$ (in radians) plane and $\theta_{13}$ (right fig.)
over  $\theta_{23}-\theta_{12}$ (in degrees) plane. The information about color coding and various horizontal, vertical lines in right fig. is given in text.}
\label{fig.12}
\end{figure}

The parameters $\beta$ and $\gamma$ enters into all mixing angles at leading order and thus show good correlations among
themselves. Figs.~\ref{fig.10}-\ref{fig.12} corresponds to TBM, BM and DC case respectively with $\theta_1 = \gamma$ and $\theta_2 = \beta$.
For all cases parameter space prefers two regions for mixing angles. In TBM and DC perturbative case, first viable region is extremely tiny with $\theta_{23} \approx 36^{\circ}$
while for 2nd region $\theta_{23} \in [51^{\circ}, 55^{\circ}]$. However for corresponding BM case, first region lies outside its $3\sigma$ band
and thus excluded while for 2nd allowed region $\theta_{23} \in [35^{\circ}, 47^{\circ}]$. 
This rotation scheme with BM case is highly favorable as it allows $\chi^2 < 3$ with fitting of all mixing angles at $1\sigma$ level. 
However in TBM and DC case the value of $\chi^2 > 10$ and favorable only at $3\sigma$ level.

\subsection{23-12 Rotation}

This case corresponds to rotations in 23 and 12 sector of  these special matrices. 
However in this case there are no corrections from perturbation parameters  to 13 mixing angle i.e. $\theta_{13}=0$. So we will not
discuss it any further.

\subsection{23-13 Rotation}

This case is much similar to 13-12 rotation with interchange of expressions for $\theta_{12}$ and
$\theta_{23}$ mixing angles. 
The neutrino mixing angles for small perturbation parameters $\beta$ and $\gamma$ are given by

\beqa
 \sin\theta_{13} &\approx&  |\gamma U_{11}|,\\
 \sin\theta_{23} &\approx& |\frac{U_{23} + \beta U_{33} + \gamma U_{21}-(\beta^2 + \gamma^2) U_{23} + \beta\gamma U_{31} }{\cos\theta_{13}}|,\\
 \sin\theta_{12} &\approx& |\frac{U_{12} }{\cos\theta_{13}}|.
 \eeqa

\begin{figure}[!t]\centering
\begin{tabular}{c c} 
\includegraphics[angle=0,width=80mm]{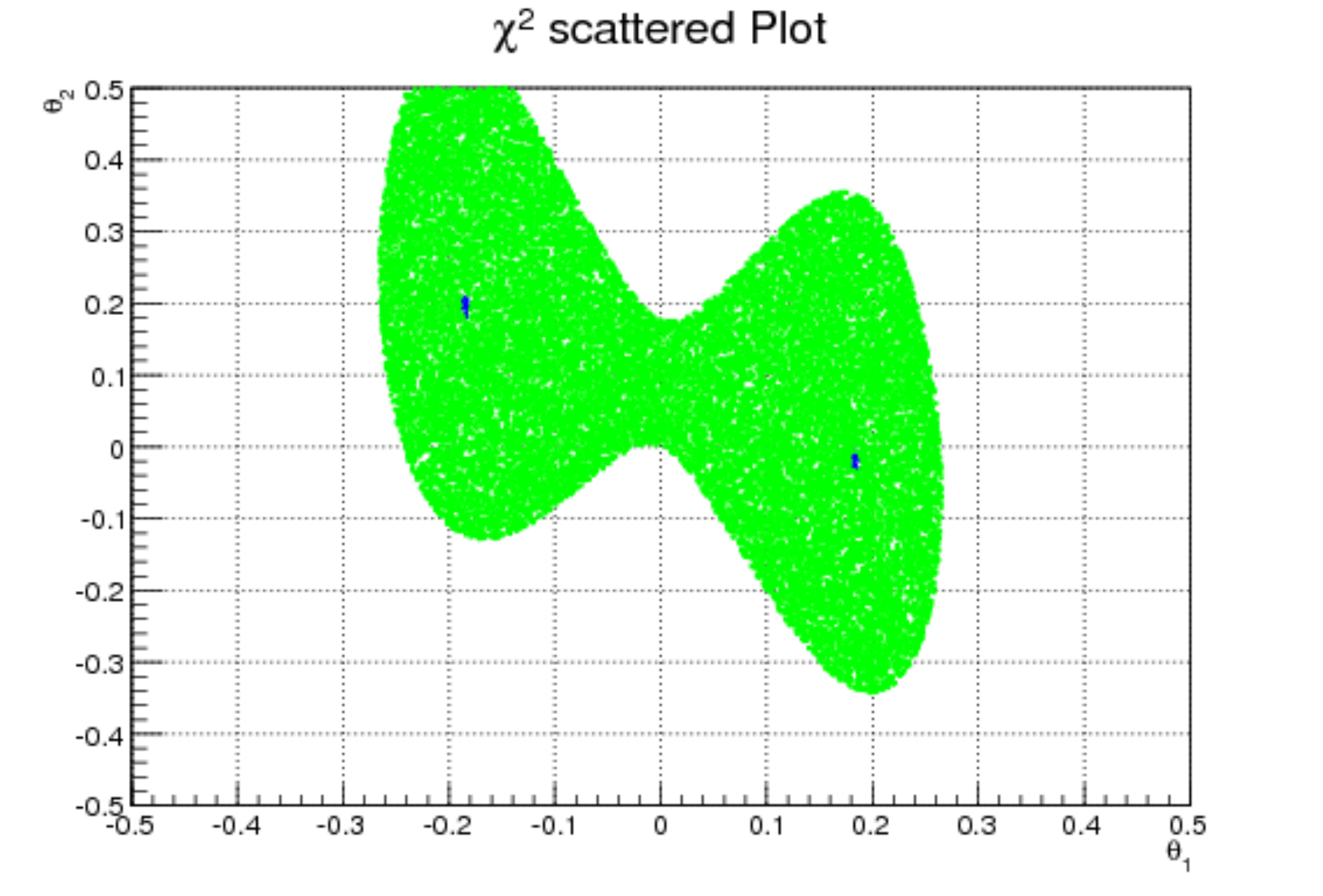} &
\includegraphics[angle=0,width=80mm]{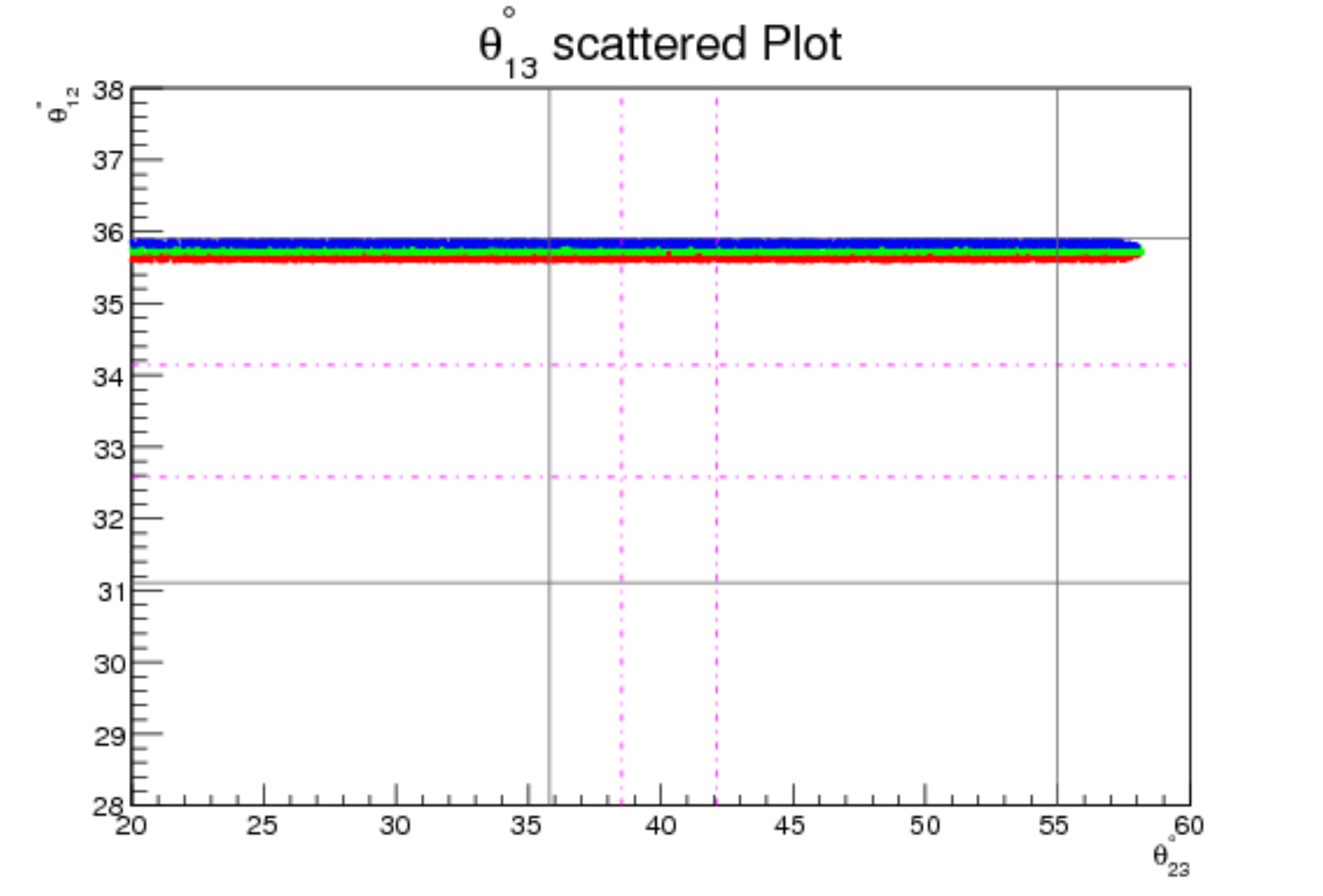}\\
\end{tabular}
\caption{$U_{TBM}^{2313}$ scatter plot of $\chi^2$ (left fig.) over $\beta-\gamma$ (in radians) plane and $\theta_{13}$ (right fig.) 
over  $\theta_{23}-\theta_{12}$ (in degrees) plane. The information about color coding and various horizontal, vertical lines in right fig. is given in text.}
\label{fig.13}
\end{figure}

\begin{figure}[!t]\centering
\begin{tabular}{c c} 
\includegraphics[angle=0,width=80mm]{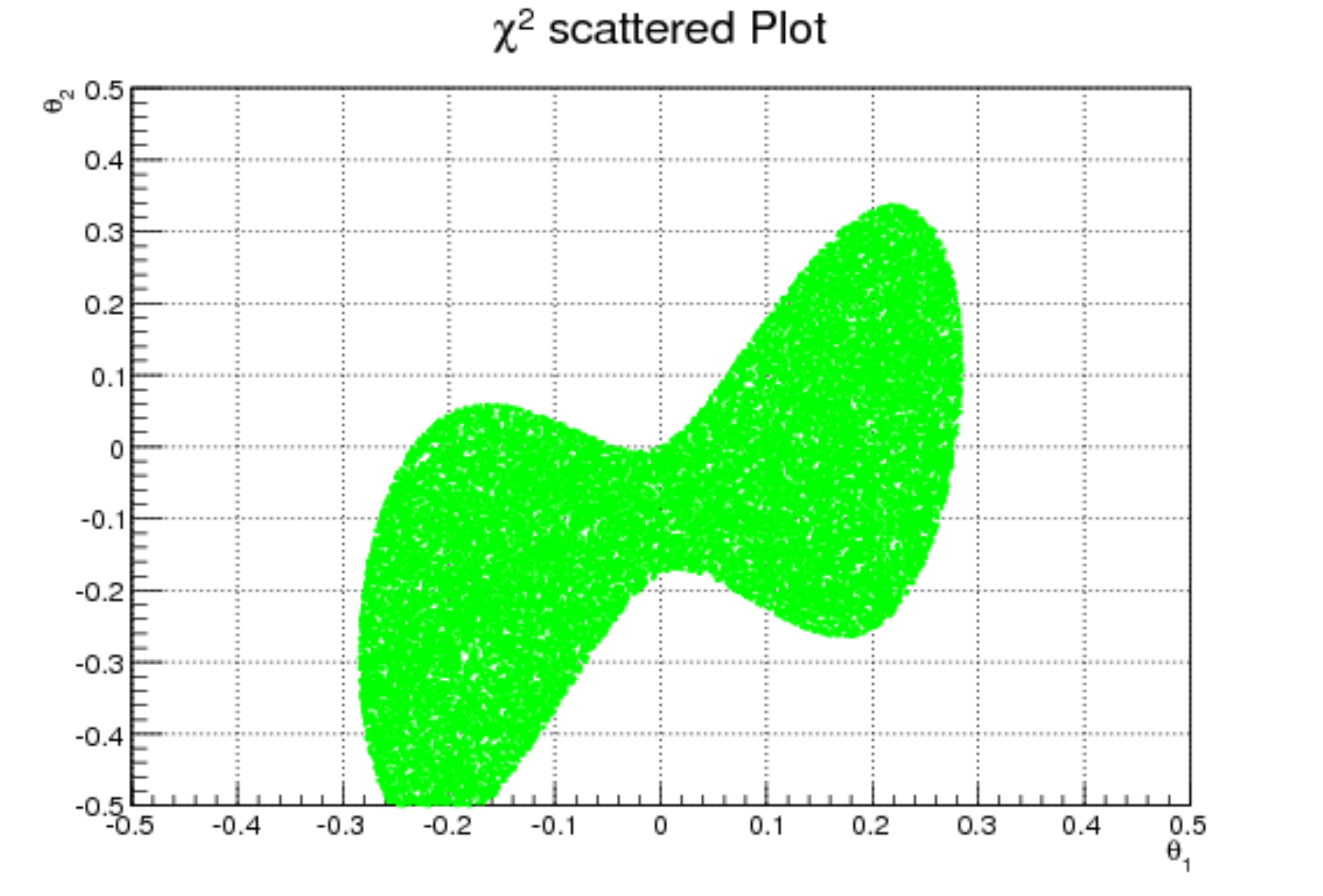} &
\includegraphics[angle=0,width=80mm]{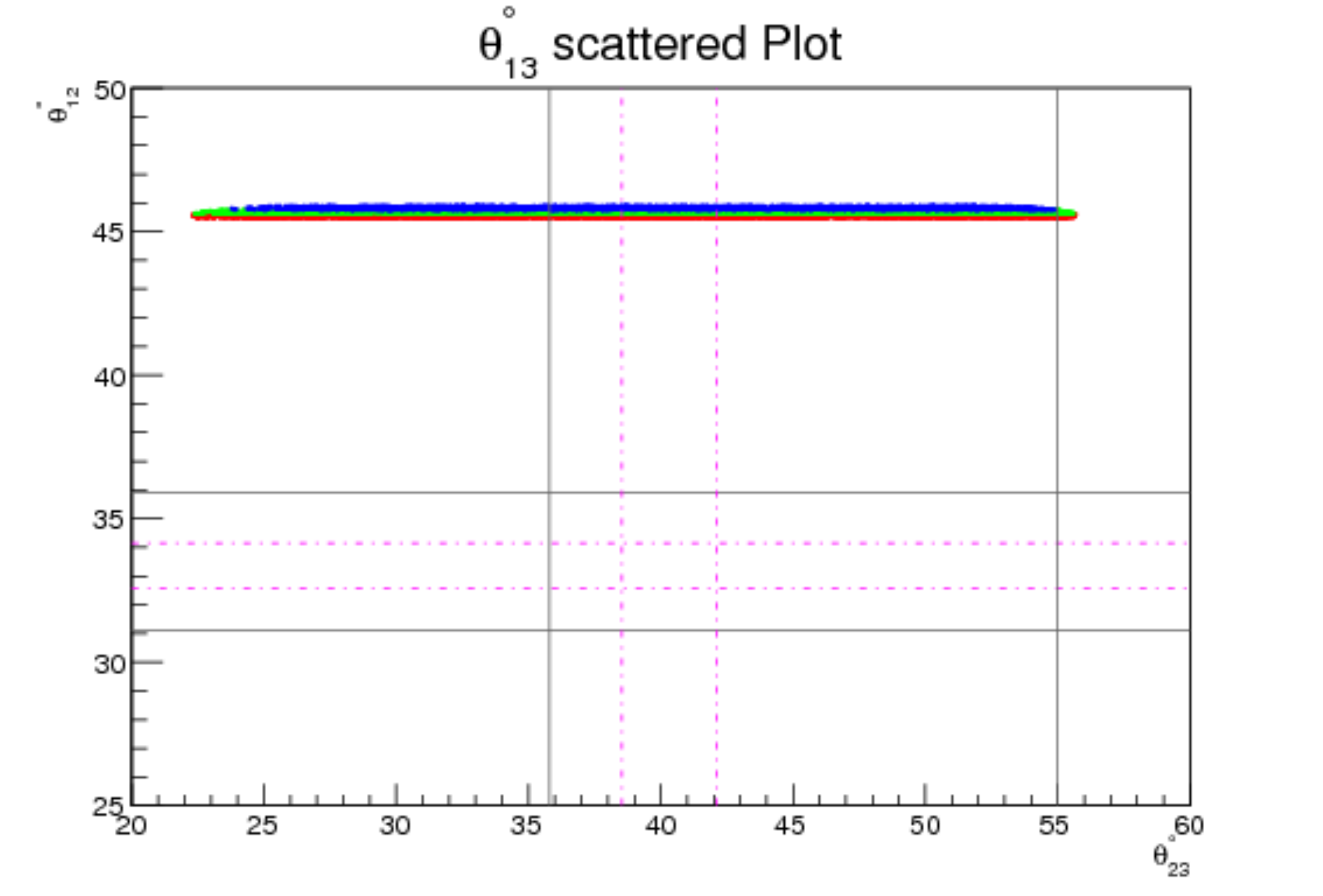}\\
\end{tabular}
\caption{$U_{BM}^{2313}$ scatter plot of $\chi^2$ (left fig.) over $\beta-\gamma$ (in radians) plane and $\theta_{13}$ (right fig.) 
over  $\theta_{23}-\theta_{12}$ (in degrees) plane. The information about color coding and various horizontal, vertical lines in right fig. is given in text.}
\label{fig.14}
\end{figure}

\begin{figure}[!t]\centering
\begin{tabular}{c c} 
\includegraphics[angle=0,width=80mm]{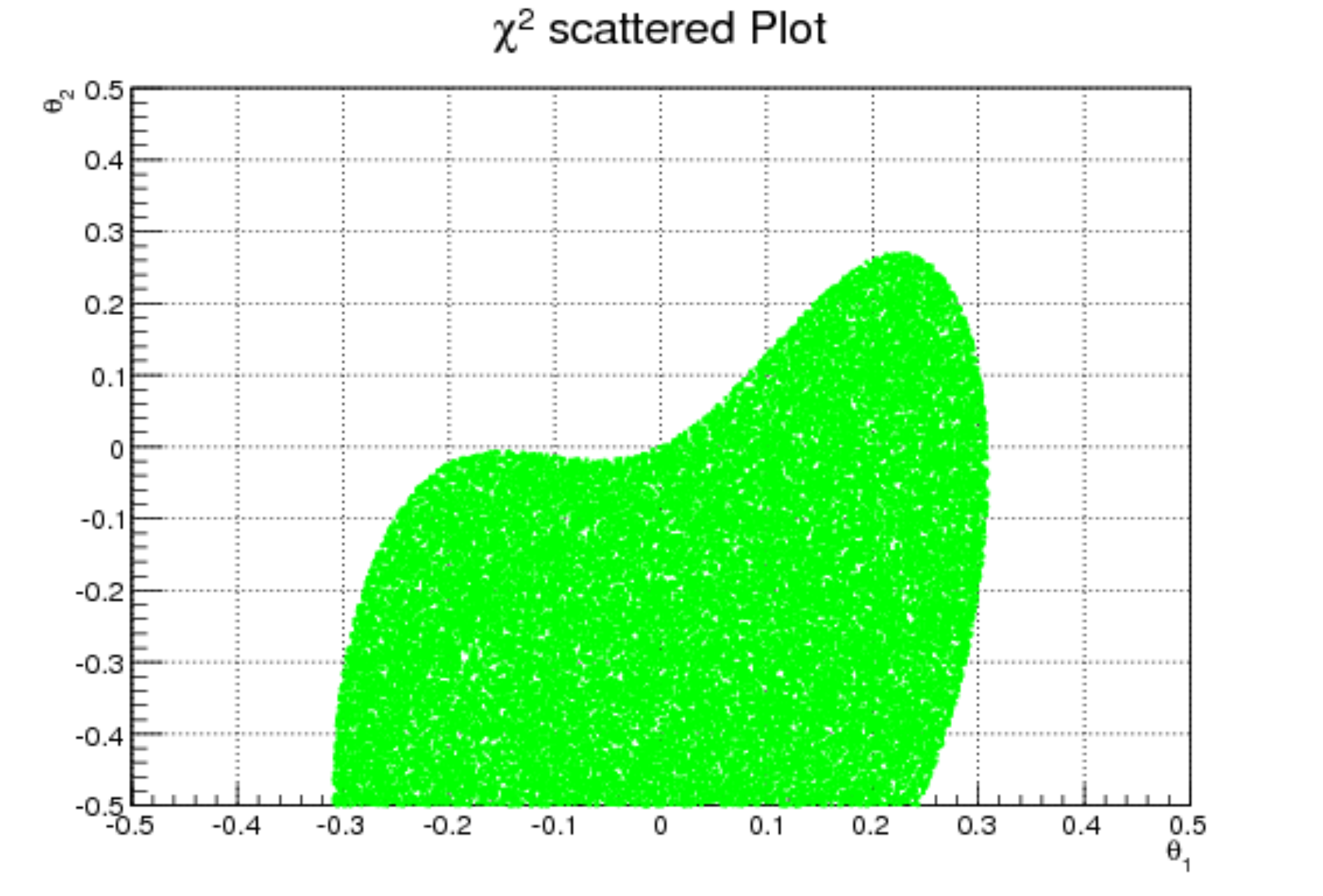} &
\includegraphics[angle=0,width=80mm]{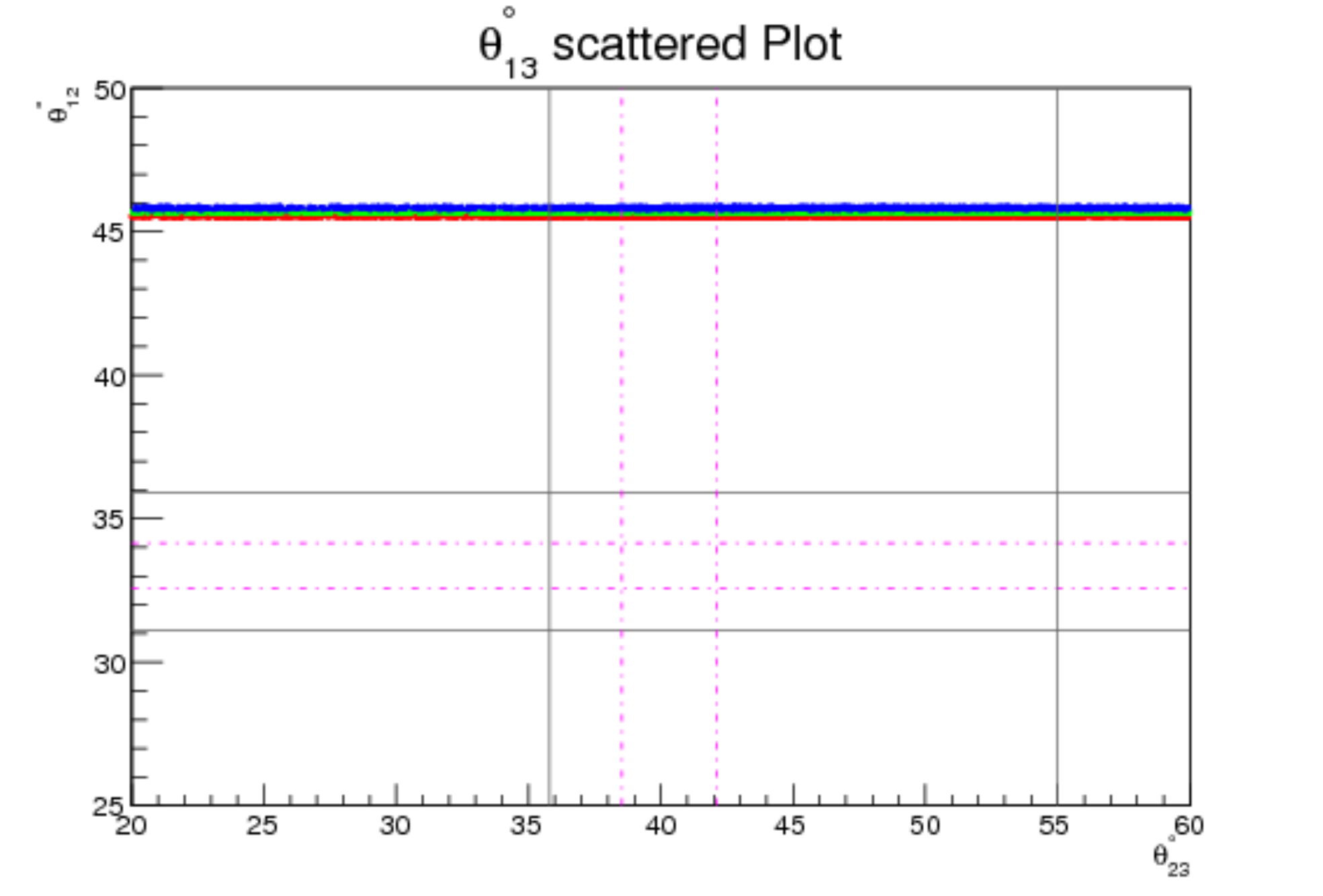}\\
\end{tabular}
\caption{$U_{DC}^{2313}$ scatter plot of $\chi^2$ (left fig.) over $\beta-\gamma$ (in radians) plane and $\theta_{13}$ (right fig.) 
over  $\theta_{23}-\theta_{12}$ (in degrees) plane. The information about color coding and various horizontal, vertical lines in right fig. is given in text.}
\label{fig.15}
\end{figure}

Figs.~\ref{fig.13}-\ref{fig.15} corresponds to TBM, BM and DC case respectively with $\theta_1 = \gamma$ and $\theta_2 = \beta$.
For this rotation scheme its not possible to get $\chi^2 < 10$ in viable parameter space for BM and DC case while for TBM the minimum
$\chi^2$ can lie in [3, 10] interval for extremely tiny parameter space. Since mixing angle 
$\theta_{12}$ doesn't receive any leading order corrections in this perturbation scheme so its value remain close to its unperturbed value. The value of $\gamma$ in BM and DC case is 
$|\gamma|\approx 0.175-0.245$ for $\theta_{13}$ to be in its 3$\sigma$ range. Thus for these cases $\theta_{12}$ lies in a vary narrow range
$[45.4^\circ, 45.8^\circ]$ which is outside its allowed 3$\sigma$ range. However corresponding value for TBM case is $|\gamma|\approx 0.154-0.213$
which implies $\theta_{12}\in [35.5^\circ, 35.8^\circ]$ that is consistent at 3$\sigma$ level. In this case its possible to fit $\theta_{13}$
and $\theta_{23}$ quite accurately to their central values which corresponds to $\chi^2 \in [3, 10]$ region.

\section{Rotations-$R_{ij}.U.R_{kl}(ij = kl)$}

Now we come to the rotations where $ij = kl$
and investigate their significance in fitting the neutrino mixing data.

\subsection{12-12 Rotation}
This case corresponds to rotations in 12 sector of  these special matrices. The neutrino mixing angles for small perturbation
parameters $\alpha_1$ and $\alpha_2$ are given by

\beqa
 \sin\theta_{13} &\approx&  |\alpha_1 U_{23} |,\\
 \sin\theta_{23} &\approx& |\frac{(\alpha_1^2-1) U_{23} }{\cos\theta_{13}}|,\\
 \sin\theta_{12} &\approx& |\frac{U_{12}+\alpha_1 U_{22}+\alpha_2 U_{11} -(\alpha_1^2+\alpha_2^2) U_{12} + \alpha_1 \alpha_2 U_{21}  }{\cos\theta_{13}}|.
\eeqa

\begin{figure}[!t]\centering
\begin{tabular}{c c} 
\includegraphics[angle=0,width=80mm]{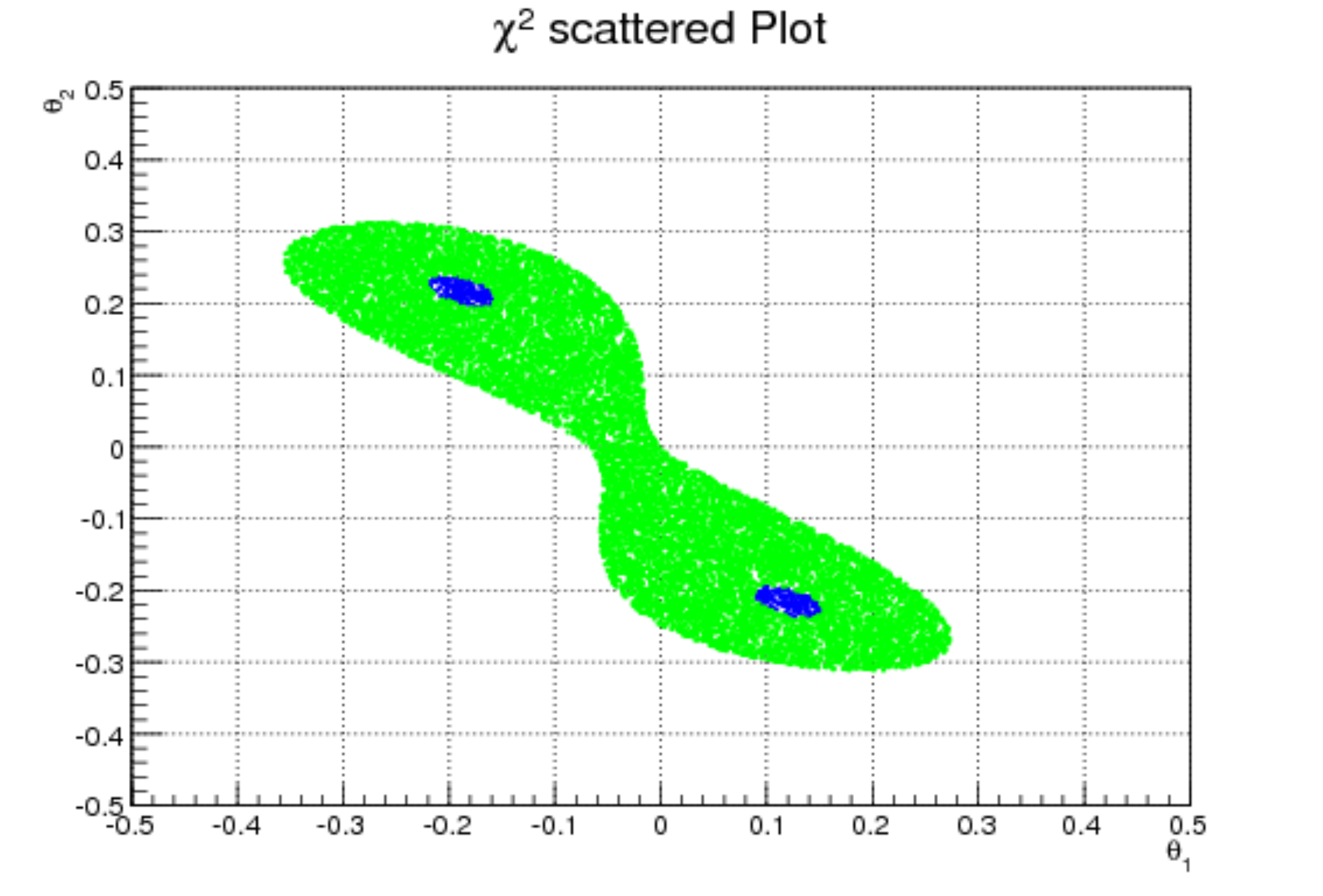} &
\includegraphics[angle=0,width=80mm]{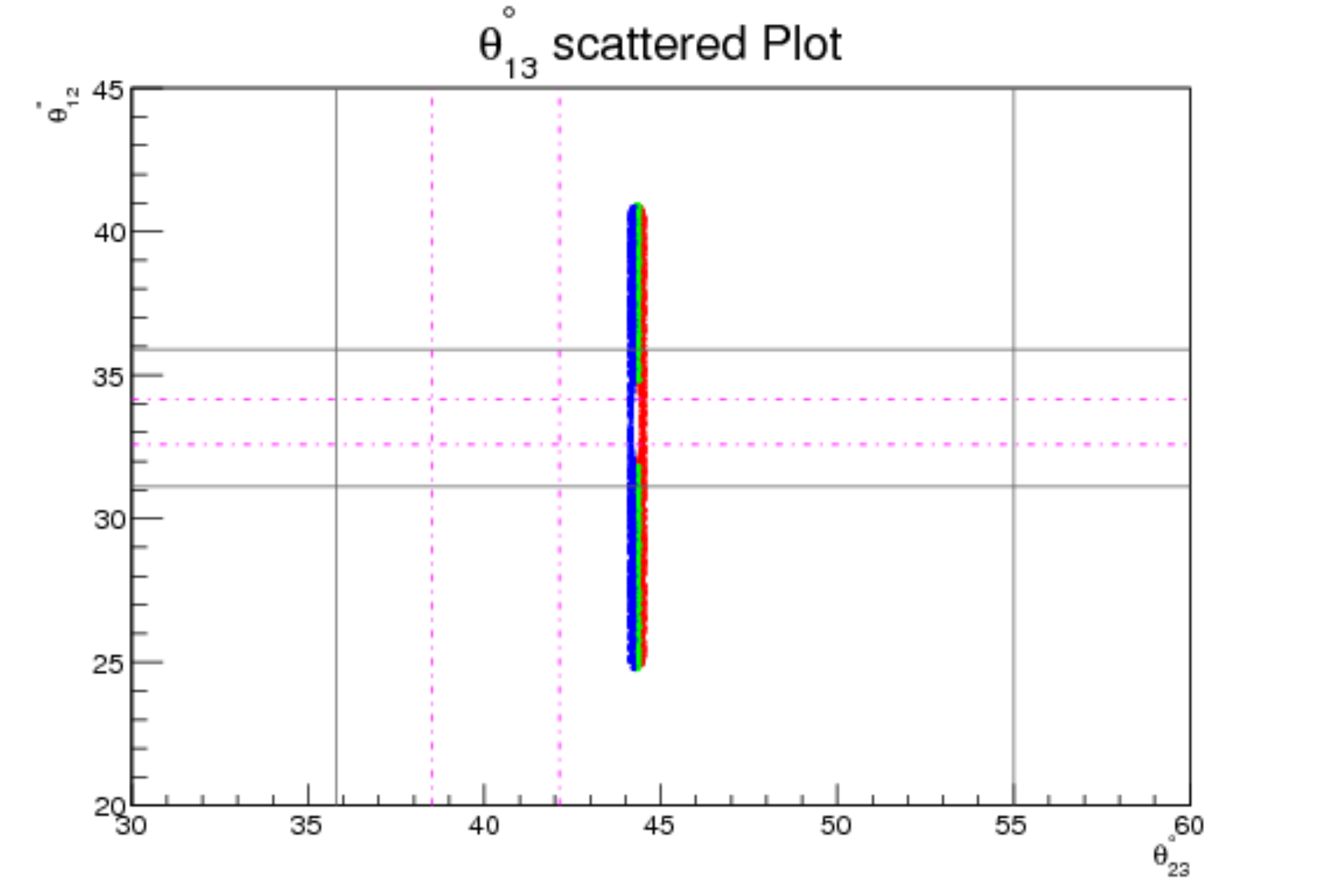}\\
\end{tabular}
\caption{$U_{TBM}^{1212}$ scatter plot of $\chi^2$ (left fig.) over $\alpha_1-\alpha_2$ (in radians) plane and $\theta_{13}$ (right fig.) 
over $\theta_{23}-\theta_{12}$ (in degrees) plane. The information about color coding and various horizontal, vertical lines in right fig. is given in text. }
\label{figa.1212}
\end{figure}

\begin{figure}[!t]\centering
\begin{tabular}{c c} 
\includegraphics[angle=0,width=80mm]{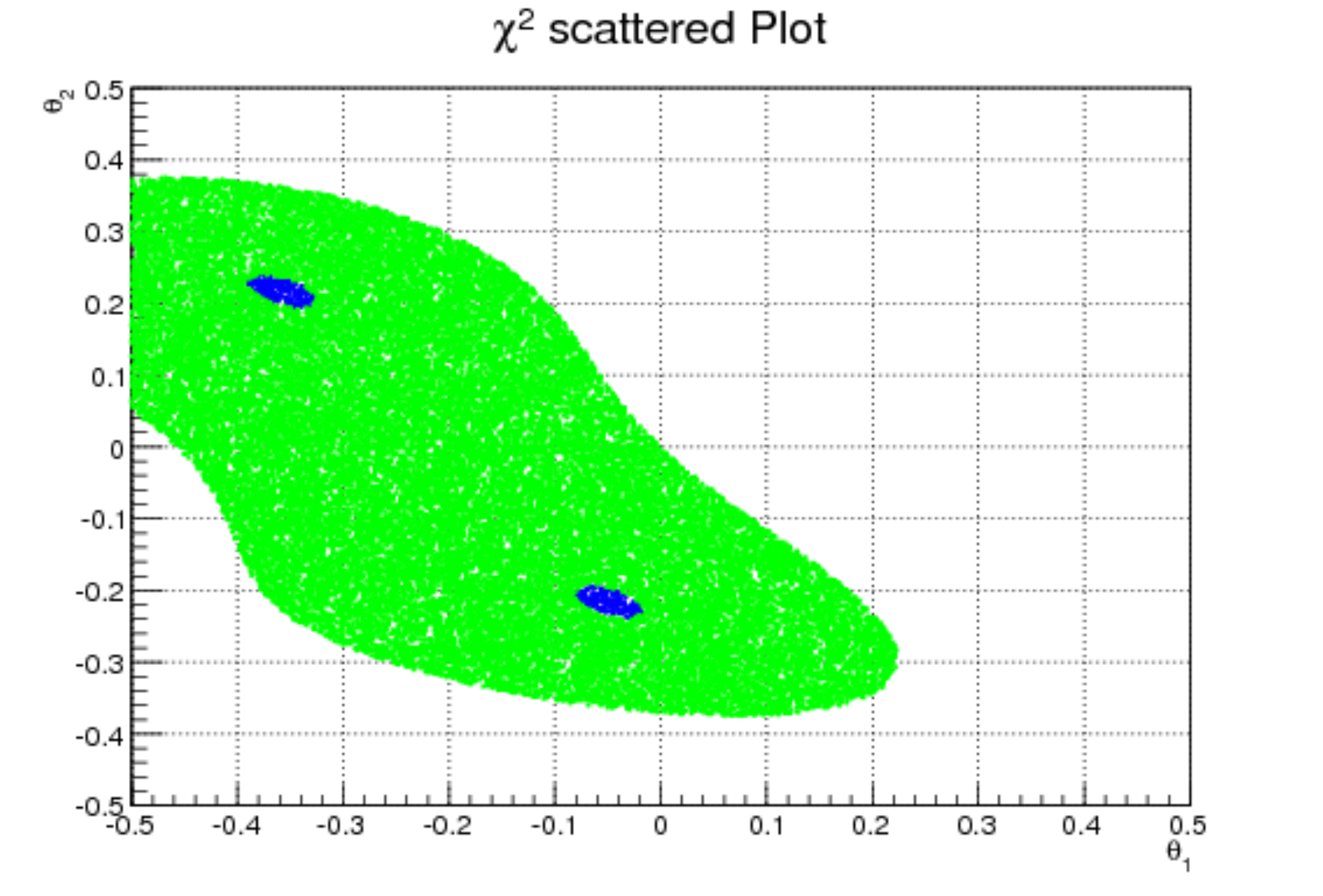} &
\includegraphics[angle=0,width=80mm]{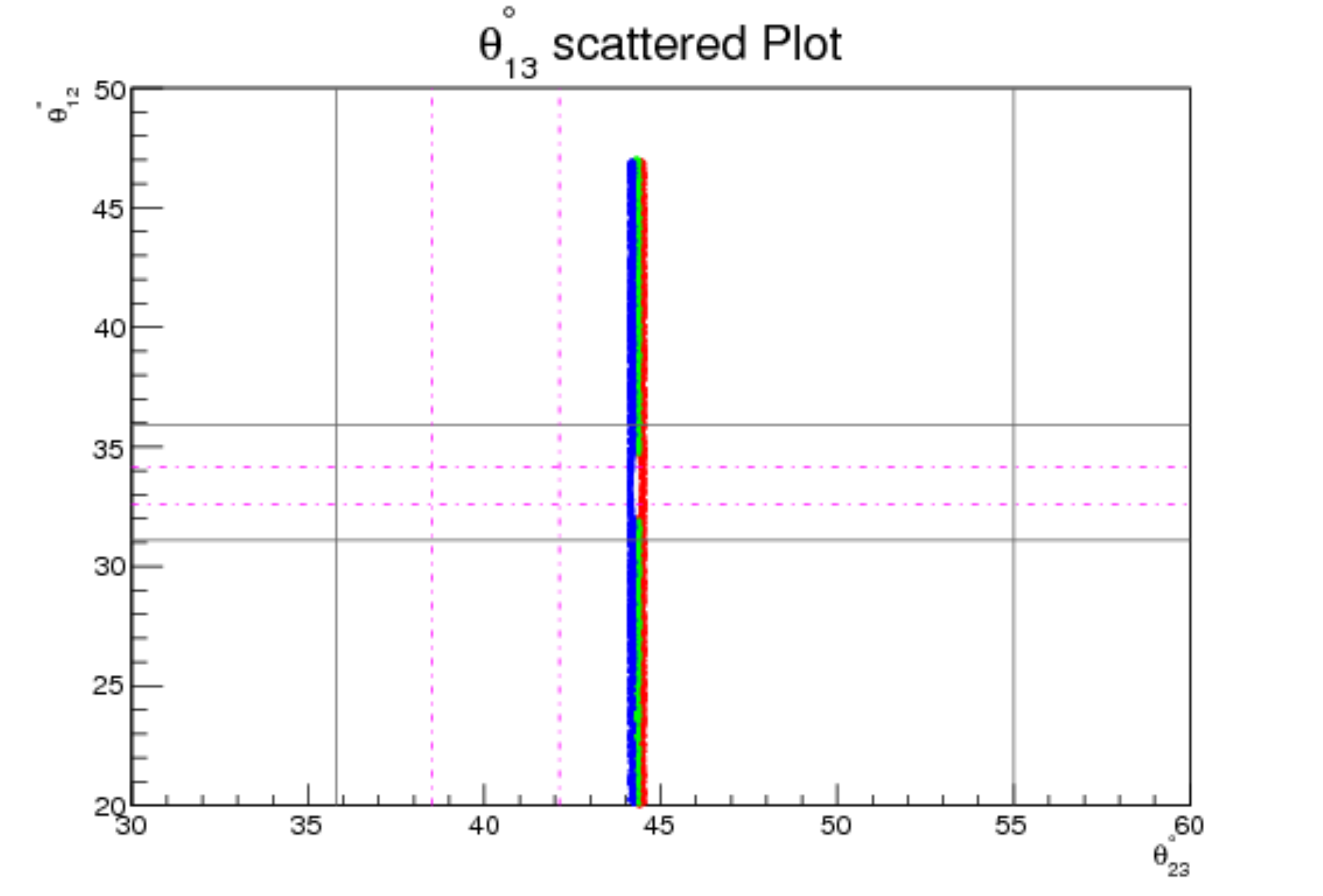}\\
\end{tabular}
\caption{$U_{BM}^{1212}$ scatter plot of $\chi^2$ (left fig.) over $\alpha_1-\alpha_2$ (in radians) plane and $\theta_{13}$ (right fig.) 
over  $\theta_{23}-\theta_{12}$ (in degrees) plane. The information about color coding and various horizontal, vertical lines in right fig. is given in text.}
\label{figb.1212}
\end{figure}

\begin{figure}[!t]\centering
\begin{tabular}{c c} 
\includegraphics[angle=0,width=80mm]{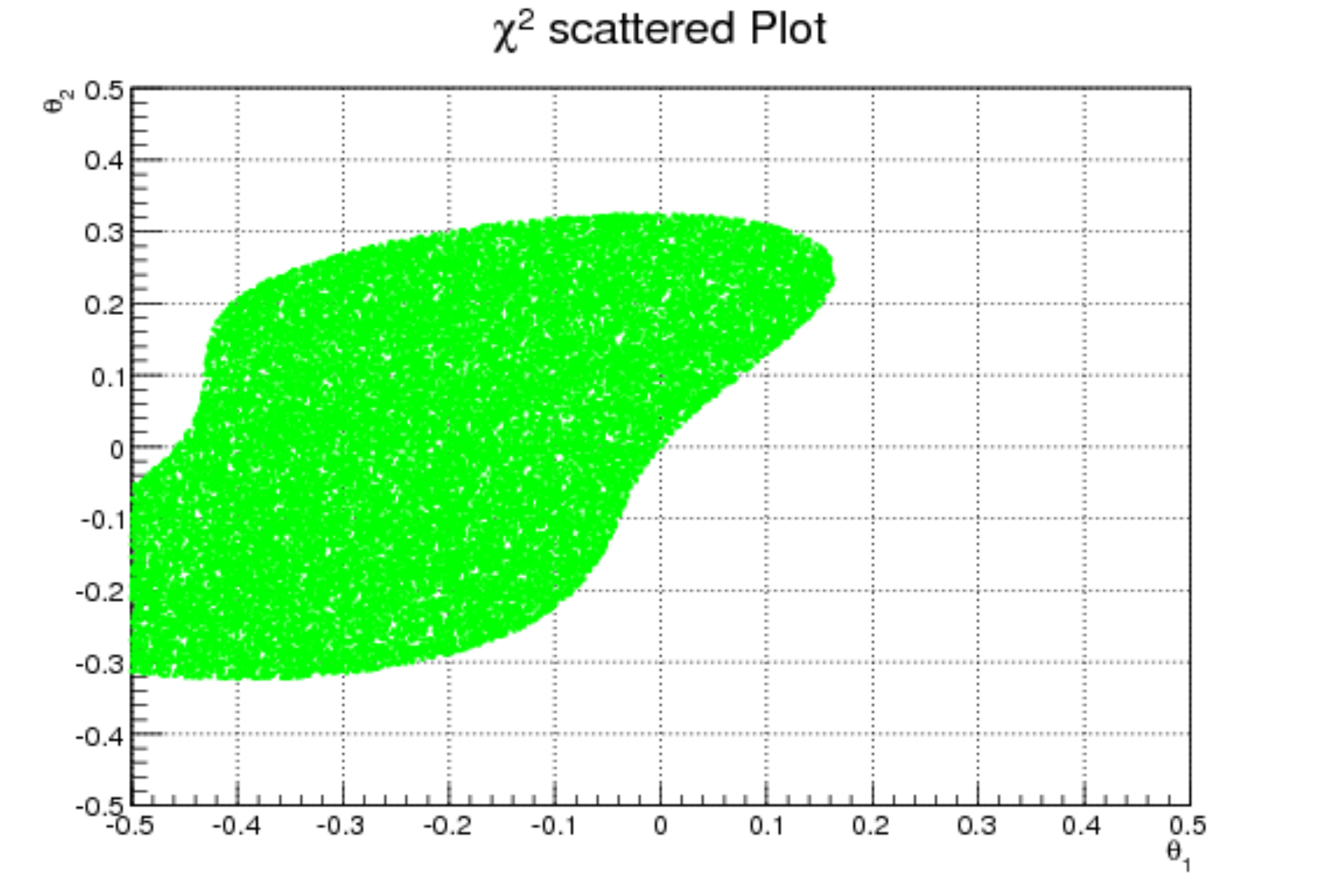} &
\includegraphics[angle=0,width=80mm]{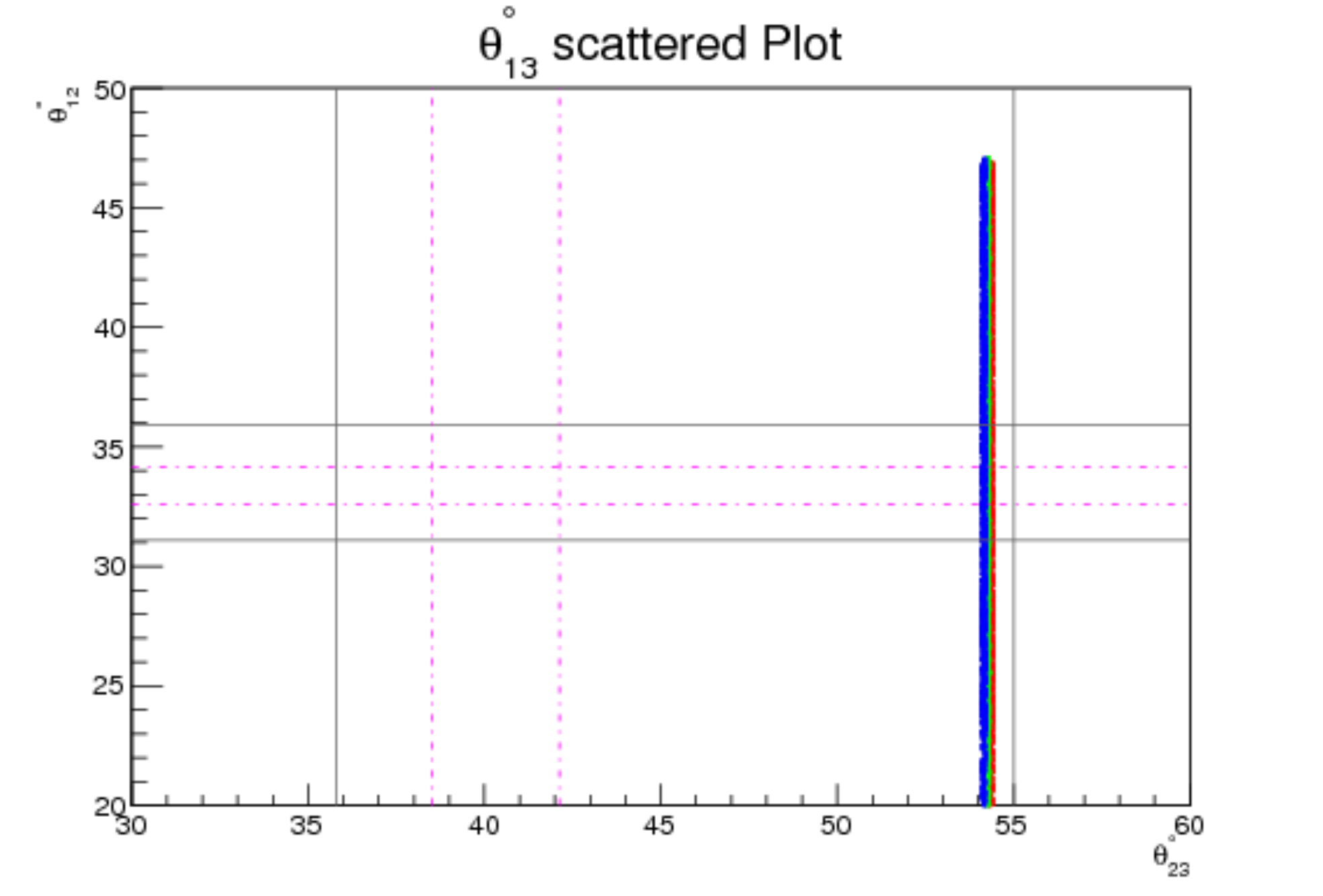}\\
\end{tabular}
\caption{$U_{DC}^{1212}$ scatter plot of $\chi^2$ (left fig.) over $\alpha_1-\alpha_2$ (in radians) plane and $\theta_{13}$ (right fig.) 
over  $\theta_{23}-\theta_{12}$ (in degrees) plane. The information about color coding and various horizontal, vertical lines in right fig. is given in text.}
\label{figc.1212}
\end{figure}

Figs.~\ref{figa.1212}-\ref{figc.1212} corresponds to TBM, BM and DC case respectively with $\theta_1 = \alpha_2$ and $\theta_2 = \alpha_1$.
For this perturbation the minimum value of $\chi^2 \in [3, 10]$ in TBM and BM case for a tiny region of parameter space. This minimum $\chi^2$
region prefers $|\alpha|\sim 0.2$ as around this parameter value $\theta_{13}$ and $\theta_{12}$ can be fitted quite accurately. However since $\theta_{23}$ mixing
angle doesn't receive corrections at leading order so it remains close to its original prediction. Thus all cases are only allowed at 3$\sigma$ level for this rotation 
scheme. The TBM and BM case lies quite close to 2$\sigma$ boundary of $\theta_{23}$ while DC is completely away from it. 

\subsection{13-13 Rotation}
This case corresponds to rotations in 13 sector of  these special matrices. The neutrino mixing angles for small perturbation
parameters $\gamma_1$ and $\gamma_2$ are given by

\beqa
 \sin\theta_{13} &\approx& | \gamma_1 U_{33} + \gamma_2 U_{11}+\gamma_1 \gamma_2 U_{31} |,\\
 \sin\theta_{23} &\approx& |\frac{U_{23}+\gamma_2 U_{21}-\gamma_2^2 U_{23}}{\cos\theta_{13}}|,\\
 \sin\theta_{12} &\approx& |\frac{U_{12}+\gamma_1 U_{32}-\gamma_1^2 U_{12}}{\cos\theta_{13}}|.
\eeqa

\begin{figure}[!t]\centering
\begin{tabular}{c c} 
\includegraphics[angle=0,width=80mm]{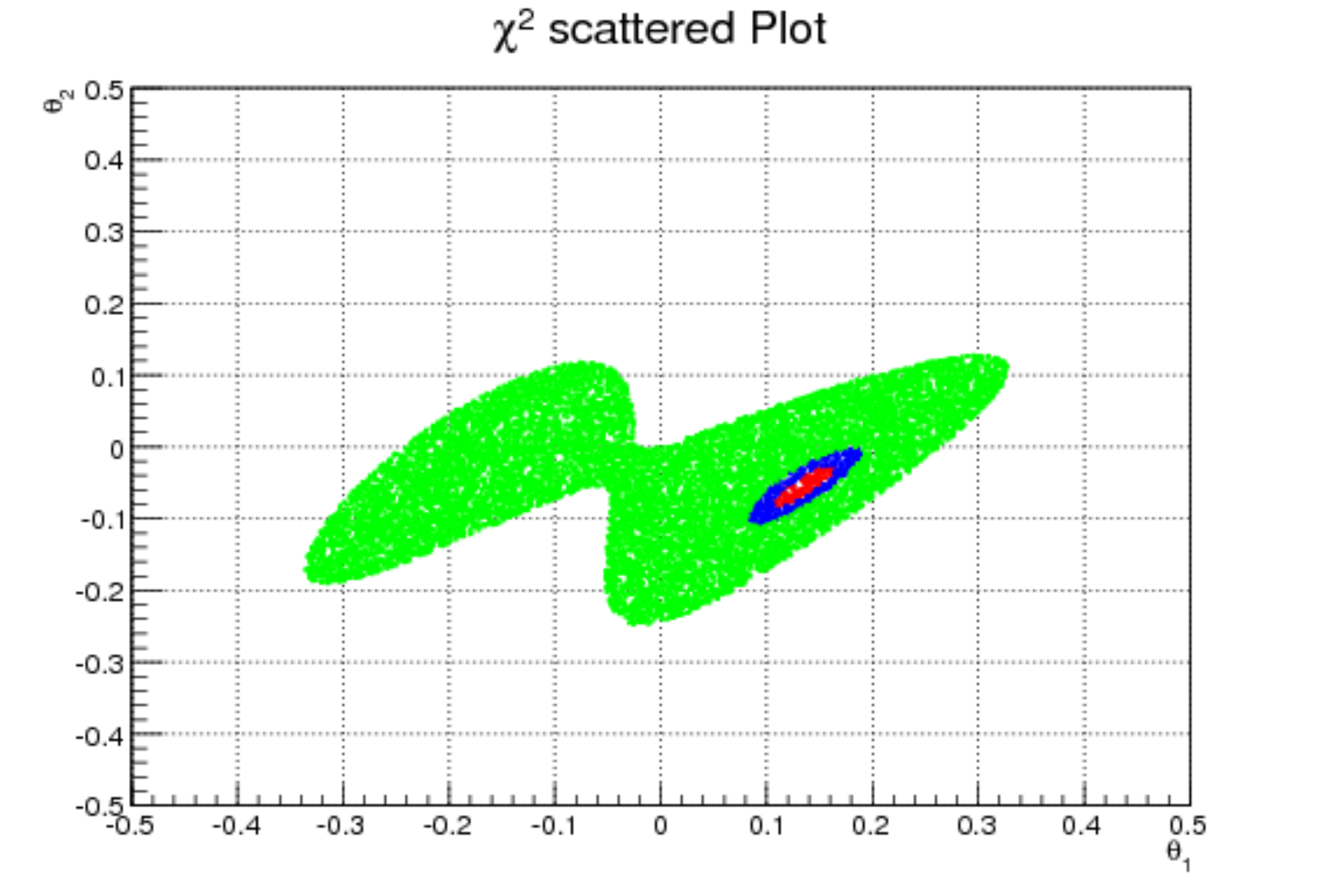} &
\includegraphics[angle=0,width=80mm]{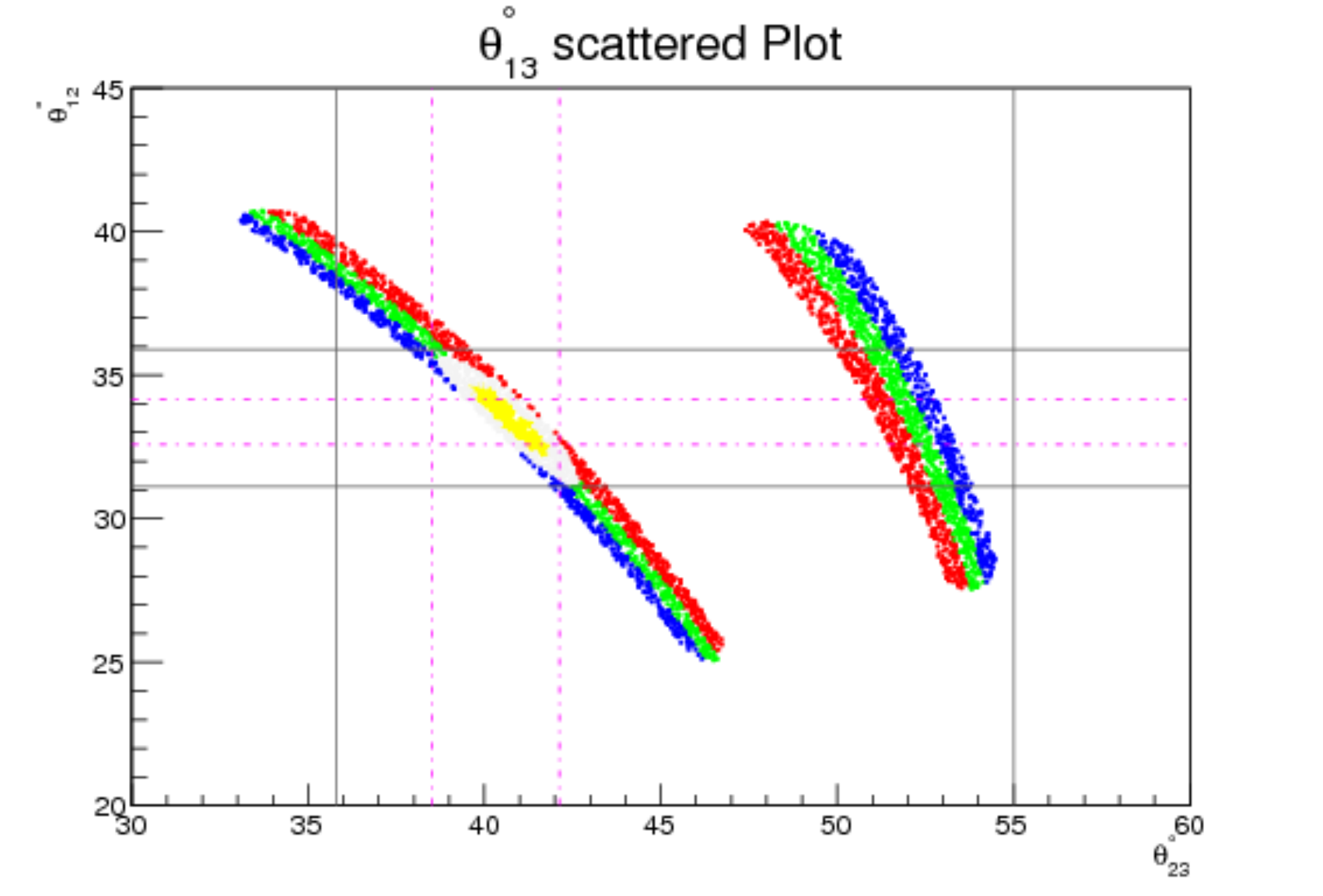}\\
\end{tabular}
\caption{$U_{TBM}^{1313}$ scatter plot of $\chi^2$ (left fig.) over $\gamma_1-\gamma_2$ (in radians) plane and $\theta_{13}$ (right fig.) 
over $\theta_{23}-\theta_{12}$ (in degrees) plane. The information about color coding and various horizontal, vertical lines in right fig. is given in text. }
\label{figa.1313}
\end{figure}

\begin{figure}[!t]\centering
\begin{tabular}{c c} 
\includegraphics[angle=0,width=80mm]{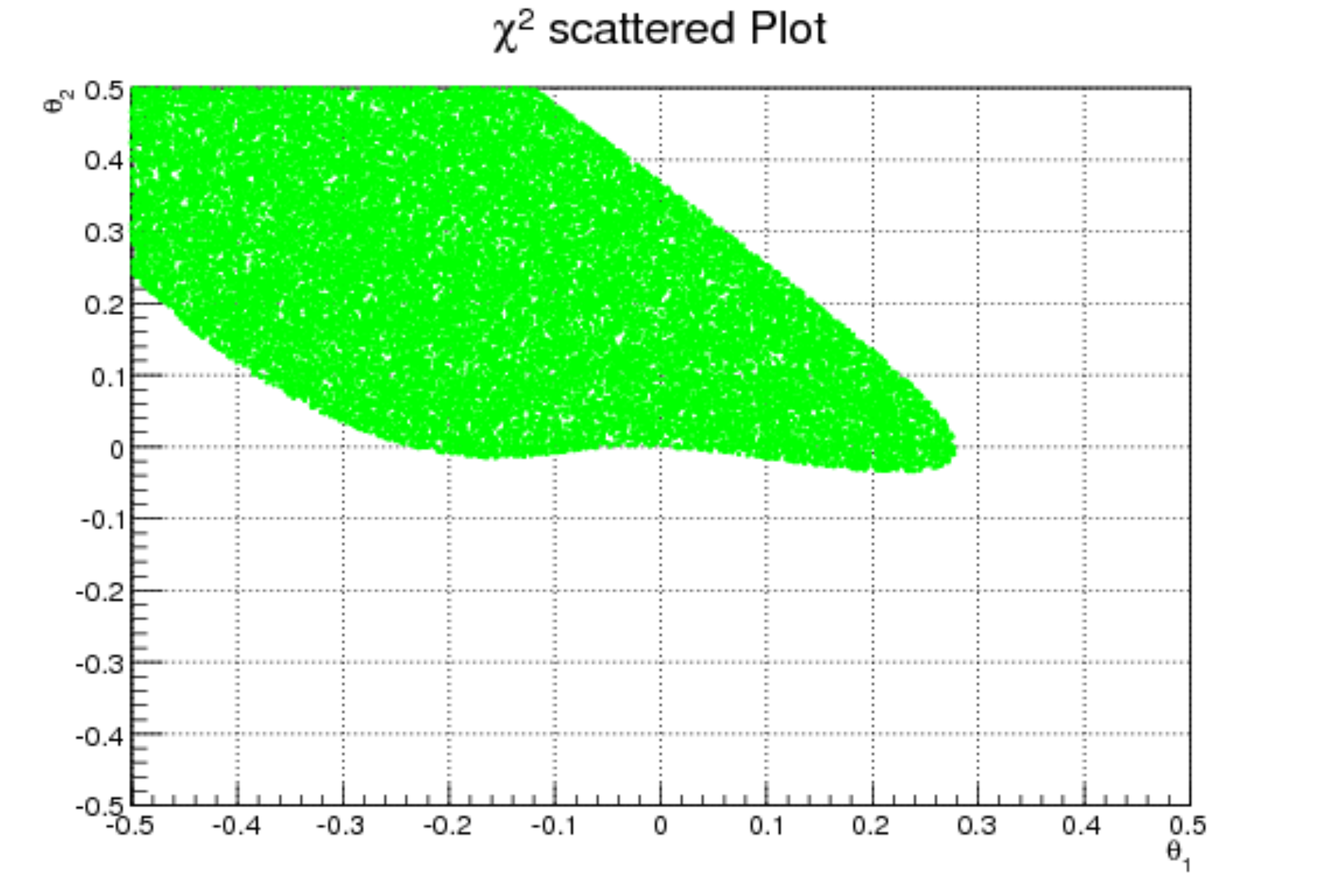} &
\includegraphics[angle=0,width=80mm]{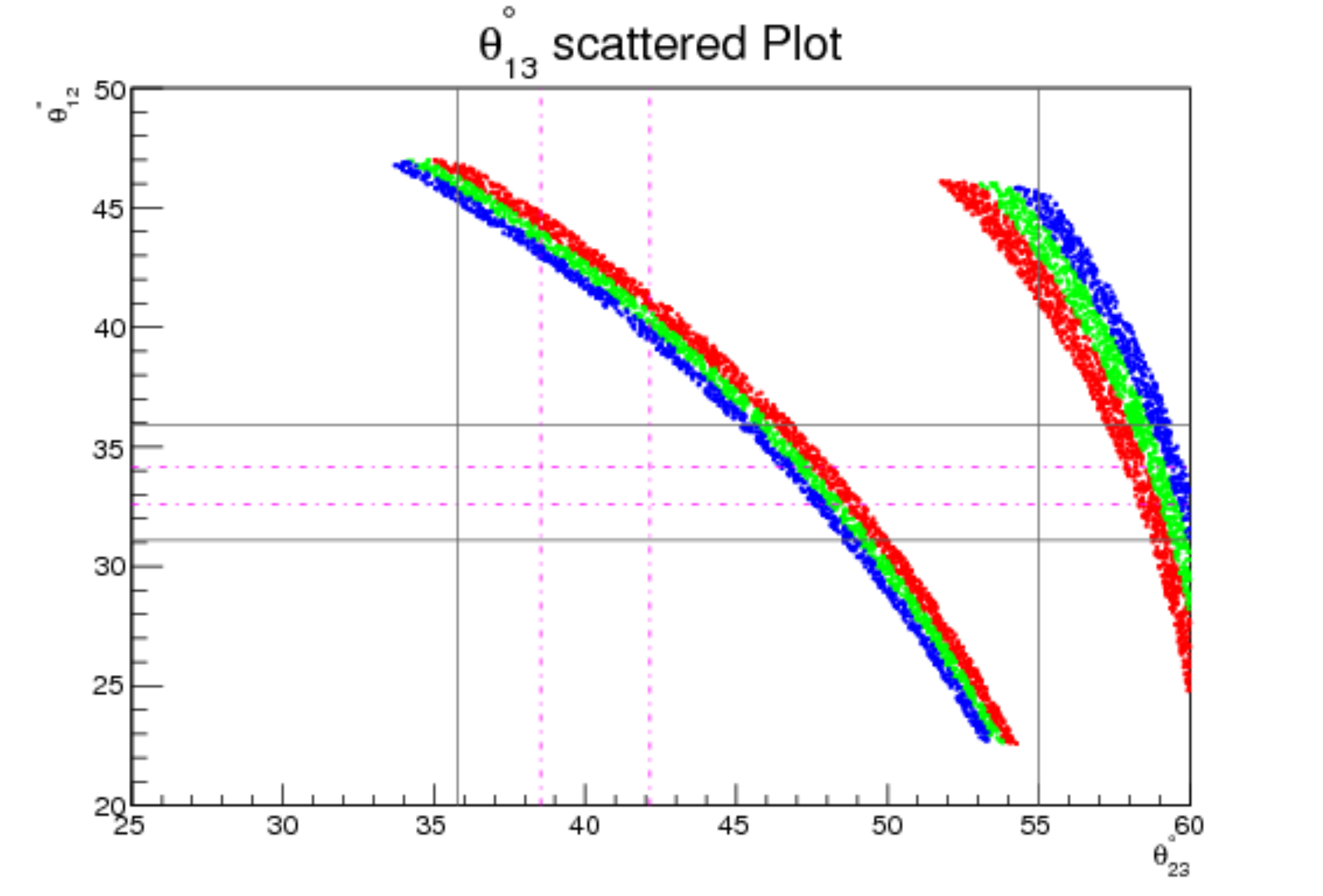}\\
\end{tabular}
\caption{$U_{BM}^{1313}$ scatter plot of $\chi^2$ (left fig.) over $\gamma_1-\gamma_2$ (in radians) plane and $\theta_{13}$ (right fig.) 
over  $\theta_{23}-\theta_{12}$ (in degrees) plane. The information about color coding and various horizontal, vertical lines in right fig. is given in text.}
\label{figb.1313}
\end{figure}

\begin{figure}[!t]\centering
\begin{tabular}{c c} 
\includegraphics[angle=0,width=80mm]{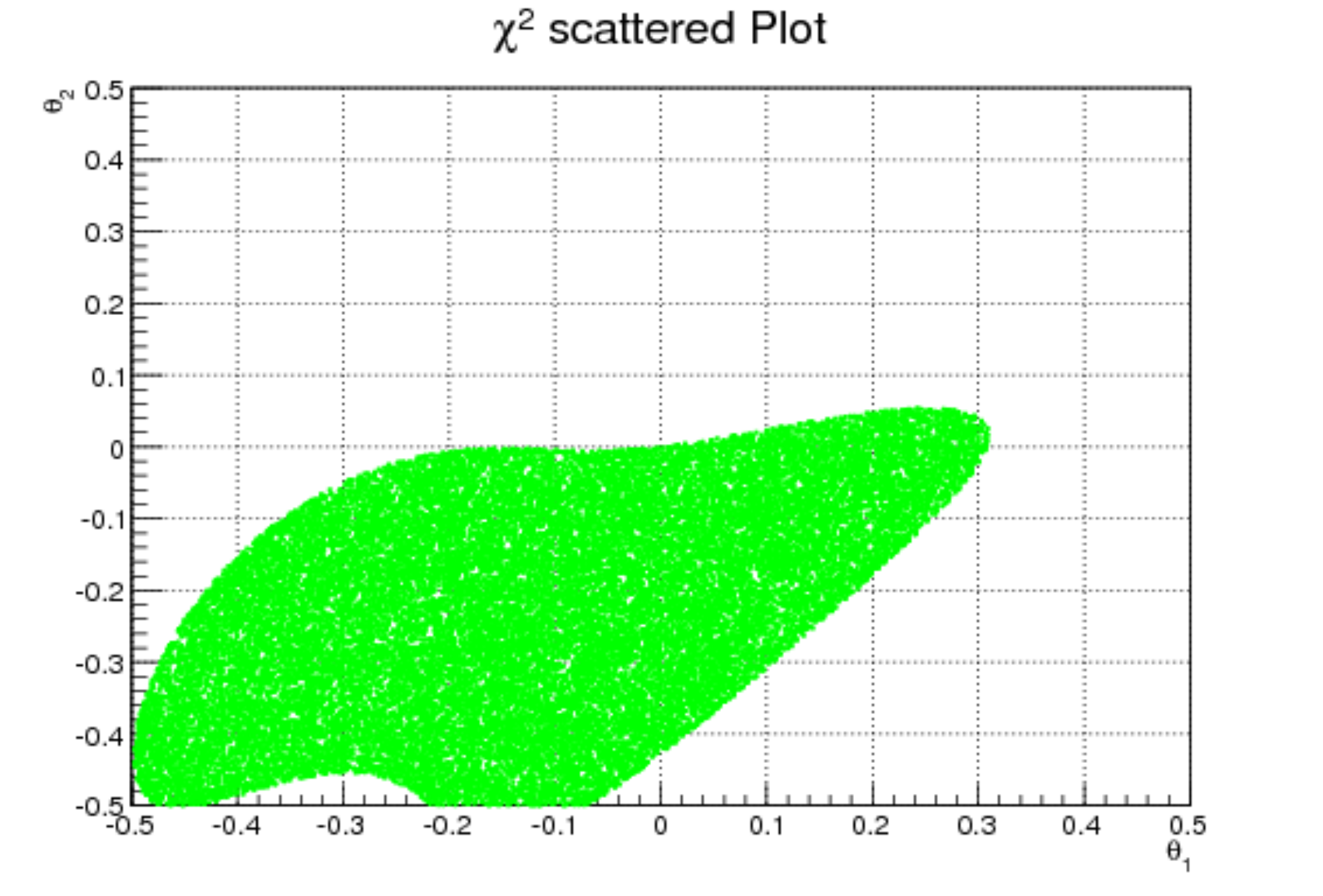} &
\includegraphics[angle=0,width=80mm]{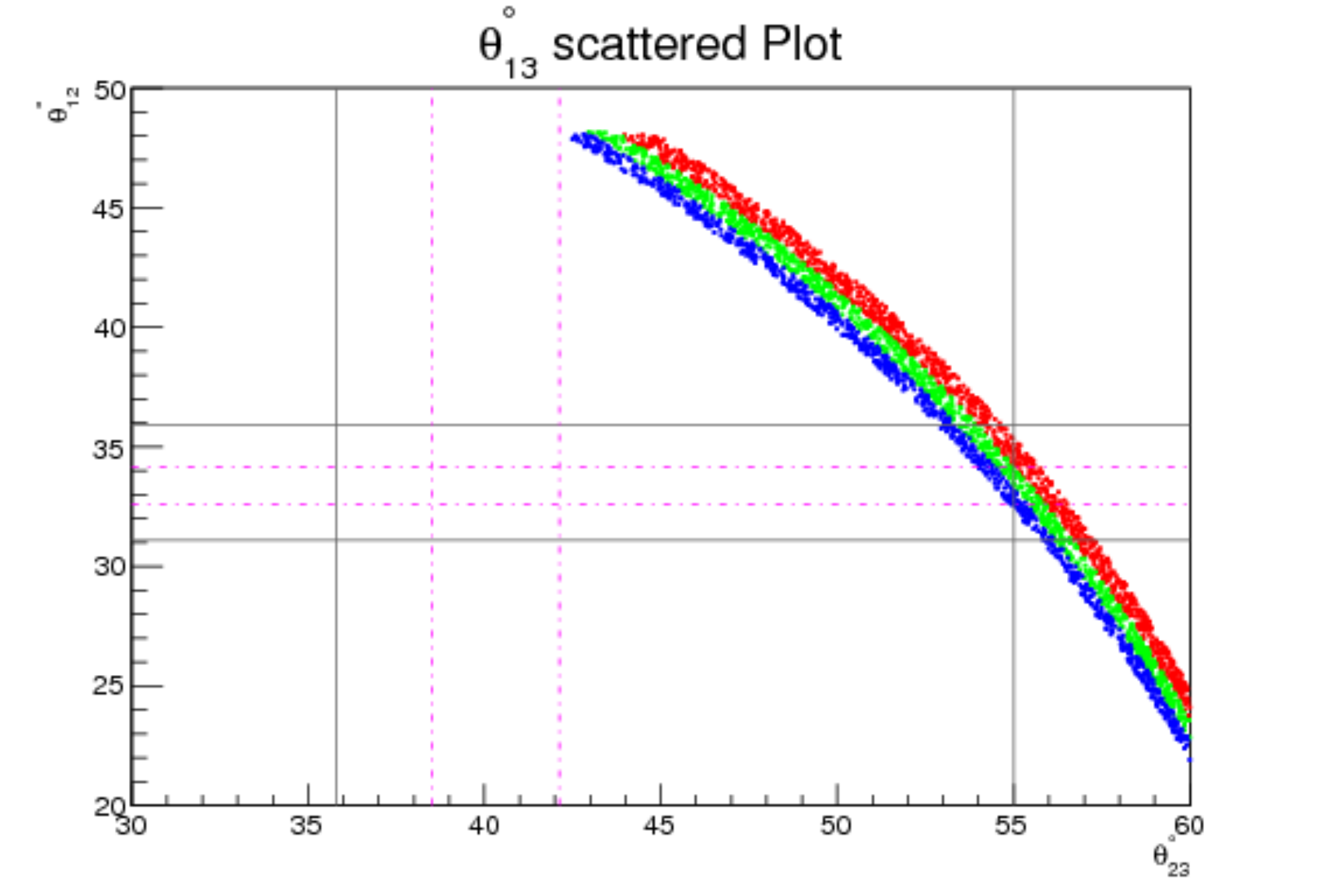}\\
\end{tabular}
\caption{$U_{DC}^{1313}$ scatter plot of $\chi^2$ (left fig.) over $\gamma_1-\gamma_2$ (in radians) plane and $\theta_{13}$ (right fig.) 
over  $\theta_{23}-\theta_{12}$ (in degrees) plane. The information about color coding and various horizontal, vertical lines in right fig. is given in text.}
\label{figc.1313}
\end{figure}

Figs.~\ref{figa.1313}-\ref{figc.1313} corresponds to TBM, BM and DC case respectively with $\theta_1 = \gamma_2$ and $\theta_2 = \gamma_1$.
Since correction parameters enter into all mixing angles at leading order so this rotation scheme exhibit nice correlations among
them. In case of TBM its possible to get $\chi^2 < 3$ while for other two cases minimum value of $\chi^2 \in [3, 10]$ always. In TBM case all mixing
angles can be fitted at $1\sigma$ level while for BM and DC case can only be fitted at 3$\sigma$ level.

\subsection{23-23 Rotation}
This case corresponds to rotations in 13 sector of  these special matrices. The neutrino mixing angles for small perturbation
parameters $\beta_1$ and $\beta_2$ are given by

\beqa
 \sin\theta_{13} &\approx& |\beta_2 U_{12}| ,\\
 \sin\theta_{23} &\approx& |\frac{U_{23} +\beta_1 U_{33} + \beta_2 U_{22} -(\beta_1^2+\beta_2^2) U_{23} +\beta_1 \beta_2 U_{32} }{\cos\theta_{13}}|,\\
 \sin\theta_{12} &\approx& |\frac{(\beta_2^2 -1) U_{12}}{\cos\theta_{13}}|.
\eeqa

\begin{figure}[!t]\centering
\begin{tabular}{c c} 
\includegraphics[angle=0,width=80mm]{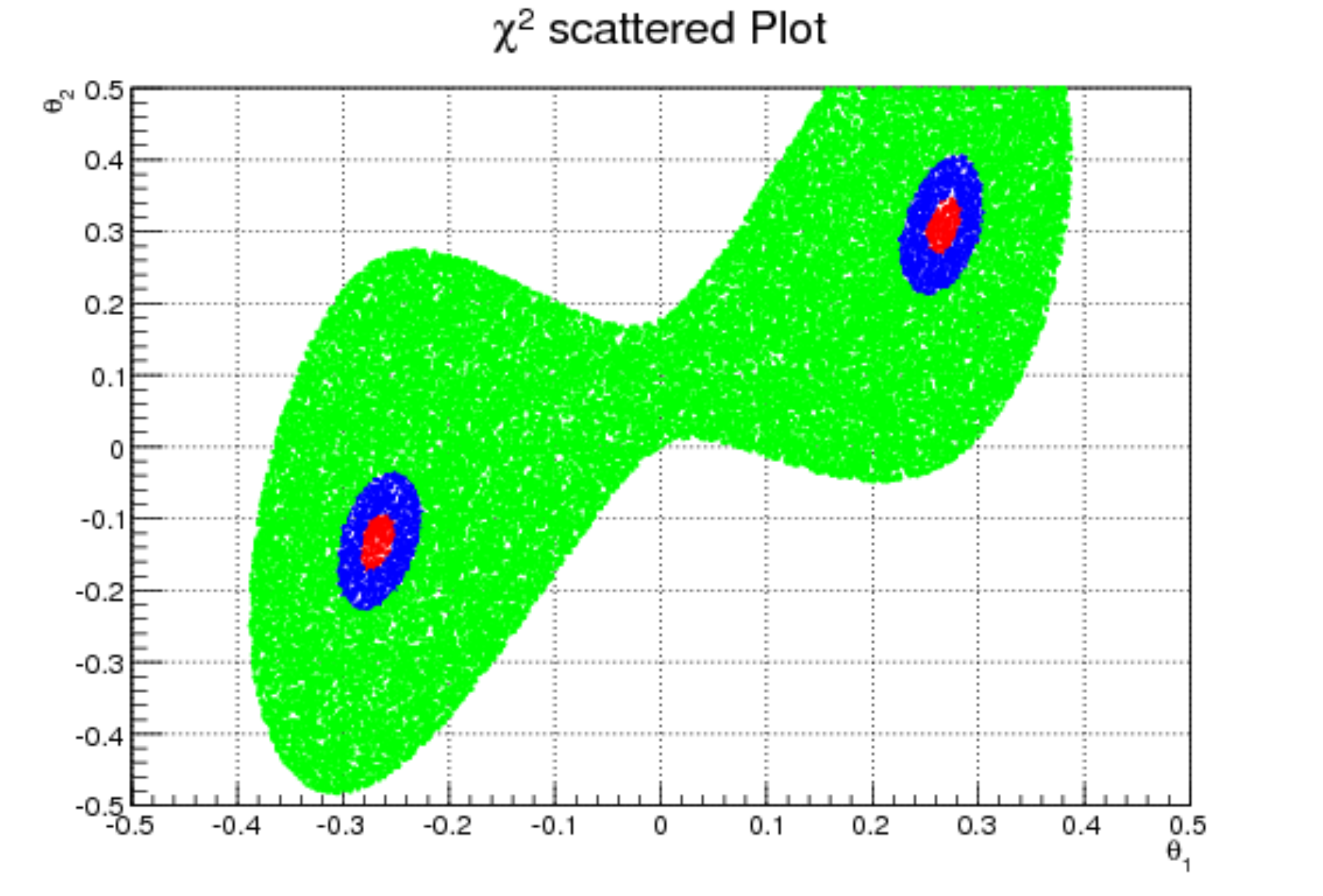} &
\includegraphics[angle=0,width=80mm]{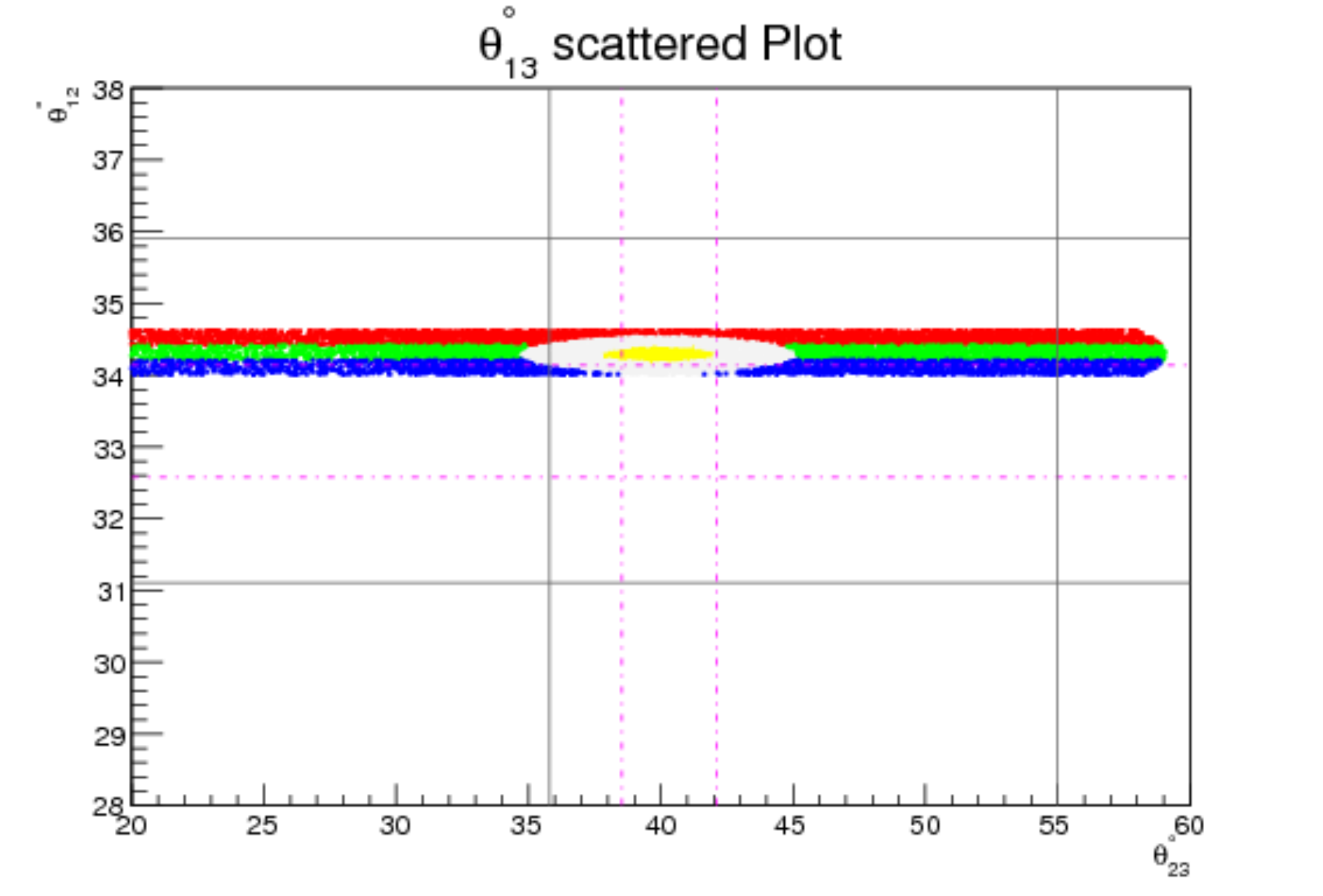}\\
\end{tabular}
\caption{$U_{TBM}^{2323}$ scatter plot of $\chi^2$ (left fig.) over $\beta_1-\beta_2$ (in radians) plane and $\theta_{13}$ (right fig.) 
over $\theta_{23}-\theta_{12}$ (in degrees) plane. The information about color coding and various horizontal, vertical lines in right fig. is given in text. }
\label{figa.2323}
\end{figure}

\begin{figure}[!t]\centering
\begin{tabular}{c c} 
\includegraphics[angle=0,width=80mm]{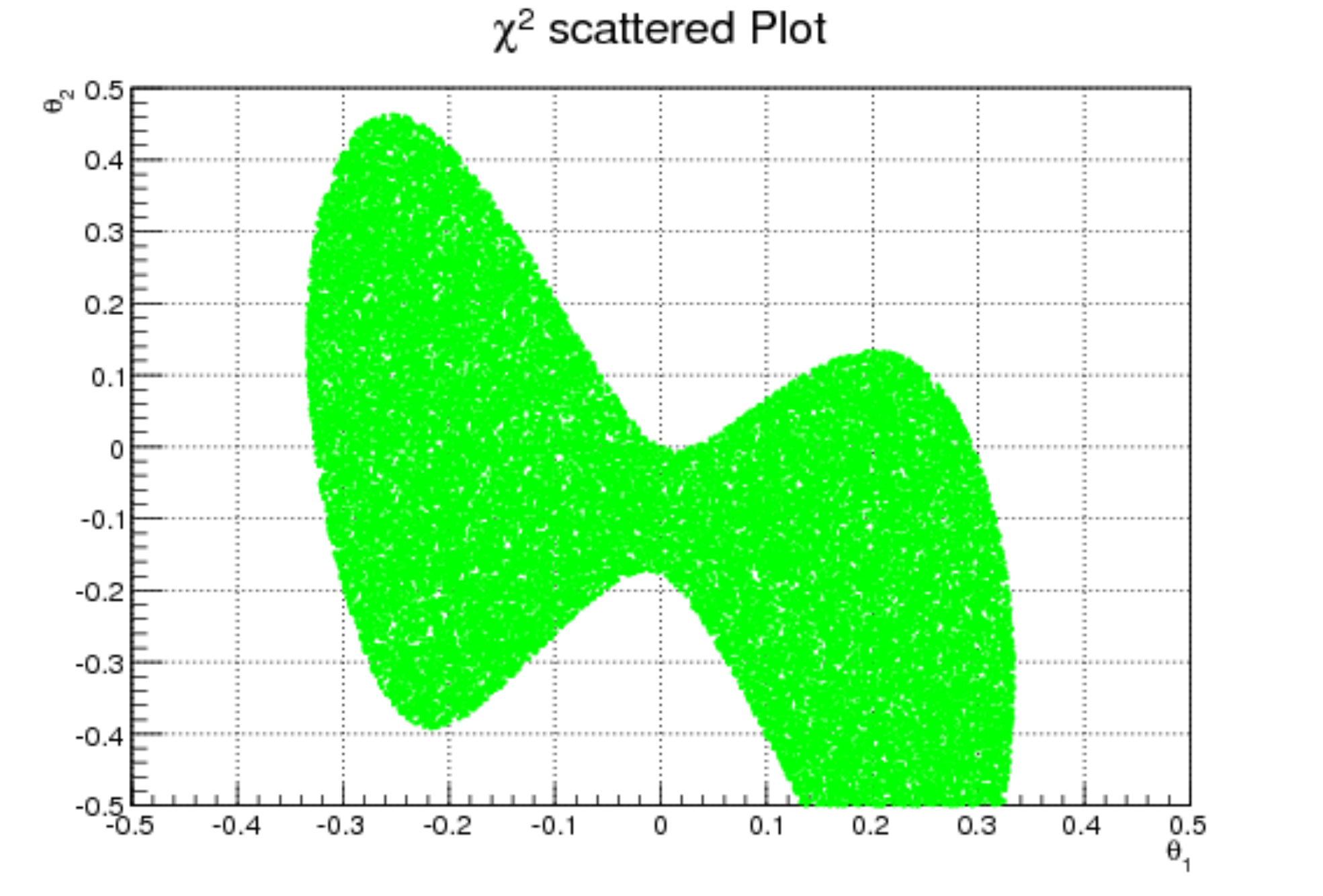} &
\includegraphics[angle=0,width=80mm]{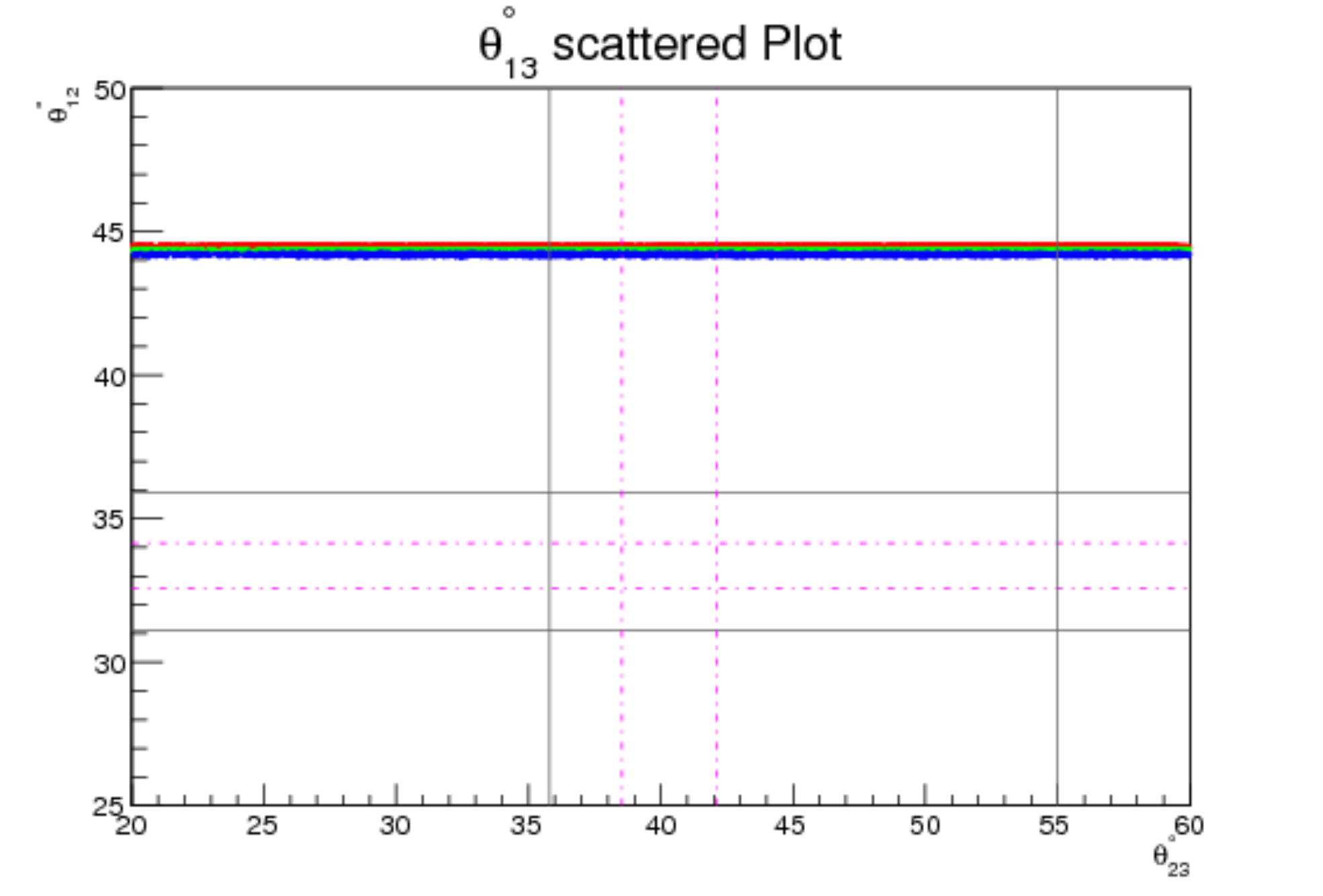}\\
\end{tabular}
\caption{$U_{BM}^{2323}$ scatter plot of $\chi^2$ (left fig.) over $\beta_1-\beta_2$ (in radians) plane and $\theta_{13}$ (right fig.) 
over  $\theta_{23}-\theta_{12}$ (in degrees) plane. The information about color coding and various horizontal, vertical lines in right fig. is given in text.}
\label{figb.2323}
\end{figure}

\begin{figure}[!t]\centering
\begin{tabular}{c c} 
\includegraphics[angle=0,width=80mm]{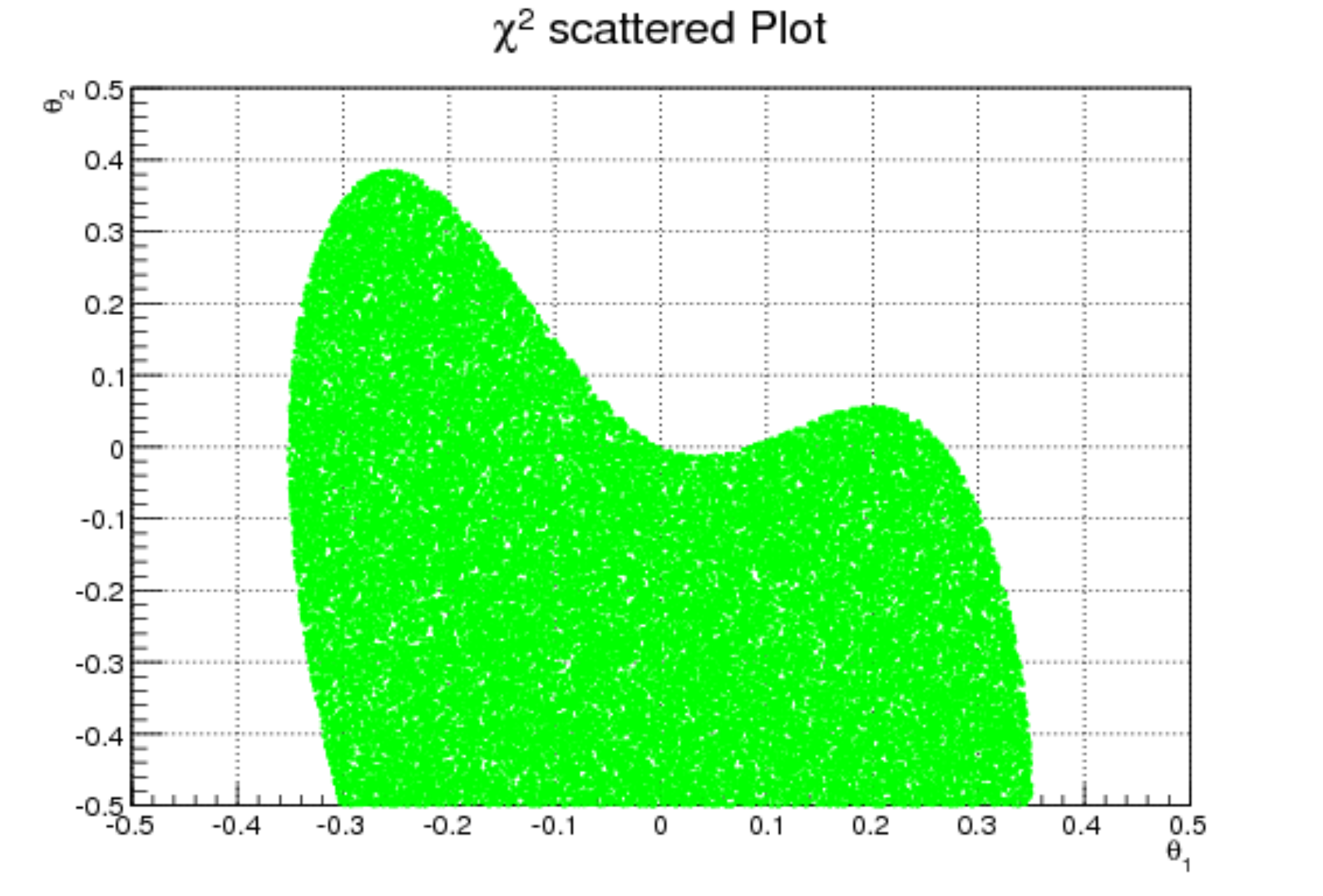} &
\includegraphics[angle=0,width=80mm]{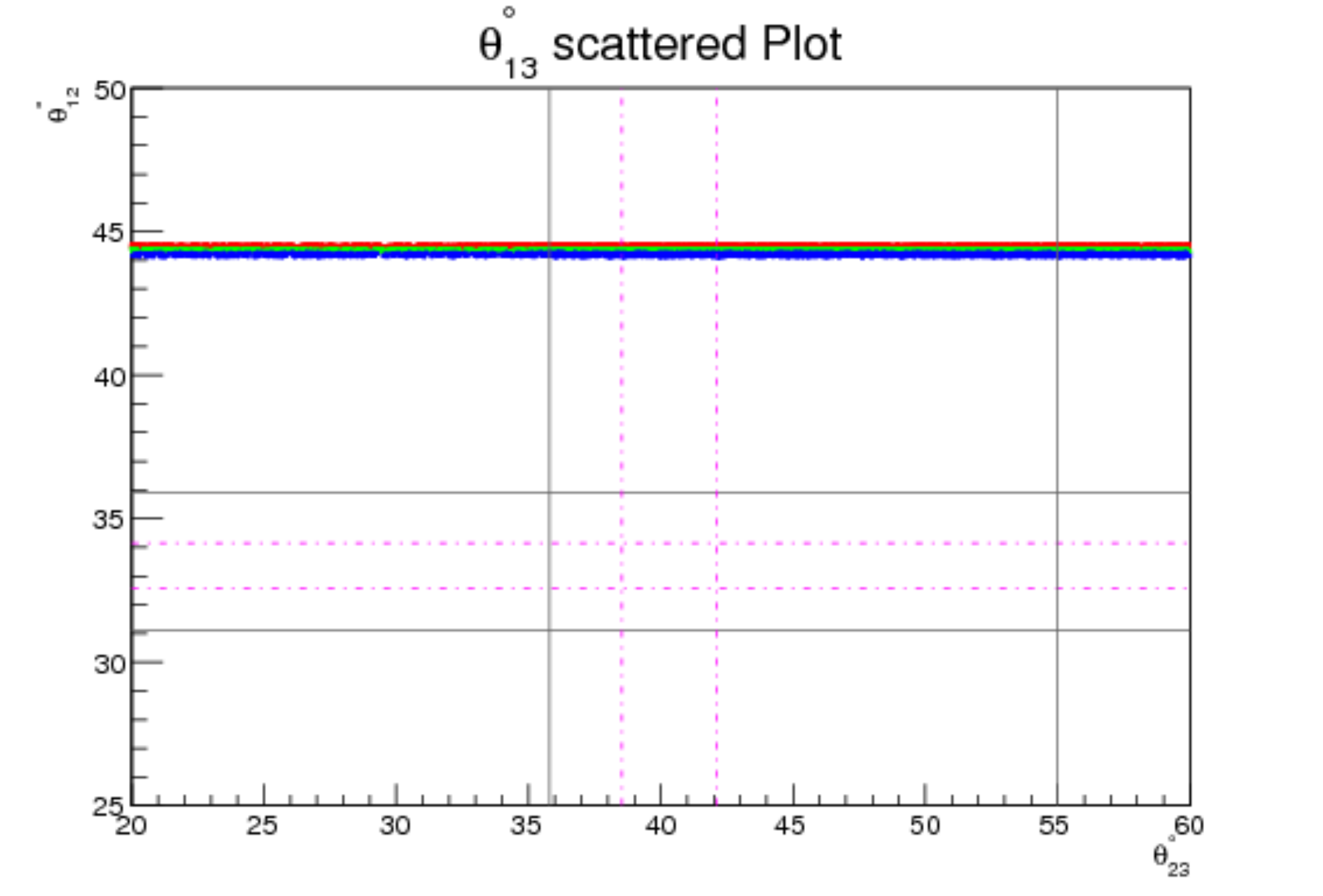}\\
\end{tabular}
\caption{$U_{DC}^{2323}$ scatter plot of $\chi^2$ (left fig.) over $\beta_1-\beta_2$ (in radians) plane and $\theta_{13}$ (right fig.) 
over  $\theta_{23}-\theta_{12}$ (in degrees) plane. The information about color coding and various horizontal, vertical lines in right fig. is given in text.}
\label{figc.2323}
\end{figure}

Figs.~\ref{figa.2323}-\ref{figc.2323} corresponds to TBM, BM and DC case respectively with $\theta_1 = \beta_2$ and $\theta_2 = \beta_1$.
For this perturbation its possible to $\chi^2 < 3$ in TBM case while for other two cases the minimum $\chi^2 \in [3, 10]$. For TBM mixing case minimum $\chi^2$
region prefers $|\beta_2|\sim 0.3$ as around this parameter value $\theta_{13}$ and $\theta_{23}$ can be fitted quite accurately. However as $\theta_{12}$ mixing
angle doesn't receive corrections at leading order so it remains close to its original prediction. Thus DC and BM case is excluded by 3$\sigma$ level
while TBM case is viable at 2$\sigma$ level for this rotation scheme. 

\section{Other Cases}

Here we also comment on few previously discussed cases \cite{Chaoetal} which were investigated by fixing one mixing angle and
reported to be  excluded at 3$\sigma$ level. Here we looked into the situation when all three angles can vary in their 3$\sigma$ range.
We marked $\chi^2 < 10$ region in mixing angle plots with large size black color dots. 

\subsection{13-12 Rotation}

This case corresponds to rotations in 13 and 12 sector and the modified PMNS matrix is given by $U_{PMNS} = R_{13}\cdot R_{12} \cdot U$. 
The neutrino mixing angles for small perturbation parameters $\alpha$ and $\gamma$ are given by

\beqa
 \sin\theta_{13} &\approx&  |\alpha U_{23} + \gamma U_{33}|,\\
  \sin\theta_{23} &\approx& | \frac{(1-\alpha^2) U_{23} }{\cos\theta_{13}}|,\\
  \sin\theta_{12} &\approx& |\frac{ U_{12} + \alpha U_{22} + \gamma U_{32} -(\alpha^2 + \gamma^2) U_{12} }{\cos\theta_{13}}|.
\eeqa

\begin{figure}[!t]\centering
\begin{tabular}{c c} 
\includegraphics[angle=0,width=80mm]{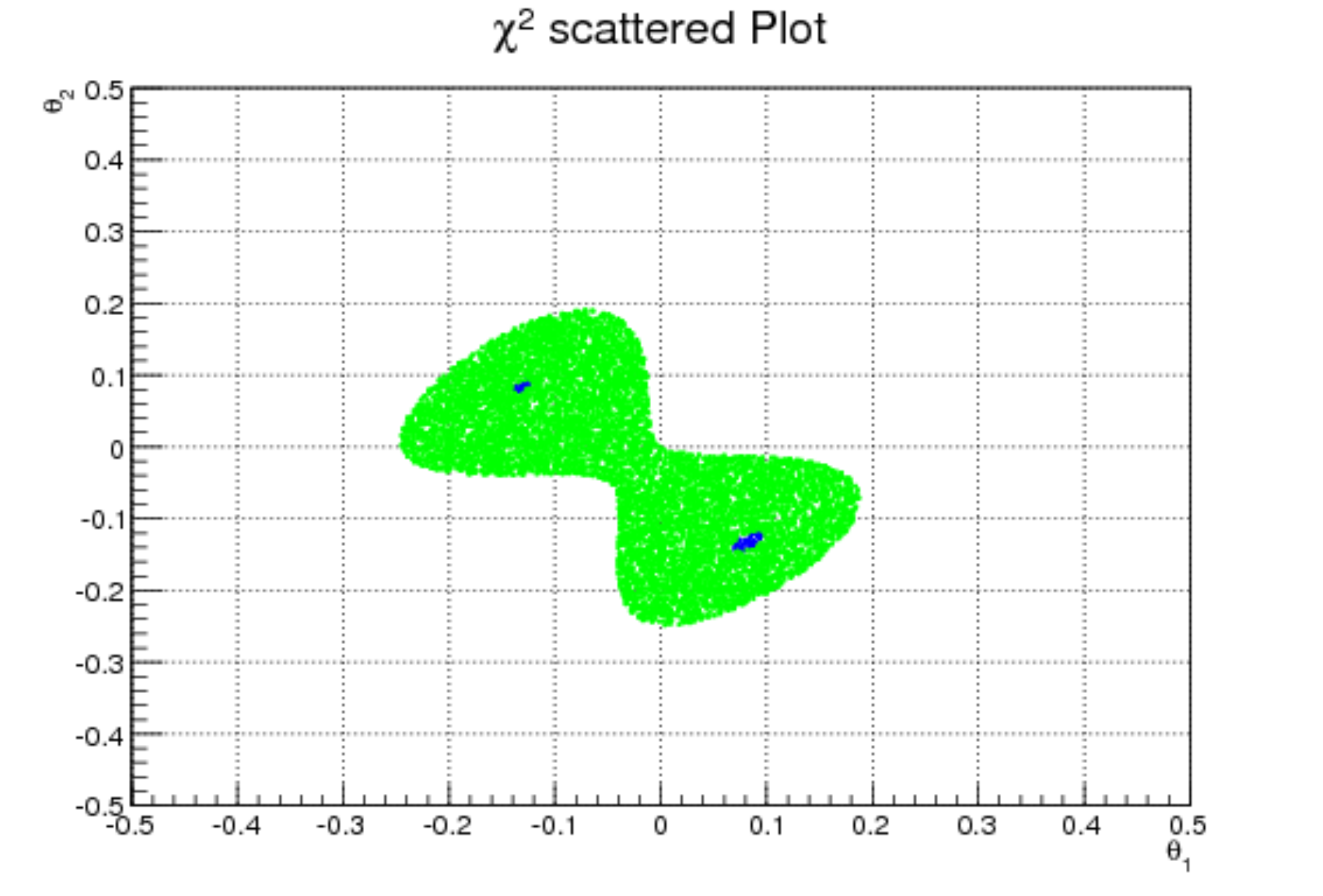} &
\includegraphics[angle=0,width=80mm]{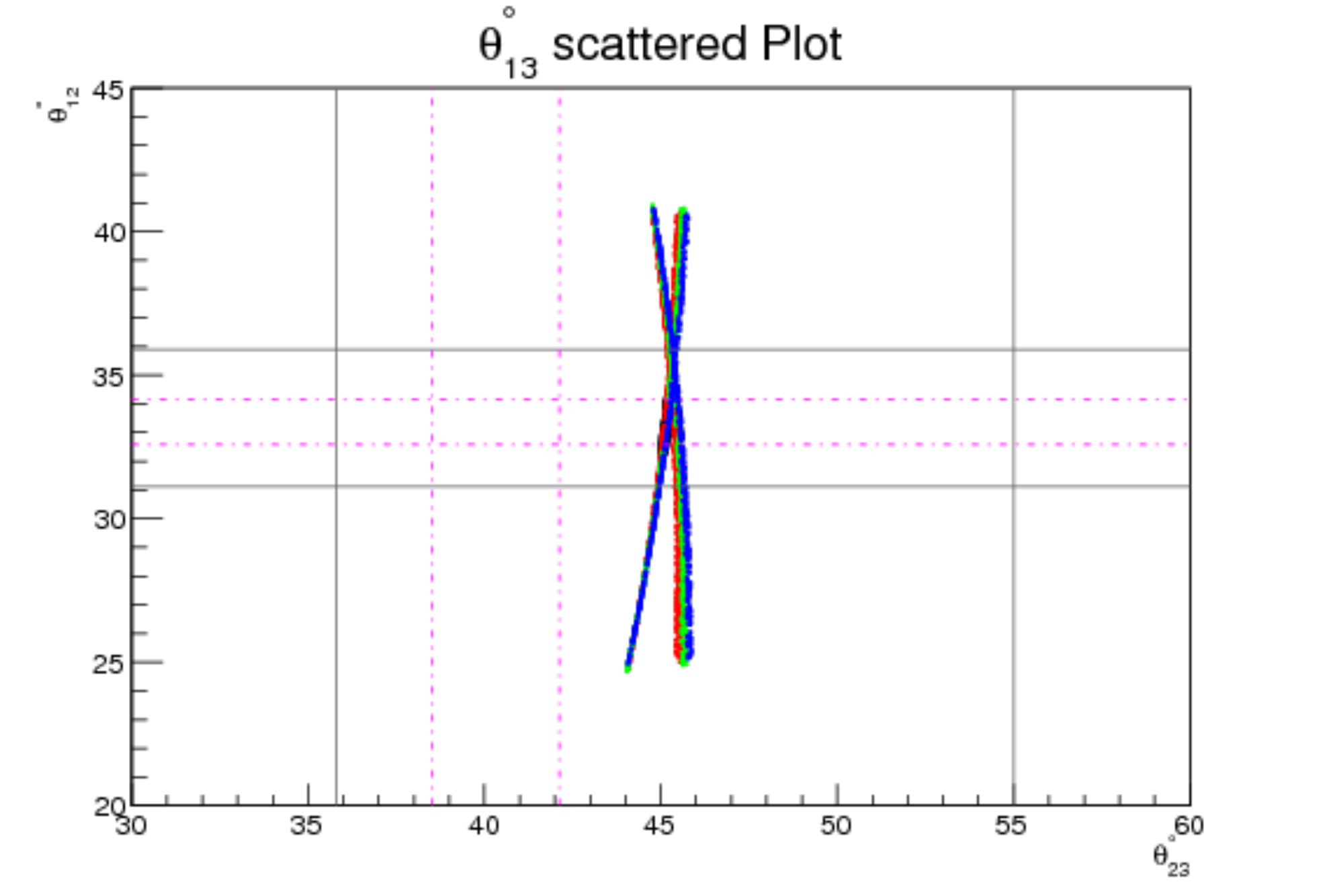}\\
\end{tabular} 
\caption{$U_{TBM}^{1312}(= R_{13}\cdot R_{12}\cdot U_{TBM})$ scatter plot of $\chi^2$ (left fig.) over $\gamma-\alpha$(in radians) plane and $\theta_{13}$ (right fig.) 
over  $\theta_{23}-\theta_{12}$ (in degrees) plane. The information about color coding and various horizontal, vertical lines in right fig. is given in text.}
\label{fig.16} 
\end{figure}

\begin{figure}[!t]\centering
\begin{tabular}{c c} 
\includegraphics[angle=0,width=80mm]{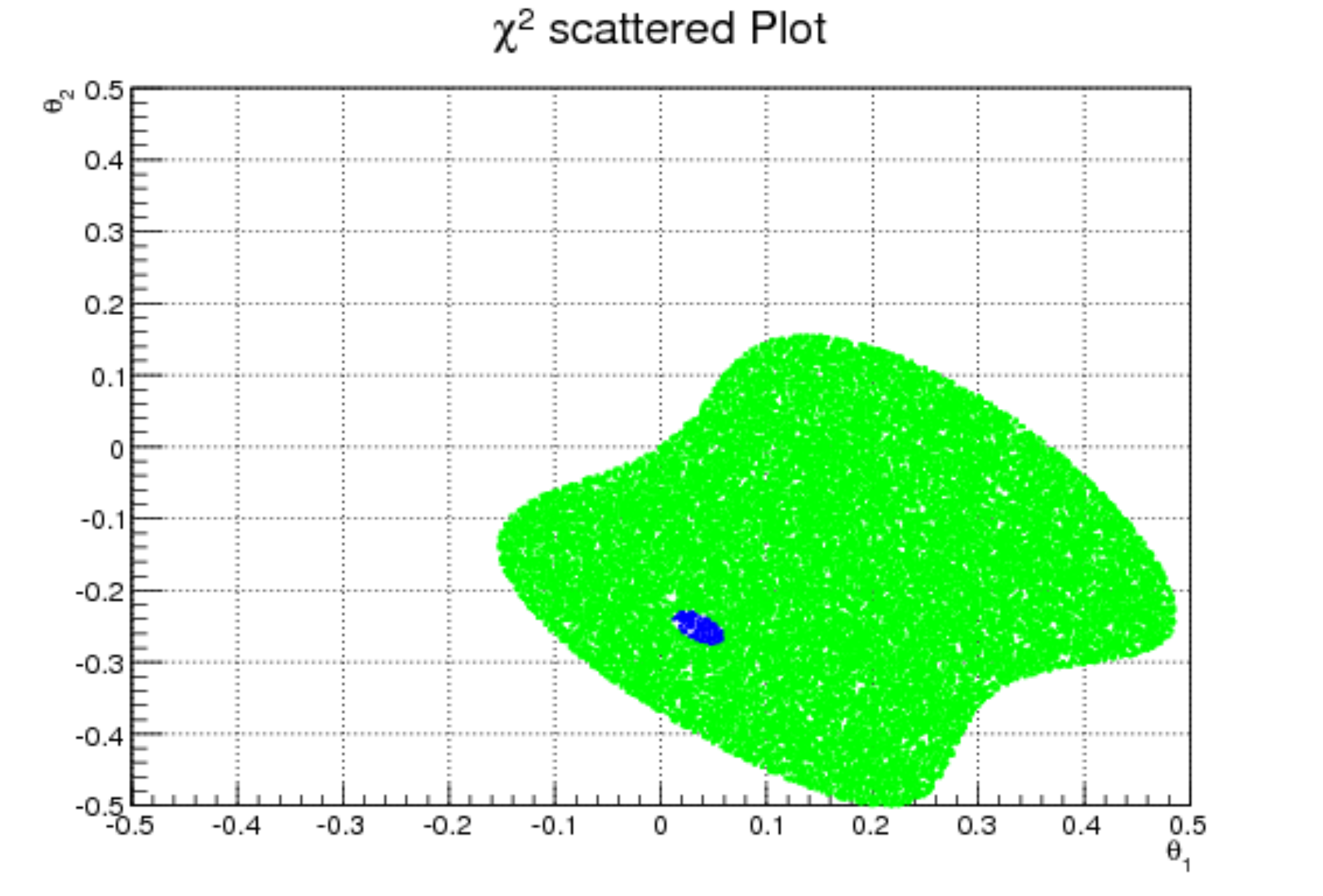} &
\includegraphics[angle=0,width=80mm]{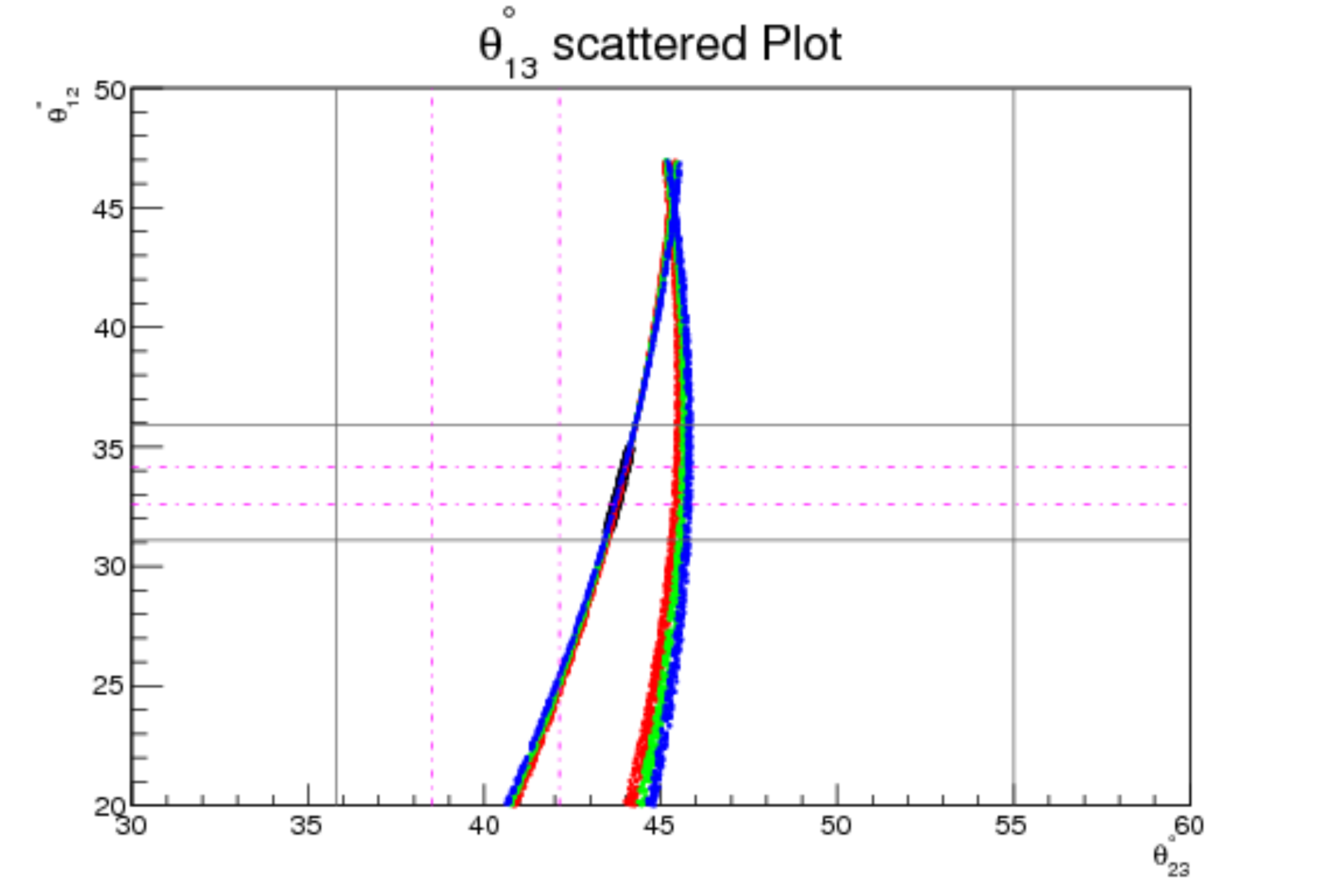}\\
\end{tabular}
\caption{$U_{BM}^{1312}(= R_{13}\cdot R_{12}\cdot U_{BM})$ scatter plot of $\chi^2$ (left fig.) over $\gamma-\alpha$(in radians) plane and $\theta_{13}$ (right fig.) 
over  $\theta_{23}-\theta_{12}$ (in degrees) plane. The information about color coding and various horizontal, vertical lines in right fig. is given in text.}
\label{fig.17}
\end{figure}

\begin{figure}[!t]\centering
\begin{tabular}{c c} 
\includegraphics[angle=0,width=80mm]{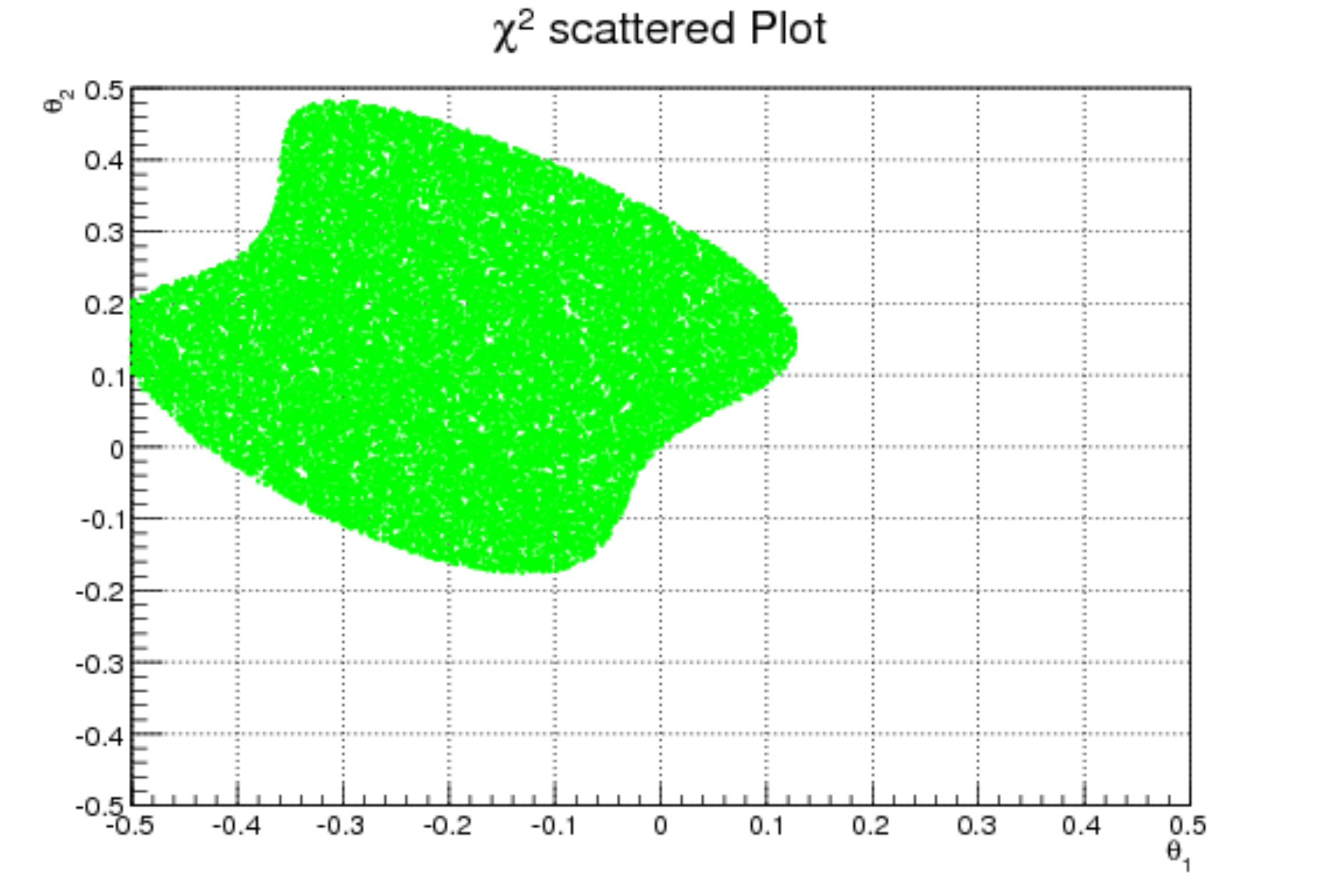} &
\includegraphics[angle=0,width=80mm]{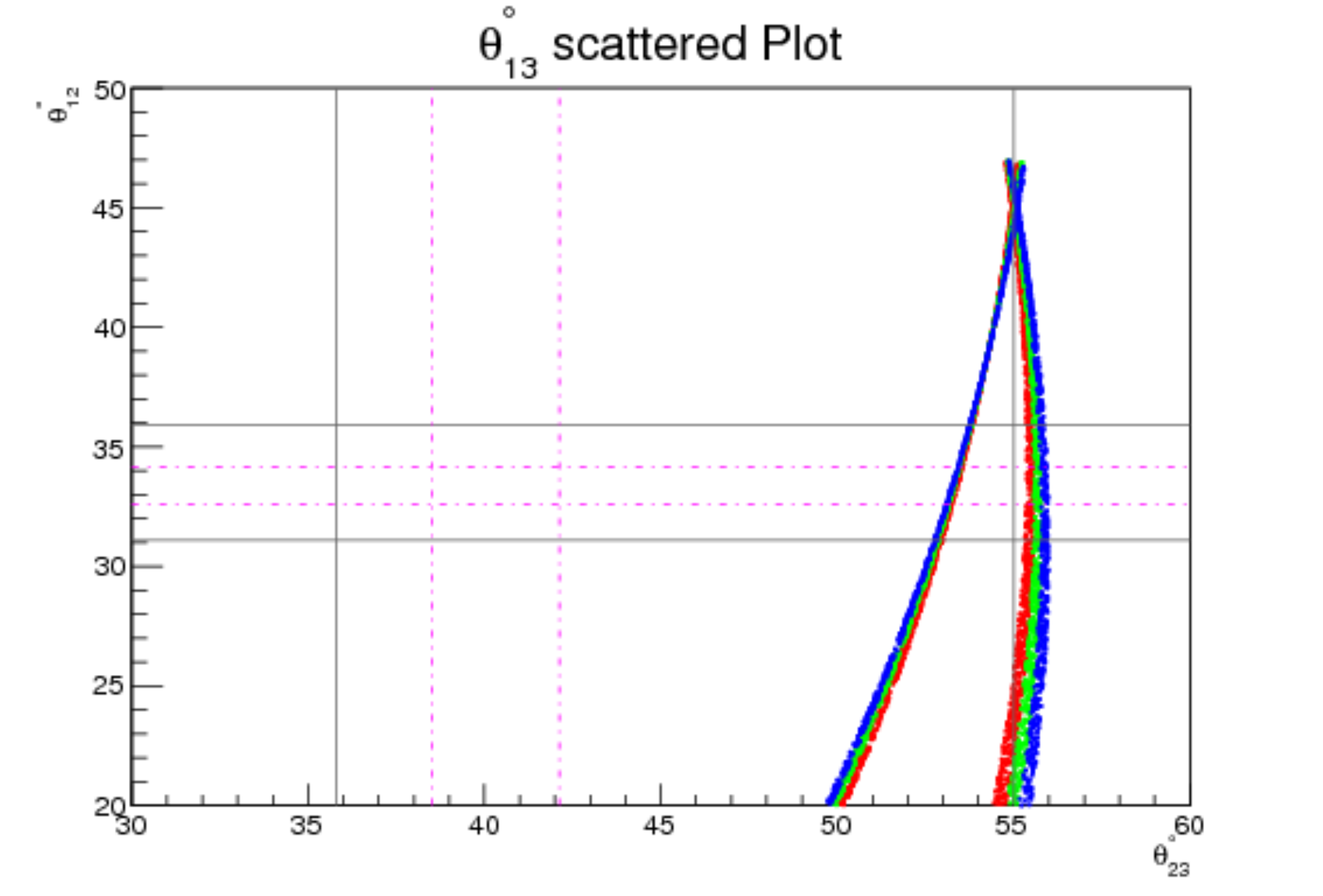}\\
\end{tabular}
\caption{$U_{DC}^{1312}(= R_{13}\cdot R_{12}\cdot U_{DC})$ scatter plot of $\chi^2$ (left fig.) over $\gamma-\alpha$(in radians) plane and $\theta_{13}$ (right fig.) 
over  $\theta_{23}-\theta_{12}$ (in degrees) plane. The information about color coding and various horizontal, vertical lines in right fig. is given in text.}
\label{fig.18}
\end{figure}

Figs.~\ref{fig.16}-\ref{fig.18} corresponds to TBM, BM and DC case respectively with $\theta_1 =  \gamma$ and $\theta_2 =\alpha$. Since the corrections
are of $O(\theta^2)$ in $\theta_{23}$ so its value remain close to that obtained from unperturbed PMNS case. The viable region prefers larger value of reactor mixing
angle($\theta_{13}^\circ$) with BM and TBM rotations can be consistent at $2\sigma$ level while DC case is only allowed at $3\sigma$ level.

\subsection{23-12 Rotation}

This case corresponds to rotations in 23 and 12 sector for BM mixing and the modified PMNS matrix is given by $U_{PMNS} = R_{23}.R_{12}.U_{BM}$. 
The neutrino mixing angles for small perturbation parameters $\alpha$ and $\beta$ are given by

\beqa
 \sin\theta_{13} &\approx&  |\alpha U_{23}|,\\
  \sin\theta_{23} &\approx& |\frac{U_{23} + \beta U_{33} -(\alpha^2 +\beta^2) U_{23} }{\cos\theta_{13}}|,\\
  \sin\theta_{12} &\approx& |\frac{ U_{12} + \alpha U_{22}  -\alpha^2  U_{12}}{\cos\theta_{13}}|.
\eeqa

\begin{figure}[!t]\centering
\begin{tabular}{c c} 
\includegraphics[angle=0,width=80mm]{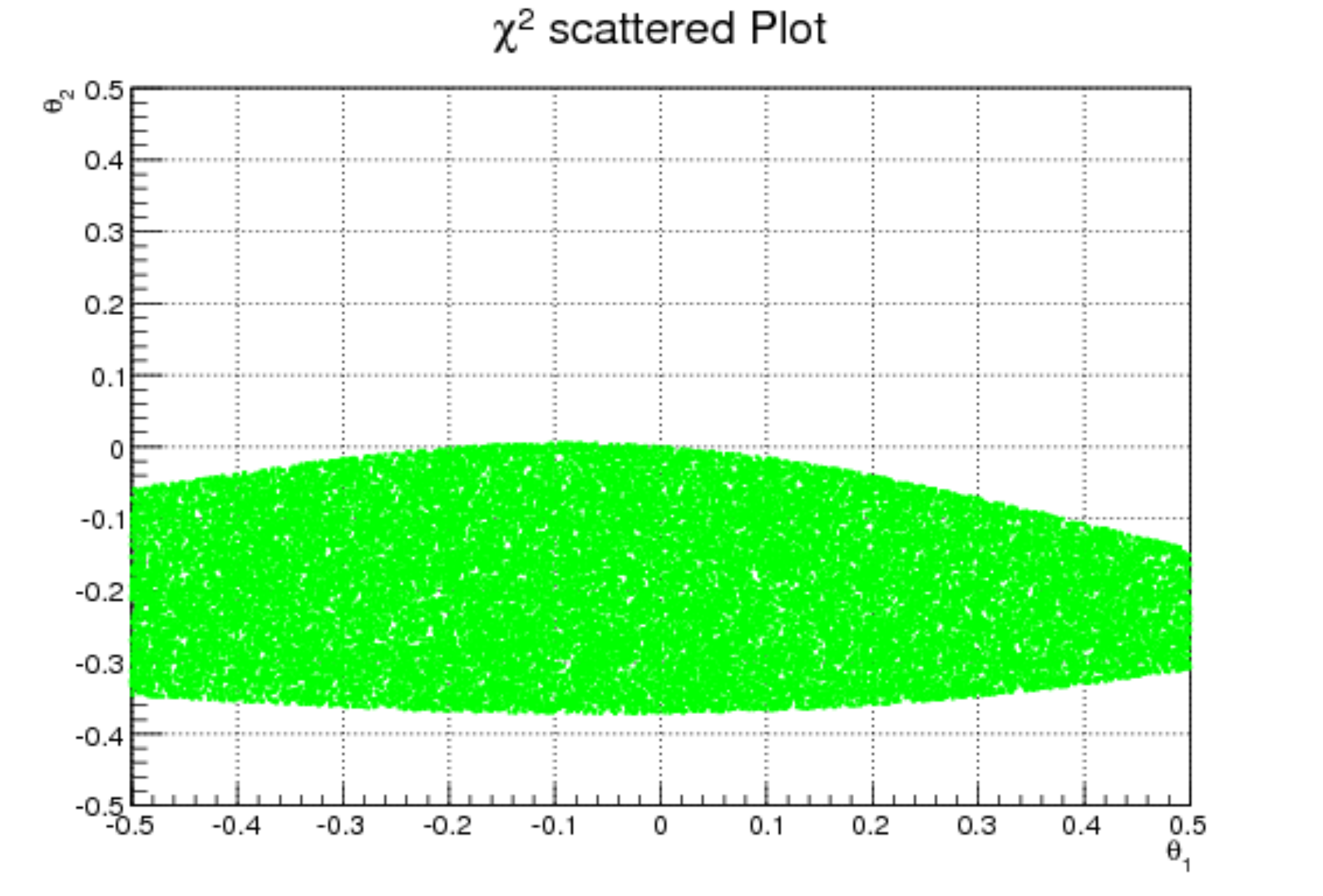} &
\includegraphics[angle=0,width=80mm]{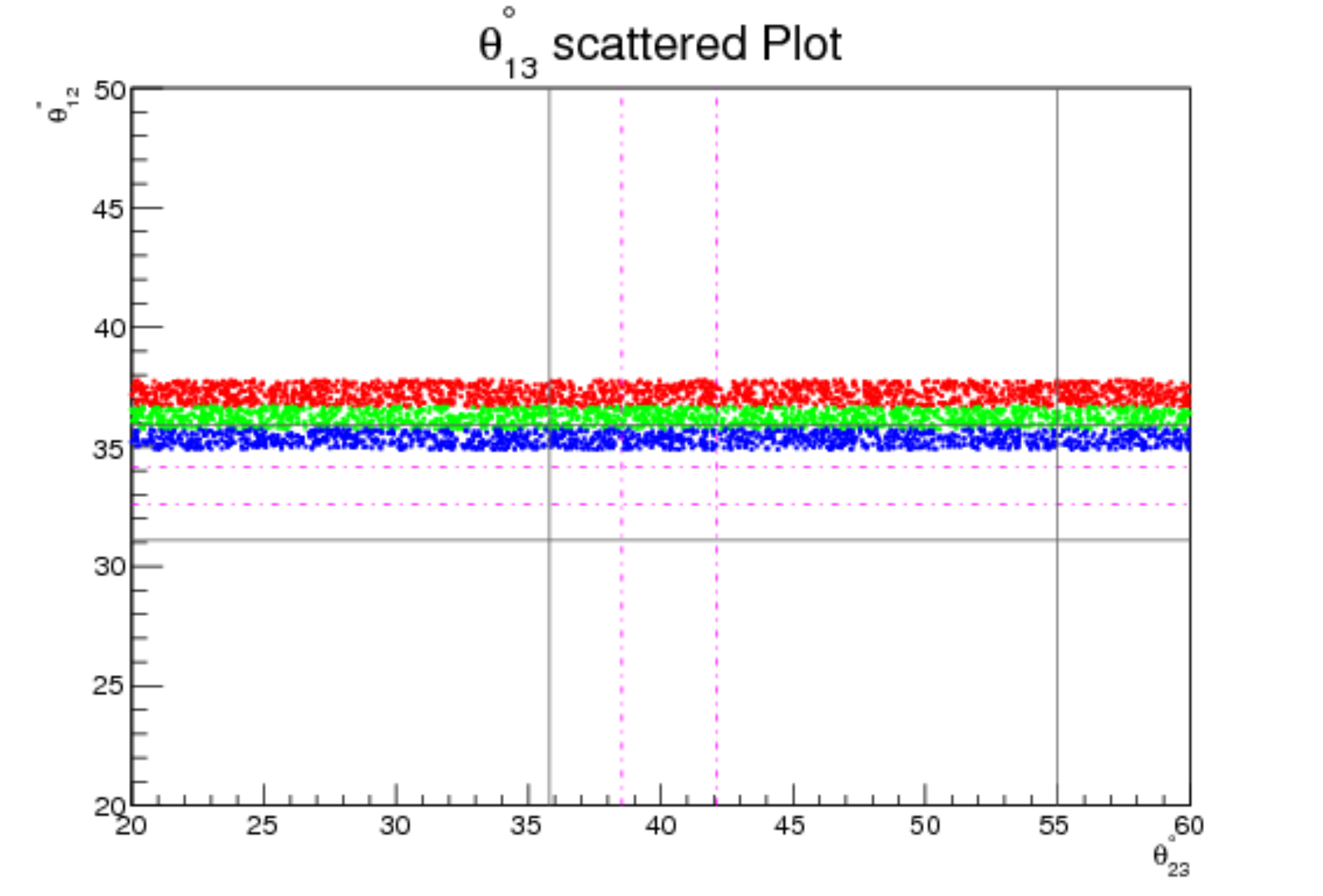}\\
\end{tabular}
\caption{$U_{BM}^{2312} (= R_{23}.R_{12}.U_{BM})$ scatter plot of $\chi^2$ (left fig.) over $\gamma-\alpha$ (in radians) plane and $\theta_{13}$ (right fig.) 
over  $\theta_{23}-\theta_{12}$ (in degrees) plane. The information about color coding and various horizontal, vertical lines in right fig. is given in text.}
\label{fig.19}
\end{figure}

Fig.~\ref{fig.19} corresponds to BM with $\theta_1 =  \beta$ and $\theta_2 =\alpha$. For fixing $\theta_{13}$
in 3$\sigma$ range $|\alpha| \in [0.176, 0.245]$, however since for BM mixing scheme  $\theta_{12} = 45^\circ$ so only negative
values of $\alpha \in [-0.245, -0.221]$ are preferred for fitting both angles. The larger values of $\alpha(> -0.221)$ are not allowed
since it pushes $\theta_{12}$ outside its 3$\sigma$ band which in turn constrain  $\theta_{13} \in [9.15^\circ, 9.97^\circ]$. For $\theta_{23}$ all values are allowed as it gets
corrections exclusively from parameter $\beta$. Thus this case is still allowed at $3\sigma$ level and prefers larger
value of $\theta_{13}$.

\section{Summary and Conclusion}

The non zero value of reactor mixing angle ($\theta_{13}\approx 9^\circ$) and departure of atmospheric mixing
angle ($\theta_{23}\approx 40^\circ(50^\circ)$) from maximality is asking for corrections in well studied mixing schemes.
In this work we discussed the perturbations around TBM, BM and DC mixing scenarios. These modifications are expressed in terms of 
three orthogonal rotation matrices R$_{12}$, R$_{13}$ and R$_{23}$ which acts on 12, 13 and 23 sector of unperturbed PMNS matrix respectively. We looked
into various possible cases that can be parametrized in terms of  corresponding modified PMNS matrix as
$R_{ij}\cdot U \cdot R_{kl}$ where U is any one of these special matrices. As the form of PMNS matrix is given 
by $U_{PMNS} = U_l^{\dagger} U_\nu$ so these corrections may originate from both charged lepton and neutrino
sector. In this study we restricted ourselves to CP conserving case by setting phases to be zero. The effects of Dirac CP violation on mixing angles will be 
reported somewhere else. For our analysis we constructed $\chi^2$ function which is a measure of deviation
from experimental best fit values of mixing angles. The numerical results for different rotations are presented in terms of $\chi^2$ vs perturbation parameters and as correlations 
among neutrino mixing angles.

The rotation $R_{12}\cdot U \cdot R_{13}$  is successful in fitting all mixing angles at 1$\sigma$ level for BM case.  
However for TBM and DC mixing scheme this rotation is favorable at 2$\sigma$ level. For $R_{12}\cdot U \cdot R_{23}$ perturbation, DC case is excluded while 
TBM and BM case are only allowed at 3$\sigma$ level of confidence.
The rotation $R_{13}\cdot U \cdot R_{12}$ introduces negligible corrections for angle $\theta_{23}$.  Since TBM and BM mixing schemes predict 
$\theta_{23}=45^\circ$ so these cases
are still allowed at 3$\sigma$ level while DC predicts $\theta_{12}= 54.7^\circ$ so its excluded. The rotation $R_{13}\cdot U \cdot R_{23}$ is most interesting 
as for BM mixing scheme it completely covers all 1$\sigma$ region and thus most favorable while TBM and DC case are only preferable at 3$\sigma$ level of confidence.
The case of rotation $R_{23}\cdot U \cdot R_{12}$ doesn't impart any corrections to reactor mixing angle i.e. $\theta_{13}=0$ even after
perturbations. Thus this case is completely excluded from the picture. Finally the rotation $R_{23}\cdot U \cdot R_{13}$ impart very
negligible corrections for solar angle $\theta_{12}$. Since TBM and DC mixing scenario predict $\theta_{12}=45^\circ$ so these cases
are excluded while BM predicts $\theta_{12}= 33.3^\circ$ so this is allowed at $3\sigma$ level of confidence. As far as the rotations for which
$ij=kl$ is concerned, TBM is the most preferred case. The rotation $R_{12}\cdot U \cdot R_{12}$ imparts negligible corrections to $\theta_{23}$ mixing angle
and thus all cases are only viable at $3\sigma$ level. The rotation  $R_{13}\cdot U \cdot R_{13}$
is highly preferable for TBM case as it can fit the mixing data at 1$\sigma$ level while BM and DC case are only allowed at 3$\sigma$ level.
Finally the rotation scheme $R_{23}\cdot U \cdot R_{23}$ produces negligible corrections for $\theta_{12}$ and thus only TBM case is allowed
at 2$\sigma$ level while other cases are disallowed at 3$\sigma$ level. Here we also investigated few previously considered rotations \cite{Chaoetal} which 
were reported to be either excluded or highly unfavorable with one of the mixing angles fixed to its central value.
We showed when all three angles vary in their permissible limits then these cases are still viable at 2-3$\sigma$ level. 

Thus to conclude the rotations $R_{13}\cdot U \cdot R_{23}$, $R_{12}\cdot U \cdot R_{13}$  for BM case and $R_{13}\cdot U \cdot R_{13}$ for TBM case are the most 
preferable among this form of perturbed PMNS as they successfully fits  all neutrino mixing angles within $1\sigma$ range.
Among various other investigated cases, some are excluded while others are only preferable at 2-3$\sigma$ level of confidence. This information about 
the perturbative rotations to get mixing schemes in favorable region of oscillation data can be a guideline for neutrino model building. 
Such type of studies can be helpful in restricting the large number of possible models in literature
and can help in selecting some concrete group which can lead to such mixing scenarios. All such issues including the origin of these perturbations
we left for future consideration.

\bigskip

\noindent {\bf{Acknowledgements}} \\
We are grateful to C S Kim for encouraging discussions, suggestions and for carefully reading the manuscript. 
We are also indebted to anonymous JHEP referee for pointing us some rotation cases which we missed in the earlier
version of this draft. The work of SKG and SG is supported by the National Research Foundation
of Korea(NRF) grant funded by Korea government of the Ministry of Education,
Science and Technology(MEST)(Grant No. 2011-0017430 and Grant No. 2011-0020333).

\end{document}